\documentclass[12pt]{article}

\AtBeginDocument{
  \addtocontents{toc}{\normalsize}
}
\pdfoutput=1

\usepackage{amsmath,amssymb,amsfonts,amsthm,amscd,mathrsfs}
\usepackage{xcolor}
\definecolor{darkblue}{rgb}{0.1,0.1,.7}
\usepackage[]{version}
\usepackage{float}
\usepackage[]{graphicx}
\usepackage[]{latexsym}
\usepackage{geometry}
\usepackage{subcaption}
\usepackage[utf8]{inputenc}
\usepackage[all,cmtip]{xy}

\usepackage{ifthen}
\usepackage{tikz}
\usepackage{array,setspace,yfonts,dsfont,bbm,colonequals}
\usepackage[colorlinks, linkcolor=darkblue, citecolor=darkblue, urlcolor=darkblue, linktocpage]{hyperref} 

\usepackage{cite}
\usepackage{xspace}
\usepackage{empheq}
\usepackage{xcolor}

\numberwithin{equation}{section}
\newcommand{\ba}{\begin{equation}\begin{aligned}}
\newcommand{\ea}{\end{aligned}\end{equation}}

\newcommand{\cO}{\mathcal O}

\newcommand{\reef}[1]{(\ref{#1})}
\newcommand{\be}{\begin{equation}}
\newcommand{\ee}{\end{equation}}

\newcommand{\bea}{\begin{equation}\begin{aligned}}
\newcommand{\eea}{\end{aligned}\end{equation}}

\newcommand{\ud}{\mathrm d}

\newcommand{\Df}{{\Delta_\phi}}

\begin{document}

\vspace*{-.6in} \thispagestyle{empty}
\begin{flushright}
\end{flushright}
\vspace{1cm} {\Large
\begin{center}
{\bf Analytic Functional Bootstrap for CFTs in $d>1$}\\
\end{center}}
\vspace{1cm}
\begin{center}
{Miguel F.~Paulos}\\[1cm] 
{
\small
{\em Laboratoire de Physique de l'\'Ecole Normale Sup\'erieure\\ PSL University, CNRS, Sorbonne Universit\'es, UPMC Univ. Paris 06\\ 24 rue Lhomond, 75231 Paris Cedex 05, France}
}\normalsize
\\
\end{center}

\begin{center}
	{\texttt{miguel.paulos@ens.fr}
	}
	\\
\end{center}

\vspace{4mm}

\begin{abstract}
We introduce analytic functionals which act on the crossing equation for CFTs in arbitrary spacetime dimension. The functionals fully probe the constraints of crossing symmetry on the first sheet, and are in particular sensitive to the OPE, (double) lightcone and Regge limits. Compatibility with the crossing equation imposes constraints on the functional kernels which we study in detail. We then introduce two simple classes of functionals. The first class has a simple action on generalized free fields and their deformations and can be used to bootstrap AdS contact interactions in general dimension. The second class is obtained by tensoring holomorphic and antiholomorphic copies of $d=1$ functionals which have been considered recently. They are dual to simple solutions to crossing in $d=2$ which include the energy correlator of the Ising model. We show how these functionals lead to optimal bounds on the OPE density of $d=2$ CFTs and argue that they provide an equivalent rewriting of the $d=2$ crossing equation which is better suited for numeric computations than current approaches.

\end{abstract}
\vspace{2in}


\newpage

{
\setlength{\parskip}{0.05in}
\tableofcontents
\renewcommand{\baselinestretch}{1.0}\normalsize
}


\setlength{\parskip}{0.1in}
\newpage

\section{Introduction}\label{sec:introduction}
In the last decade, a ruthless siege of the crossing equation has yielded a number of detailed insights into the structure of conformal field theories (CFTs) in general spacetime dimension~$d$.
The surprising observations that even simple proddings of crossing  equations~\cite{Rattazzi:2008pe,Hellerman2011}  yield strong constraints on CFT spectra have since led to the development of numerical (see \cite{Poland:2018epd} for an extensive, but still far from complete review) and analytical methods to bound and determine CFT data. An incomplete list of the latter includes applications of Tauberian theory \cite{Pappadopulo:2012jk,Mukhametzhanov:2019pzy,Mukhametzhanov:2018zja,Qiao:2017xif,Pal2019a}, large spin expansions and systematics \cite{Fitzpatrick2014,Komargodski2013,Collier2019a,Alday:2015ewa,Alday:2016njk,Caron-Huot:2017vep,Simmons-Duffin:2016wlq,Simmons-Duffin:2017nub} and the Polyakov bootstrap \cite{Polyakov:1974gs,Sen:2015doa,Gopakumar:2016wkt,Gopakumar:2016cpb,Gopakumar:2018xqi}.

In recent work \cite{Mazac2019a}, it has been proposed that it is possible to unify at least some of these approaches in a single framework.\footnote{For precursors to that work see \cite{Mazac:2016qev,Mazac:2018}. See also  \cite{Kaviraj2018,MazacOPE2018,Mazac:2018biw,Paulos:2019fkw,Hartman2019b} for related developments.} The main idea is to consider appropriate complete bases of linear {\em functionals} which act on the crossing equation. The functionals have access to all the information that is contained in this equation, and in particular they treat the Euclidean and Lorentzian regimes equally. The guiding principle which distinguishes these bases is that they should be dual (in a precise sense) to sparse or extremal solutions to crossing~\cite{ElShowk:2012hu,Gliozzi2013,El-Showk:2016mxr}. Extremality of functionals not only leads to tight bounds on CFT data, but also implies that it is possible to ``flow'' from one extremal solution to another, numerically and analytically.

The main achievement of \cite{Mazac2019a} was to recast the constraints of crossing symmetry on a line (or for $d=1$ CFTs) into an equivalent form:
\bea
\sum_{\Delta} a_{\Delta} F_{\Delta}(z)=0\qquad &\mbox{for all}\quad z\in(0,1)\\&\Leftrightarrow \\ \sum_{\Delta} a_{\Delta} \alpha_{n}(\Delta)=0,\qquad \sum_{\Delta} a_{\Delta} \beta_{n}(\Delta)&=0,\quad \text{for all}\ n \in \mathbb N_{\geq 0}\,.\label{eq:completeness1d}
\eea
The first line is the ordinary formulation of crossing symmetry for a correlator of identical operators. It defines a continuous set of constraints, labeled by $z$, on the OPE density $a_{\Delta}$, with $F_{\Delta}(z)$ a known function determined in terms of $SL(2,\mathbb R)$ conformal blocks. The second formulation can be obtained from the first by acting with a countably infinite set of linear functionals $\alpha_n, \beta_n$ on the crossing equation. Choosing these functionals appropriately leads to not only necessary but also sufficient conditions for crossing to hold. In this sense we say that such functionals form a complete basis, and the resulting set of {\em functional bootstrap equations} is then equivalent to the original crossing equation.

Following our guiding principle, it turns out that we can choose functionals to be dual to a particular sparse solution to crossing in $d=1$, namely the fundamental field correlator of a generalized free field (GFF). In this case, duality means essentially that the functionals bootstrap the generalized free solution, as well as arbitrary small deformations away from it, see equations \reef{eq:ONboson} and \reef{eq:ONfermion} below. These duality properties lead in turn to nice positivity properties of the functional actions. The functional bootstrap equations become sum rules which determine stringent bounds on the OPE density of unitary theories. This can be done analytically but also numerically. In the latter approach, by considering finite subsets of the functional bootstrap equations, we can determine rapidly convergent numerical bounds \cite{Paulos:2019fkw}, making contact with and (finally) improving on the original methodology of \cite{Rattazzi:2008pe}.

Furthermore, the functional actions $\alpha_n(\Delta)$, $\beta_n(\Delta)$ turn out to compute the conformal block expansion of (crossing-symmetric sums of) Witten exchange diagrams. The functional bootstrap equations are then essentially the same as those proposed in the so-called Polyakov bootstrap \cite{Sen:2015doa,Gopakumar:2016wkt,Gopakumar:2016cpb,Gopakumar:2018xqi}. More precisely, they rigorously define what is meant by Polyakov bootstrap, at least in $d=1$. Thus, these functionals unite in a single framework analytic and numerical bounds, and the Polyakov bootstrap, which is often used for exact computations of CFT data. In $d=1$ there is no spin, so we cannot really make contact with the large spin methods which exist for $d\geq 2$. Nevertheless, we point out that the functionals can also be understood as arising from a crossing-symmetric $d=1$ version \cite{MazacOPE2018} of the Lorentzian OPE inversion formula of Caron-Huot \cite{Caron-Huot:2017vep}. It remains to be understood if and how Tauberian methods can also be understood from a functional perspective.

These results are encouraging, but the restriction to the line is severe and one may well wonder if similar constructions are possible for the significantly more complicated crossing equation in higher dimensions. To be clear, we wish to know if complete bases of functionals can be constructed for this equation, if they are dual to special solutions to crossing and if they lead to tight bounds on CFT data. The goal of the present work is to make the case that all these statements are true, while creating the appropriate formalism along the way. Specifically, we will propose a general framework for functionals which act on the full crossing equation. The functionals probe the Euclidean OPE, (double) lightcone and Regge limits, and are constrained by the properties of CFT four-point functions in those limits. After studying these constraints in detail, we then introduce two simple classes of interesting functionals which satisfy them. One of these, we claim, has the desired properties in the special case of $d=2$.

The first class, which we name HPPS functionals in honor of \cite{Heemskerk2009}, is defined by simple meromorphic functional kernels. In spite of this simplicity, the resulting functional actions are very interesting: they have finite support on the generalized free field spectrum, and furthermore are non-negative above a certain twist for each spin channel. This means that such functionals can be used to bootstrap generalized free fields in any spacetime dimension.\footnote{Up to some important caveats as we shall see, namely that this functional set is not complete.}  Furthermore, just as in \cite{Heemskerk2009}, they can also be used to bootstrap certain deformations away from generalized free fields exactly, namely those that arise from considering contact interactions in AdS space. Unfortunately however, they are not suitable for bootstrapping deformations of generalized free fields involving an infinite number of spins. Hence, they come close but are not quite sufficient to define the higher dimensional version of the Polyakov bootstrap.

The second class is obtained by tensoring holomorphic and antiholomorphic copies of 1d functionals, some of which have appeared in the literature \cite{Mazac:2018,Mazac2019a,Mazac:2016qev,Hartman2019b} (including the ones mentioned above), others which we construct in detail. These product functionals exist in $d=2$ thanks to the special form that conformal blocks take in that dimension.\footnote{A similar construction can be done, less usefully, in $d=4$ for the same reasons, see appendix \ref{app:4dprod}.} The functionals are dual to a simple class of solutions to crossing symmetry which includes the energy correlator of the 2d Ising model. The functional basis again has nice positivity properties following from this duality, and leads to a rigorous upper bound on the OPE density for any 2d CFT. Numerical explorations give stringent bound on operator dimensions, which indicates that the basis is complete, although we will not rigorously prove it. In particular, we show that one special functional basis element gives an optimal scaling dimension bound which is saturated by the 2d Ising model correlator mentioned above.

The plan for this work is as follows:

In section \ref{sec:review}, we will briefly summarize what is known about 1d functional bases, leaving a detailed review and construction of such functionals to appendix \ref{sec:bases}. These bases will be used when considering product functionals in $d=2$. Furthermore, they give us a flavor of how to define general functionals for the higher dimensional case.

In section \ref{sec:ansatz} we introduce our general functional ansatz and the associated kernels. Demanding that the functionals are well defined on the crossing equation constrains the functional kernels in a way that is determined by the behaviour of correlation functions in various limits, notably the double lightcone limit which we will consider in some detail. This leads to a set of boundary conditions on the functional kernels. Several details of the computation are left for appendix \ref{app:constraints}. 

In section \ref{sec:hpps} we introduce one of the simplest possible class of functionals which has nice positivity properties, the HPPS functionals. We show that choosing functional kernels to be simple meromorphic functions leads to functional actions which are generically positive and have double zeros on the generalized free field spectrum. An important result is that these functionals do not form a complete set, in the sense that they do not fully capture the constraints of crossing symmetry. Nevertheless, they do constrain possible solutions to crossing enormously, and in particular are sufficient to bootstrap general contact interactions in AdS for any dimension. We show how the infinity of such solutions to crossing is compatible with the functional equations by understanding the interplay between the Regge behaviour of contact interactions and boundedness conditions on the functional kernels.

Section \ref{sec:d2prod} defines a class of functionals obtained by tensoring two copies of $d=1$ functionals, suitable for acting on the $d=2$ crossing equation. The resulting functionals can be used to bootstrap certain tensor product correlators, which include the energy four-point function in the 2d Ising model. The functionals imply bounds on scaling dimensions of operators, and in particular an optimal bound which is saturated by that correlator. More generally they can be used to determine numerical bounds, which we compare against the more traditional derivative functional basis, finding substantial improvements at least in some regimes. Finally, we show that the product functional basis leads to a universal upper bound on the fixed spin OPE density which is exactly saturated by the tensor product solutions (and hence optimal in those cases).

After a discussion in section \ref{sec:discussion} of the lessons learned in this work and future prospects for the functional bootstrap, a few appendices follow. In appendix \ref{sec:bases} we provide a detailed construction of the 1d functional bases, as well as a study of special cases and asymptotics of functional actions. Appendix \ref{app:constraints} provides details on the determination of the constraints on the higher dimensional functional ansatz. In appendix \ref{sec:1duplift} we show how 1d functionals may be made compatible with such ansatz. Appendix \ref{app:hpps} explain how the HPPS functionals that we introduce relate to the original work \cite{Heemskerk2009}. Finally we comment on product functionals for $d=4$ in appendix \ref{app:4dprod}.

\section{Bases of 1d functionals}
\label{sec:review}

In this section we consider functionals appropriate for the simplified setting of a CFT on a line, or equivalently where we take into account only an $SL(2,\mathbb R)$ subgroup of the full conformal group. These 1d functionals form a distinguished subset of general functionals which act on the full crossing equation. Below we will present complete bases of such functionals, where completeness means that they fully capture the constraints of crossing symmetry on the line. Some of these bases have already appeared in print, while the existence of others was only indicated \cite{Mazac:2018,Mazac2019a,Mazac:2016qev,Hartman2019b}. In any case, all of them are reviewed and/or constructed (accordingly) in appendix \ref{sec:bases}, to which we refer the reader for further details.

\subsection{Basic kinematics}
In one dimension a CFT four-point function is a piecewise real-analytic function on the three intervals $(-\infty,0)$, $(0,1)$ and $(1,\infty)$, which depends on a single cross-ratio $z$:
\bea
\langle \phi(x_1)\phi(x_2)\phi(x_3)\phi(x_4)\rangle=\frac{\mathcal G(z)}{x_{13}^{2\Df} x_{24}^{2\Df}},\qquad z:=\frac{x_{12}x_{34}}{x_{13}x_{24}}\,\qquad x_{ij}:=x_i-x_j\,.
\eea
For $z\in (0,1)$ we can express the correlator in terms of $SL(2,\mathbb R)$ conformal blocks as:
\bea
\mathcal G(z)=\sum_{\Delta} a_{\Delta} G_{\Delta}(z|\Df),\qquad G_{\Delta}(z|\Df):=z^{\Delta-2\Df}\, _2F_1(\Delta,\Delta,2\Delta,z)\,.\label{eq:sl2block}
\eea
One should think of the sum over scaling dimensions $\Delta$ as a sum over operators $\cO_{\Delta}$ appearing in the OPE $\phi\times \phi$. The $a_{\Delta}$ correspond then to values of the OPE coefficients squared and are positive for unitary theories. Crossing symmetry of the four-point function is the statement that $\mathcal G(z)=\mathcal G(1-z)$. Let us define the {\em crossing vectors}:
\bea
F_{\pm,\Delta}(z|\Df):=G_{\Delta}(z|\Df)\pm G_{\Delta}(1-z|\Df)\,,
\eea
in terms of which crossing becomes:%
\bea
\sum_{\Delta} a_{\Delta} F_{-,\Delta}(z)=0 \label{eq:crossing1d}\,.
\eea
The $F_{+,\Delta}$ functions appear in more general crossing equations, for instance in the context of correlators with global symmetries.

A simple but important example of a four-point correlator satisfying crossing corresponds to generalized free fields, for which we have:
\bea
\mathcal G(z)=\pm 1+\frac{1}{z^{2\Df}}+\frac{1}{(1-z)^{2\Df}}\,,\qquad z\in (0,1)\,,
\eea
where the $+(-)$ sign corresponds to a free boson/fermion respectively. In this case the spectrum appearing in the four point function includes the identity operator with $\Delta=0$, as well as ``double-trace'' operators with dimensions $\Delta_n^B=2\Df+2n$ and $\Delta_n^F=1+2\Df+2n$ for a boson and fermion respectively. The corresponding OPE coefficients are $a_{\Delta_n^{B,F}}^{\mbox{\tiny gff}}$ with
\bea
a_{\Delta}^{\mbox{\tiny gff}}=\frac{2 \Gamma(\Delta)^2}{\Gamma(2\Delta-1)}\, \frac{\Gamma(\Delta+2\Df-1)}{\Gamma(2\Df)^2\Gamma(\Delta-2\Df+1)} \label{eq:ope1d}\,.
\eea

\subsection{Functionals}
We want to consider linear functionals $\omega_{\pm}$, which act on $F_{\pm,\Delta}$ and which are {\em crossing-compatible}. By this we mean that their action should commute with crossing equations, 
\bea
\omega_{\pm}\left[\sum_{\Delta} a_{\Delta} F_{\pm,\Delta}(z|\Df)\right]=\sum_{\Delta} a_{\Delta} \omega_{\pm}(\Delta|\Df)\,,
\eea
where we defined the shorthand
\bea
\omega_{\pm}(\Delta|\Df):=\omega_{\pm}\left[F_{\pm,\Delta}(z|\Df)\right]\,.
\eea
Using the analyticity properties of the blocks, a very general class of functionals is defined by:
\bea
\omega_{\pm}(\Delta|\Df)=\int_1^{\infty} \frac{\ud z}{\pi} h_{\pm}(z)\, \mathcal I_z F_{\pm,\Delta}(z|\Df)
\eea
with the discontinuity defined as:
\bea
\mathcal I_z F(z):=\lim_{\epsilon \to 0^+} \frac{F(z+i\epsilon)-F(z-i\epsilon)}{2i}.\label{eq:discdef}
\eea
By making a suitable choice of kernels $h_{\pm}$, it is possible to construct two sets of functionals which are dual to the 1d bosonic and fermionic generalized free field solutions to crossing, $\omega_\pm=\alpha_{\pm,n}^{B,F},\beta_{\pm,n}^{B,F}$. Duality means the following relations:
\bea
\alpha_{\pm ,n}^{F}(\Delta_m^F)&=\delta_{nm},&\partial \alpha_{\pm,n}^{F}(\Delta_m^F)&=0,\\
\beta_{\pm ,n}^{F}(\Delta_m^F)&=0,&\partial \beta_{\pm,n}^{F}(\Delta_m^F)&=\delta_{nm} \label{eq:ONboson}
\eea
for the fermionic basis, and
\bea
\alpha_{\pm ,n}^{B}(\Delta_m^B)&=\delta_{nm},&\partial \alpha_{\pm,n}^{F}(\Delta_m^B)&=-c_{\pm,n} \delta_{m0},\\
\beta_{\pm ,n}^{B}(\Delta_m^B)&=0,&\partial \beta_{\pm,n}^{B}(\Delta_m^B)&=\delta_{nm}-d_{\pm,n} \delta_{m0} \label{eq:ONfermion}
\eea
for the bosonic one, for some coefficients $c_{\pm,n}, d_{\pm,n}$ which may be determined explicitly and $\beta_{\pm,0}^B\equiv 0$. 

These functionals allow for the basis decompositions:
\bea
F_{\pm, \Delta}(z|\Df)=\sum_{n=0}^{+\infty}\left[\alpha_{\pm,n}^F(\Delta) F_{\pm,\Delta_n^F}(z)+\beta_{\pm,n}^F(\Delta) \partial_\Delta F_{\pm,\Delta_n^F}(z) \right] \label{eq:basesdecomp}\,,
\eea
with a similar expression for the bosonic case. In the $(-)$ case, it is natural to plug in this decomposition into the crossing equation \reef{eq:crossing1d} and commute series to arrive at the result:
\bea
\sum_{\Delta} a_{\Delta} F_{-,\Delta}(z)=0\qquad &\mbox{for all}\quad z\in(0,1)\\&\Leftrightarrow \\ \sum_{\Delta} a_{\Delta} \alpha_{-,n}(\Delta)=0,\qquad \sum_{\Delta} a_{\Delta} \beta_{-,n}(\Delta)&=0,\quad \text{for all}\ n \in \mathbb N_{\geq 0}\,.\label{eq:completeness1d2}
\eea
We have omitted the $B,F$ label on the second line since we may choose the one we wish, each set of functionals being complete independently. The statement that the series can be commuted is not trivial, but it can be proved \cite{Mazac2019a}. The set of functional bootstrap equations in the second line are thus a rephrasing of the original crossing equation and must be satisfied by any hypothetical solution to crossing. One can for instance check that the GFF solutions satisfy the equations trivially as long as we use the correct dual functionals. This follows from the duality conditions \reef{eq:ONfermion}, \reef{eq:ONboson} together with
\bea
\beta_{-,n}^{B,F}(0)=0,\qquad \alpha_{-,n}^{B,F}(0)=-a^{\mbox{\tiny gff}}_{\Delta_n^{B,F}}.\label{eq:idaction}
\eea
Similarly, it can be checked (in practice, numerically) that the generalized free fermion (say) also satisfies the equations of the bosonic functional basis.

We should point out that the functional bootstrap equations also provide valid constraints on the OPE density for CFTs in any dimension, by restricting them to the line and decomposing blocks with respect to the $SL(2,\mathbb R)$ subgroup. In this process we lose a lot of information, including information about spin. Remarkably, this does not mean such constraints are weak: in fact they can even be optimal.\footnote{For instance, as pointed out in \cite{Paulos:2019fkw}, the $\beta_0^F$ equation applied to $d=2$ CFTs implies the existence of a spin-0 primary of dimension less than $1+2\Df$ when $\Df<1/2$, a constraint which is saturated (and hence optimal) as $\Df\to 1/2$ by a $c=1$ vertex operator correlator.} However, the main application of these functionals in this work is that they will allow us, by making use of holomorphic factorization, to construct simple tensor product functionals which act on the 2d crossing equation (and less usefully in 4d, see appendix \ref{app:4dprod}).

\section{Functionals in general dimension}
\label{sec:ansatz}
In this section we will propose a general framework for functionals acting on the higher dimensional crossing equation. Our ansatz now involves functional kernels which depend on the two available independent cross-ratios $z,\bar z$. An important simplifying assumption will be that the overall analytic structure is essentially a product, in that analyticity properties in one cross-ratio are independent of the other one. One way to lift this assumption is to allow for poles in one cross-ratio whose position depends on the other one. It would be easy to generalize the discussion in this section to that case. As a simple example, in appendix \ref{sec:1duplift} we show how such poles allow for the 1d functionals of the previous section to be incorporated into the general formalism that will be described below. 

A key condition that functionals must satisfy is that they should be compatible with the crossing equation, i.e. that their action should commute with the infinite sums of crossing vectors involved in that equation. This imposes constraints on the functional kernels appearing in the ansatz, as we will examine in detail. The constraints depend on knowledge of the behaviour of correlators in various limits: OPE, Regge and (double) lightcone. In particular, the latter is not completely understood and we will have to make some reasonable assumptions to make progress.

\subsection{Basics}
Let us quickly review some basic kinematics. We have in mind four-point correlators of scalar operators which depend on two independent cross-ratios $z,\bar z$, as follows:
\bea
\langle \phi(x_1)\phi(x_2)\phi(x_3)\phi(x_4)\rangle=\frac{\mathcal G(z,\bar z)}{x_{13}^{2\Df} x_{24}^{2\Df}},\qquad z \bar z=\frac{x_{12}^2 x_{34}^2}{x_{13}^2 x_{24}^2}\,,\quad (1-z)(1-\bar z)=\frac{x_{14}^2 x_{23}^2}{x_{13}^2 x_{24}^2}\,.
\eea
Such correlators satisfy $s$-channel conformal block decompositions
\bea
\mathcal G(z,\bar z)=\sum_{\Delta,\ell} a_{\Delta,\ell} G_{\Delta,\ell}(z,\bar z|\Df)\,,\qquad a_{\Delta,\ell}=\lambda_{\phi \phi \mathcal O_{\Delta,\ell}}^2\,,
\eea
where the sum runs over primary operators $\mathcal O_{\Delta,\ell}$ in the OPE $\phi \times \phi\sim \sum_{\Delta,\ell} \lambda_{\phi \phi \mathcal O_{\Delta,\ell}} \mathcal O_{\Delta,\ell}$, with dimension $\Delta$ and transforming under the traceless symmetric $O(d)$ irreducible representation of spin $\ell$. Note that we have introduced the useful notation
\bea
G_{\Delta,\ell}(z,\bar z|\Df)=\frac{G_{\Delta,\ell}(z,\bar z)}{z^\Df \bar z^\Df}\,,
\eea
with $G_{\Delta,\ell}(z,\bar z)$ the ordinary conformal block. The conformal blocks themselves are not known in closed form in general spacetime dimension, although many of their properties are, see e.g. \cite{Dolan:2011dv,Dolan:2003hv,Dolan:2000ut}. In particular it is known how to compute them numerically efficiently \cite{Kos:2013tga}, at least when $z,\bar z$ are not too close to unity. A useful result is the lightcone expansion:
\bea
G_{\Delta,\ell}(z,\bar z|\Df)=\sum_{a=0}^{\infty}\sum_{b=\mbox{\tiny max} \{-a,-\ell\}}^{a} c_{a,b} z^{\frac{\tau}2+a-\Df} G_{\frac{\rho}2+b}(\bar z|\Df)\,,\label{eq:lightconeexp}
\eea
where we have introduced the twist $\tau=\Delta-\ell$, the conformal spin $\rho=\Delta+\ell$ and $G_h(z|\Df)$ are the $SL(2,\mathbb R)$ blocks introduced in the previous section. The coefficients $c_{a,b}$ may be computed by using the fact that conformal blocks are eigenfunctions of the Casimir operator of the conformal group \cite{Dolan:2003hv}. The first few coefficients are given by
\bea
\frac{c_{1,-1}}{c_{0,0}}&=\frac{\ell (d-2)}{d+2\ell-4}\,, \qquad
\frac{c_{1,0}}{c_{0,0}}=\frac{\Delta-\ell}4\,,\\
\frac{c_{1,1}}{c_{0,0}}&=\frac{1}{16} \frac{(d-2)(\Delta-1)(\Delta+\ell)^2}{(\Delta+\ell-1)(\Delta+\ell+1)(2+2\Delta-d)}\,,
\eea
and in our normalisation conventions,\footnote{Our choice matches entry 3 of table I in \cite{Poland:2018epd}.}
\bea
c_{0,0}=\frac{2^{d-3} \Gamma\left(\frac{d-1}2\right)\Gamma\left(\frac{d+2\ell-2}2\right)}{\sqrt{\pi}\,\Gamma(d+\ell-2)}\,.
\eea

Crossing symmetry follows from Bose symmetry, implying $\mathcal G(z,\bar z)=\mathcal G(1-z,1-\bar z)$. In conjunction with the OPE decomposition, this leads to the crossing equation:
\bea
\sum_{\Delta,\ell} a_{\Delta,\ell}F_{\Delta,\ell}(z,\bar z|\Df)=0\,,
\eea
where we have introduced the {\em crossing vectors}:
\bea
F_{\Delta,\ell}(z,\bar z|\Df)=G_{\Delta,\ell}(z,\bar z|\Df)-G_{\Delta,\ell}(1-z,1-\bar z|\Df)\,.
\eea
The crossing equation establishes equality of the OPE decompositions of the correlator in the $s$ and $t$-channels, where $x_1\to x_2$ and $x_1\to x_4$ respectively.
 
To conclude this subsection, we point out that in general spacetime dimension the generalized free boson correlator $\langle \phi \phi \phi \phi\rangle$ takes the form:
\bea
\mathcal G(z,\bar z)=1+\frac{1}{[(1-z)(1-\bar z)]^{\Df}}+\frac{1}{(z \bar z)^{\Df}}
\eea
Decomposing into conformal blocks we find the identity operator, together with double trace operators of dimension $\Delta_{n,\ell}=2\Df+2n+\ell$ for all $n\in \mathbb N_{\geq 0}$ and even spins $\ell$. The corresponding OPE coefficients are (in our normalization conventions) \cite{Fitzpatrick2012}:
\bea
&a_{n,\ell}^{\mbox{\tiny gff}}:=a_{\Delta_{n,\ell},\ell}\\
&=\frac{2^{4-d}\sqrt{\pi}}{l!n!}\,\frac{\Gamma\left(d+\ell-2\right) (\Df)_{l+n}^2 \left(1+\Df-\frac d2\right)_n^2}{\Gamma\left(\frac{d-1}2\right)\Gamma(\frac d2+\ell-1)\left(\frac d2+\ell\right)_n \left(1+2\Df-d+n\right)_n \left(2\Df+\ell+2n-1\right)_\ell}\,.\label{eq:gffope}
\eea

\subsection{General ansatz}

Our goal is to construct a class of functionals which act on the crossing equation and in particular on the crossing vectors $F_{\Delta,\ell}(z,\bar z)$. The analytic properties of the crossing vectors  follow from those of the conformal blocks: they are holomorphic in the subset of $\mathbb C^2$ defining the {\em crossing region} $\mathcal R=R\otimes \bar R$, with $R$ and $\bar R$ both given by $\mathbb C\backslash (-\infty,0]\cup [1,\infty)$. This is incidentally the same region where the OPE converges in both $s$ and $t$-channels, and hence the crossing equation.

It will be useful to define:
\bea
\int_-:= \int_{-\infty}^0\,,\qquad \int_0:=\int_0^1\,,\qquad \int_+:=\int_{1}^{\infty}\,\qquad \int_{\partial}:=\int_{-\infty}^0+\int_1^{\infty}\,.
\eea
Consider a function $\mathcal F(z,\bar z)$ with  the same analyticity properties and symmetries as crossing vectors. We can use Cauchy's theorem to write down a dispersion relation:\footnote{Arcs at infinity drop out as long as $\Df>0$, since for fixed $\bar z$ we have $G_{\Delta,\ell}(z,\bar z)=O(z^{\epsilon})$ for any $\epsilon>0$ as $z\to \infty$.}
\bea
\mathcal F(z,\bar z)=\int_\partial \frac{\ud w}\pi\, \frac{\mathcal I_{w} \mathcal F(w,\bar z)}{w-z}
\eea
where the discontinuity has been defined in equation \reef{eq:discdef}.
The analyticity properties of $\mathcal F(z,\bar z)$ in $\bar z$ now imply that $I_z \mathcal F(z,\bar z)$ is also analytic for $\bar z$ in $\mathcal C\backslash \bar R$. For crossing vectors this can also be argued from the fact that discontinuities of conformal blocks must also satisfy the Casimir equation and satisfy the same analyticity properties. In any case, we can write a double dispersion relation:
\bea
\mathcal F(z,\bar z)=\int_\partial \frac{\ud w}\pi\int_\partial\frac{\ud \bar w}\pi\, \frac{\mathcal I_{\bar w} \mathcal I_{w}  \mathcal F(w,\bar w)}{(w-z)(\bar w-\bar z)}\,.
\eea

From this representation it follows that very general linear functionals that act on $\mathcal F(z,\bar z)$ can be defined by integrating the double discontinuities against some kernels. There are a priori four different boundaries to consider and accordingly four different kernels. For crossing vectors however, we have the symmetry $\mathcal F(z,\bar z)=-\mathcal F(1-z,1-\bar z)$, which reduces this down to two. Explicitly, such functionals are given by
\bea
\omega\left[ \mathcal F\right]=\int_{++} \frac{\ud z \ud \bar z}{\pi^2} h_{++}(z,\bar z) \mathcal I_z \mathcal I_{\bar z}\mathcal F(z,\bar z)+\int_{+-}\frac{\ud z \ud \bar z}{\pi^2} h_{+-}(z,\bar z) \mathcal I_z \mathcal I_{\bar z}\mathcal F(z,\bar z)\,.\label{eq:funcactionh}
\eea
We will have in mind kernels $h_{++}, h_{+-}$ which are analytic in the regions of integration. Since we are interested in the action of the functional on crossing vectors, we can demand
\bea
h_{++}(z,\bar z)=h_{++}(\bar z,z),\qquad h_{+-}(z,\bar z)=-h_{+-}(1-\bar z,1-z)\,.
\eea

In the above we made a specific choice in the ordering of the discontinuities, but it is easy to see this is irrelevant. Thinking of the integrals above as contour integrals, it follows that commuting the discontinuities will have a non-trivial effect only if $\mathcal F(z,\bar z)$ has singularities for $z=\bar z$, which is never the case. 

\begin{figure}%
\begin{center}
\begin{tabular}{cc}
\includegraphics[width=8cm]{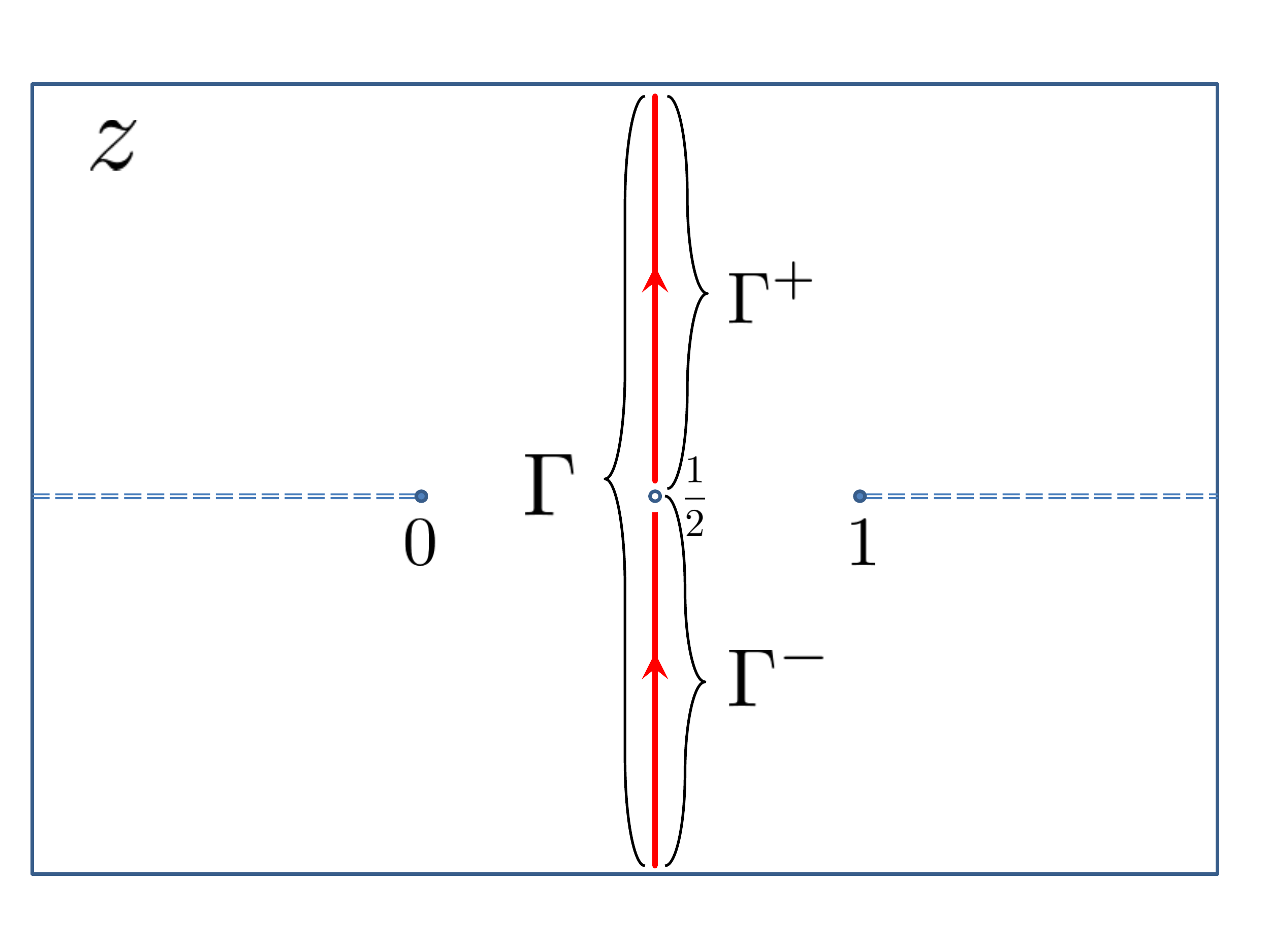}%
&
\includegraphics[width=8cm]{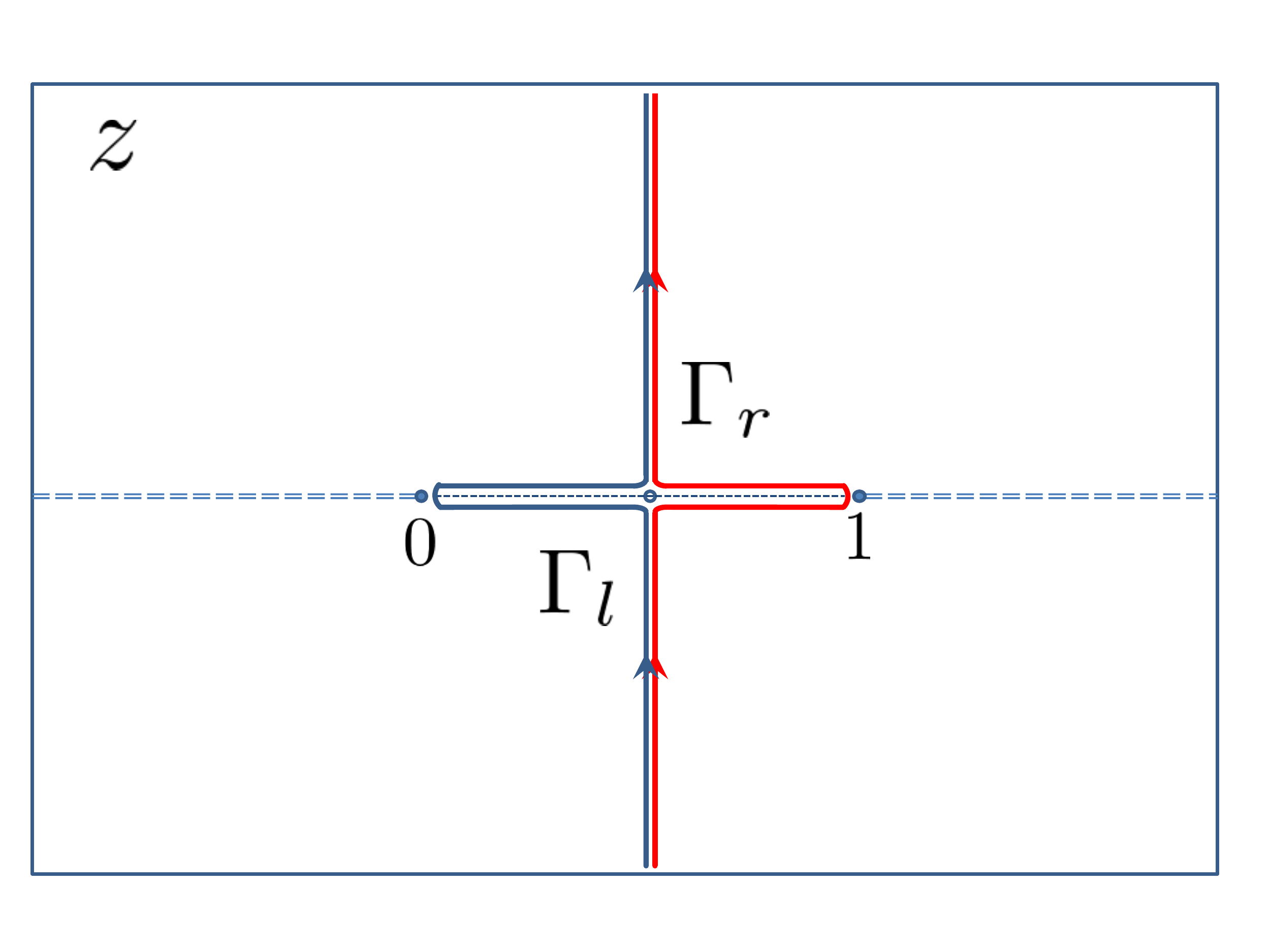}%
\end{tabular}
\caption{Contours of integration defined throughout this work. The vertical sections run along the line $\frac 12+i \mathbb R$. The horizontal sections wrap around the real line segments $(0,1/2)$ and $(1/2,1)$}%
\label{fig:contours}%
\end{center}
\end{figure}

Let us furthermore assume that $h_{++}(z,\bar z)$ and $h_{+-}(z,\bar z)$ are both separately holomorphic in $z$ and $\bar z$ away from the real line. We also assume that these kernels decay sufficiently fast at infinity. Then by deforming the contours of integration (as in the 1d case \cite{Mazac:2018}, see our appendix \ref{sec:bases}) this allows us to derive an alternative representation of the functional action:
\begin{multline}
\omega\left[\mathcal F\right]=\int_{\Gamma^+}\frac{\ud z}{2\pi i}\int_{\Gamma^+}\frac{\ud \bar z}{2\pi i}\left[ f(z,\bar z) \mathcal F(z,\bar z)+f(z,1-\bar z) \mathcal F(z,1-\bar z)\right]\\
-\int_{\Gamma} \frac{\ud z}{2\pi i} \int_{\frac 12}^1 \frac{\ud \bar z}{\pi} \,\bar e(z,\bar z) \mathcal F(z,\bar z)
-\int_{\Gamma} \frac{\ud \bar z}{2\pi i} \int_{\frac 12}^1 \frac{\ud z}\pi \,e(z,\bar z) \mathcal F(z,\bar z) \\
+\int_{\frac 12}^1 \frac{\ud z}\pi \int_{\frac 12}^1 \frac{\ud \bar z}\pi \left[ g(z,\bar z) \mathcal F(z,\bar z)+\tilde g(z,\bar z) \mathcal F(z,1-\bar z)\right]. \label{eq:fgerep}
\end{multline}
The contours of integration above are explained in figure \ref{fig:contours}. 
The various functions above are determined in terms of $h_{++},h_{+-}$ as
\bea
\label{eq:fgedefs}
f(z,\bar z)&=h_{++}(z,\bar z)-h_{+-}(z,\bar z)-h_{++}(1-z,1-\bar z)+h_{+-}(1-z,1-\bar z)\,,\\
e(z,\bar z)&\underset{0<z<1}{=} \mathcal I_{z}\left[h_{++}(z,\bar z)-h_{+-}(z,\bar z)\right]\,,\\
\bar e(z,\bar z)&\underset{0<\bar z<1}{=} \mathcal I_{\bar z} \left[h_{++}(z,\bar z)+h_{+-}(1-z,1-\bar z)\right]\,, \\
g(z,\bar z)&\underset{0<z,\bar z<1}{=} \mathcal I_z \mathcal I_{\bar z} h_{++}(z,\bar z)\,,\\
\tilde g(z,\bar z)&\underset{0<z,\bar z<1}{=}\mathcal I_z \mathcal I_{\bar z} h_{+-}(z,\bar z)\,.
\eea
Note that there is a redundancy, in that $e(\bar z,z)=\bar e(z,\bar z)$. 

This representation has two advantages. Firstly, the kernels above are simpler than $h_{++},h_{+-}$, since they involve for the most part discontinuities of the latter. Secondly, this representation makes clearer what boundary conditions to impose on the kernels. We should note that having arrived at this representation, we no longer have to discuss $h_{++}, h_{+-}$ at all. Instead, the kernels $f,e,\bar e,g,\tilde g$ remember their origins by satisfying the following ``gluing'' constraints:
\bea
\mathcal I_z f(z,\bar z)&\underset{0<z<1}{=} e(z,\bar z)+e(1-z,1-\bar z),& \mathcal I_{\bar z} f(z,\bar z)&\underset{0<z<1}{=} \bar e(z,\bar z)+\bar e(1-z,1-\bar z),\\
\mathcal I_{\bar z} e(z,\bar z)&\underset{0<z,\bar z<1}{=}g(z,\bar z)-\tilde g(1-z,\bar z),&
\mathcal I_{z} \bar e(z,\bar z)&\underset{0<z,\bar z<1}{=}g(z,\bar z)+\tilde g(z,1-\bar z),\label{eq:fgeprops1}
\eea
as well as the symmetries
\bea
e(z,\bar z)&=\bar e(\bar z,z),& g(z,\bar z)&=g(\bar z,z),\\
\tilde g(z,\bar z)&=-\tilde g(\bar z,z)& f(z,\bar z)&=-f(1-z,1-\bar z)=f(\bar z,z).\label{eq:fgeprops2}
\eea
The gluing constraints are so called because they are what allow us to put these kernels back together to reconstruct $h_{++}$ and $h_{+-}$.

\subsection{Constraints on functional kernels}
\label{sec:constraints}

Quite generally, well-defined functionals should satisfy two conditions, as emphasized in \cite{Rychkov:2017tpc}. Firstly, they should be finite when acting on all crossing vectors $F_{\Delta,\ell}$ allowed by unitarity. Secondly, they should be {\em crossing-compatible}: one should be able to commute their action with infinite sums of crossing vectors, or more precisely those infinite sums consistent with crossing symmetry. That is:
\bea
\omega\left[\sum_{\Delta,\ell} a_{\Delta,\ell} F_{\Delta,\ell}(z,\bar z|\Df)\right]=\sum_{\Delta,\ell} a_{\Delta,\ell}\,\omega(\Delta,\ell|\Df), 
\eea
with $\omega(\Delta,\ell|\Df):=\omega\left[F_{\Delta,\ell}(z,\bar z|\Df)\right]$. Since the functional actions are given by integrals which probe the boundary of the region where the OPE converges, the latter condition is non-trivial and in general more constraining than the first.

These requirements impose constraints on the kernels defining the functional action. It turns out to be more convenient to use the form of this action which depends on $f,g,e,\bar e, g$, and $\tilde g$ kernels, i.e. equation \reef{eq:fgerep}. Our strategy is to demand that the various integrals which depend on these kernels converge absolutely when applied to infinite sums of crossing vectors. Schematically, we demand
\bea
\int\!\!\int \ud z \ud \bar z |m(z,\bar z)| \sum a_{\Delta,\ell} |F_{\Delta,\ell}(z,\bar z)|<\infty
\eea
for each such integral. By the Fubini theorem this guarantees that the integrals may be swapped with the series, and in this way both our requirements will be satisfied. 

We will demand and analyse the above for each  piece of the functional action separately. For each of these the non-trivial constraint arises from examining the behaviour of the integrand near the boundary of integration, where we approach OPE, Regge or lightcone limits. As previously mentioned, the behaviour of the infinite sum may be different from that of each individual term. In each case we will be able to relate the infinite series above to an associated correlation function $\mathcal G(z,\bar z)$, which will make the behaviour in these limits more transparent, with
\bea
\mathcal G(z,\bar z)=\sum_{\Delta,\ell} \frac{G_{\Delta,\ell}(z,\bar z)}{(z \bar z)^{\Df}}\,\qquad \mathcal G(z,\bar z)=\mathcal G(1-z,1-\bar z) \Leftrightarrow \sum_{\Delta,\ell} a_{\Delta,\ell} F_{\Delta,\ell}(z,\bar z|\Df)=0\,.
\eea

To establish finiteness we must probe the boundary of the integration region by scaling $z$ and $\bar z$ in all possible ways. Intuitively, the worry is that the integral may be dominated by regions where both $z$ and $\bar z$ approach the boundary but not necessarily at the same rate. To make this more concrete,  consider $\mathcal K(z,\bar z)=\mathcal K(\bar z,z)$ a positive analytic function on $z,\bar z\in (1,\infty)$. Consider its integral,
\bea
\int_1^\infty \ud z \int_1^{\infty} \ud \bar z\, \mathcal K(z,\bar z)=2\int_1^\infty \ud z \int_1^{z} \ud \bar z \,\mathcal K(z,\bar z)\,.
\eea
We would like to determine appropriate asymptotic conditions on this function to ensure convergence. If we parametrize
$
z=\Lambda,\bar z=\Lambda^\alpha\,,
$
the integral becomes
\bea
2 \int_1^{\infty} \ud \Lambda \int_0^1 \ud \alpha \log(\Lambda) \Lambda^{\alpha} \mathcal K(\Lambda,\Lambda^\alpha)
\eea
Let us demand that:
\bea
\mathcal K(\Lambda,\Lambda^\alpha)\underset{\Lambda \to \infty}= O(\Lambda^{-1-\alpha-\epsilon})\qquad \mbox{for all}\quad \alpha\in[0,1]\,,
\eea
for some $\epsilon>0$. This implies that for any $\alpha\in [0,1]$ we can bound
\bea
\mathcal K(\Lambda,\Lambda^\alpha)\leq \frac{C}{\Lambda^{1+\alpha+\epsilon}}
\eea
for sufficiently large $C$ independent of $\alpha, \Lambda$. Hence:
\bea
2 \int_1^{\infty} \ud \Lambda \int_0^1 \ud \alpha \log(\Lambda) \Lambda^{\alpha} \mathcal K(\Lambda,\Lambda^\alpha)\leq 2 C\int_1^{\infty} \frac{\ud \Lambda}{\Lambda^{1+\epsilon}} \log(\Lambda)<\infty\,.
\eea

In appendix \ref{app:constraints} we apply this logic to obtain constraints on the kernels $f,e, g$ and $\tilde g$ ($\bar e$ is redundant). The results can be summarized as follows:
\bea
f(\Lambda^\alpha x,\Lambda \bar x)&=O(\Lambda^{-1-\alpha-\epsilon})\\
e(1-\mbox{$\frac 1{\Lambda^{\alpha} x}$},\Lambda \bar x)&= O(\Lambda^{\alpha(1-\Df)-1-\epsilon})\\
g\left(\mbox{$1-\frac{1}{\Lambda^\alpha x},1-\frac{1}{\Lambda \bar x}$}\right)&=O\left(\Lambda^{(1+\alpha)(1-\Df)-\epsilon}\right)\\
\tilde g\left(\mbox{$1-\frac{1}{\Lambda^\alpha x},1-\frac{1}{\Lambda \bar x}$}\right)&=\left\{\begin{array}{lcll}
O\left(\Lambda^{(1+\alpha)(1-\Df)-\epsilon}\right)\,,& d=2\,,& \alpha\in[0,1]\,,& \Df\geq 0\\
O\left(\Lambda^{1+\alpha-\Df-\epsilon}\right)\,,& d>2\,,& \alpha\in[0,1)\,,&\Df\geq \frac{d-2}2\\
O\left(\Lambda^{2-\Df-\epsilon}\right)\,,& d>2\,,& \alpha=1\,,& \Df\leq d-2\\
O\left(\Lambda^{d-2\Df-\epsilon}\right)\,,& d>2\,,& \alpha=1\,,& \Df\geq d-2
\end{array}
\right.
\
\label{eq:constraintskernels}
\eea
These constraints concern the limit of large positive $\Lambda$, and must hold for each $\alpha\in [0,1]$ for some $\epsilon>0$. In each case $x,\bar x$ should be chosen such that each kernel is evaluated on its respective integration contour. 

The peculiar form of the constraints on $\tilde g$ follows from the properties of correlators in the double lightcone limit. More precisely,  they rely to some extent on certain assumptions about this limit, which could turn out in practice to be overly restrictive. To better understand this, we will now study this limit in some detail.

\subsubsection{The double lightcone limit}
\label{sec:lightcone}
In order to understand the constraints on the functional kernels, we must understand the behaviour of the correlator $\mathcal G(z,\bar z)$ in various limits. As we will see in appendix \ref{app:constraints}, all such limits can be simply understood in terms of the OPE in various channels, with the exception of the double lightcone limit, which will be our focus here. This limit is defined by
\bea
\lim_{\Lambda \to \infty} \mathcal G(\mbox{$1-\frac 1{\Lambda^\alpha x },\frac{1}{\Lambda \bar x}$})\,,\qquad \alpha\in [0,1]\,,
\eea
and we will have to study separately the cases $\alpha<1$ and $\alpha=1$. 

Let us then set $\alpha<1$ such that the limit corresponds to $\bar z\ll z$. We can use the small $\bar z$ expansion of conformal blocks \reef{eq:lightconeexp} to obtain
\bea
\mathcal G(1-z,\bar z)\underset{\bar z\ll z \ll 1}{\sim} \sum_{\ell=0,2,\ldots} c_{0,0}\, a_{\tau_0(\ell)+\ell,\,\ell}\, \bar z^{\frac{\tau_0(\ell)}2-\Df} \, G_{\frac{\tau_0(\ell)}2+\ell}(1-z |\Df)\,.
\eea
The sum runs over those operators in the OPE with lowest twist $\tau_0(\ell)$ for each spin. 
If $d>2$ the lowest twist operator is always the identity. Hence in the limit above we simply obtain
\bea
\mathcal G(1-z,\bar z)\underset{\bar z\ll z \ll 1}{\sim} \frac{1}{\bar z^{\Df}},\qquad \mbox{for}\quad d>2\,.
\eea
For $d=2$ there are an infinite number of possible twist zero operators which in general must be summed over, 
\bea
\mathcal G(1-z,\bar z)\underset{\bar z\ll z \ll 1}{\sim} \frac{1}{\bar z^{\Df}}\sum_{\ell} a_{\ell,\ell} \, G_{\ell}(1-z |\Df)\,.
\eea
Each term in the sum diverges logarithmically for small $z$, but the full sum could have a stronger divergence. The limit is controlled by the tail of the sum over spins, and we have used that
$c_{0,0}=\frac 12(1+\delta_{0,\ell})$ for $d=2$. 

To say something about the small $z$ behaviour of the sum we would need to be able to bound the OPE coefficients $a_{\ell,\ell}$ for large $\ell$. The (suboptimal) $d=1$ bounds on OPE coefficients of $SL(2,\mathbb R)$ blocks \cite{Mazac2019a} tell us that $a_{\ell,\ell}$ is bounded by the 1d generalized free OPE density $a_{\ell}^{\mbox{\tiny gff}}$ given in \reef{eq:ope1d}. From this it follows that
\bea
\sum_{\ell} a_{\ell,\ell} \, G_{\ell}(1-z |\Df)\underset{z\to 0}{=}O(z^{-2\Df})\,.
\eea
However, this is likely too weak a bound. For $d=2$ we can think of the presence of infinite twist zero operators as signalling the existence of a chiral algebra. In this context, understanding the limit above amounts to studying the $t$-channel OPE limit of the $s$-channel chiral algebra identity block\footnote{We thank Scott Collier for discussions regarding these points.}. Recently this has been studied for the simplest case of the Virasoro algebra in \cite{Kusuki2019,Collier2019a}, when $c>1$. Equations 2.27 and 2.29 of that reference are compatible with the stronger result
\bea
\sum_{\ell} a_{\ell,\ell} \, G_{\ell}(1-z |\Df)\underset{z\to 0}{=}O(z^{-\Df}).\label{eq:2dassumption}
\eea
When $c<1$ the $s$-channel identity block can be written as a finite linear combination of $t$-channel blocks, leading to the same result. This bound is also compatible with the double lightcone behaviour of the $c=1$ vertex operator correlator, which contains a piece of the form $[z \bar z(1-z)(1-\bar z)]^{-\Df}$.
We will henceforth assume the above is correct for any 2d CFT.

Now let us consider the case where $\alpha=1$. The reason why this is more subtle is that terms in the block expansion that are naively suppressed in $\bar z$ for having higher twist, can be reinforced by the divergence at $z\to 0$ arising from the infinite sum over spin. The simplest example of this phenomenon occurs in the GFF correlator,
\bea
\mathcal G(1-z,\bar z)=1+\frac{1}{(1-z)^\Df\bar z^\Df}+\frac{1}{z^\Df(1-\bar z)^\Df}\,.
\eea
When we take $\bar z\ll z\ll 1$ above (i.e. $\alpha<1$), a single term dominates, which can be thought of as the contribution of the identity in one channel. By symmetry the same is true when $z\ll \bar z\ll 1$. However, for $z,\bar z\ll 1$ with $z/\bar z$ fixed ($\alpha=1$), two terms contribute. The extra contribution comes from summing over an infinite number of blocks of twist $\tau=2\Df$:
\bea
\mathcal G(1-z,\bar z)\underset{z,\bar z\ll 1}{\sim}\frac{1}{\bar z^{\Df}}+\sum_{\ell=0,2,\ldots}c_{0,0} a_{2\Df+\ell,\ell} G_{\Df+\ell}(1-z)\sim \frac{1}{\bar z^{\Df}}+\frac{1}{z^\Df}
\eea
The infinite sum thus reproduces the identity of the cross-channel, and both terms are of the same order. 
Note that this is consistent with crossing symmetry here, which implies invariance under $z\leftrightarrow \bar z$. In fact because of crossing, as long as the identity contribution dominates, the result above must hold for any CFT correlation function. However, the limit may be dominated by a different set of operators with higher twist. For instance we can have
\bea
\sum_{\ell=0,2,\ldots} c_{0,0} a_{\tau_0+2\ell,\,\ell} \bar z^{\frac{\tau_0}2-\Df} G_{\frac{\tau_0}2+\ell}(1-z)\underset{z,\bar z\to 0}\sim \frac{1}{z^{\Delta_\phi-\frac{\tau_0}2} \bar z^{\Delta_\phi-\frac{\tau_0}2}}\,.\label{eq:strongest}
\eea
This is indeed dominant over the identity contribution if $\tau_0< \Df$. Together with the unitarity bound, this implies that there is a window $d-2\leq \tau\leq \Df$ where this behaviour is available. As we have mentioned above, for $d=2$ we have twist zero operators and this behaviour is indeed observed. In particular this corroborates our previous assumption \reef{eq:2dassumption}. But what about $d>2$? We are not aware of any examples where terms such as the above dominate over the identity. Going back to the GFF theory, if we consider correlators of composite operators $\phi^n$ we can indeed find terms like the one above, but they are always comparable to the identity contribution:
\bea
\mathcal G(1-z,\bar z)\underset{z\sim \bar z\ll 1}{\sim} 
\sum_{k=0}^n \frac{c_k}{z^{(1-k/n) \Df} \bar z^{k \Df /n}}
\eea
for some coefficients $c_k=c_{n-k}$ and where we have still used $\Df$ for the dimension of $\phi^n$.  In particular for even $n$ the term $c_{n/2}$ takes the above form, but $\tau_0=\Df$. 

Let us set $z=1/\Lambda^\alpha, \bar z=1/\Lambda$. Then we can summarize the $d=2$ results by:
\bea
\mathcal G(\mbox{$1-\frac{1}{\Lambda^\alpha},\frac{1}{\Lambda}$})&\underset{\Lambda\to \infty}=O(\Lambda^{-(1+\alpha)\Df})\,,&\qquad d&=2\,,& \Df&\geq 0\,, & \alpha&\in[0,1]\,.\\
\eea
As for $d>2$, the worse possible divergence corresponds to:
\bea
\mathcal G(\mbox{$1-\frac{1}{\Lambda^\alpha},\frac{1}{\Lambda}$})&\underset{\Lambda\to \infty}=O(\Lambda^{\Df})\,,&\qquad d&>2\,,& \Df&\geq \frac{d-2}2\,, & \alpha&\in[0,1)\,,\\
\mathcal G(\mbox{$1-\frac{1}{\Lambda^\alpha},\frac{1}{\Lambda}$})&\underset{\Lambda\to \infty}=O(\Lambda^{\Df})\,,&\qquad d&>2\,,& \Df&\leq d-2\,, & \alpha&=1\,,\\
\mathcal G(\mbox{$1-\frac{1}{\Lambda^\alpha},\frac{1}{\Lambda}$})&\underset{\Lambda\to \infty}=O(\Lambda^{2\Df+2-d})\,,&\qquad d&>2\,,& \Df&\geq d-2\,, & \alpha&=1\,.\\
\eea
If we postulate however that the behaviour can be no worse than that of generalized free fields, we would instead have continuity in $\alpha$:
\bea
\mathcal G(\mbox{$1-\frac{1}{\Lambda^\alpha},\frac{1}{\Lambda}$})&\underset{\Lambda\to \infty}=O(\Lambda^{\Df})\,,&\qquad d&>2\,,& \Df&\geq \frac{d-2}2\,, & \alpha&\in[0,1]\,.
\eea
We have no way of checking this postulate. 
If it is false and  we allow for a behaviour such as \reef{eq:strongest}, in general this means that the double lightcone limit is very different according to whether we approach it with $z, \bar z$ of the same order or not. Alternatively, if the postulate is true and general CFT correlators always fall into the ``GFF class'', then the identity contribution is never subdominant. In this case the double lightcone limit still presents a certain discontinuity, but not by a parametrically large factor. Just as for $d=2$, in order to more precisely determine what happens we would need to determine a bound on OPE coefficients at large spin and finite twist.

For the purposes of this paper this is somewhat irrelevant, as we will propose functionals which either act on the $d=2$ crossing equation or for which $\tilde g=0$ anyway. While for aesthetic reasons we believe the GFF like constraint is most likely the correct one, we have chosen to present the strongest possible constraints in equation \reef{eq:constraintskernels}, with the caveat that this might be too restrictive. It would obviously be important to settle this question in the future.

\section{HPPS functionals}
\label{sec:hpps}

In this section we will introduce a simple set of functionals suitable for bootstrapping generalized free fields and perturbations around them with finite support in spin. In a sense they provide a functional perspective on the seminal work \cite{Heemskerk2009}, hereby referred to as HPPS. For pedagogical reasons, we find it is better to discuss here the functionals on their own terms without making reference to the HPPS procedure, but the connection between the two approaches is clarified in appendix \ref{app:hpps}. 

Our motivation is to construct a basis of functionals which is dual to generalized free fields, much as was possible to do in $d=1$. Such a basis should allow us not only to bootstrap the generalized free field solution itself but also small deformations around it. Ideally, we would like to obtain functionals $\alpha_{n,J}, \beta_{n,J}$ satisfying
\bea
\alpha_{m,J}(\Delta_{n,\ell},\ell)&=\delta_{m,n}\delta_{J,\ell},& \qquad \partial_{\Delta}\alpha_{m,J}(\Delta_{n,\ell},\ell)&=0\\
\beta_{m,J}(\Delta_{n,\ell},\ell)&=0,& \qquad \partial_{\Delta}\beta_{m,J}(\Delta_{n,\ell},\ell)&=\delta_{m,n}\delta_{J,\ell}\label{eq:gffduality}
\eea
Such functionals can be applied to  crossing equations arising from deformations of generalized free fields,
\bea
\sum_{n=0}^\infty\sum_{\ell=0}^{\infty}\left[ a^{(1)}_{n,\ell} F_{\Delta_{n,\ell},\ell}(z,\bar z|\Df)+a^{(0)}_{n,\ell} \gamma_{n,\ell} \partial_{\Delta} F_{\Delta_{n,\ell},\ell}(z,\bar z|\Df)\right]\label{eq:gffpert}=S(z,\bar z)\,,
\eea
(where $S(z,\bar z)$ parametrizes the deformation), to read off the various coefficients in the perturbation uniquely.
We will {\em not} be able to obtain such functionals, and as we discuss in section \ref{sec:discussion}, it's not even clear to us that the above even is achievable.\footnote{In general, we expect as for bosonic functionals in $d=1$ cf. equation \reef{eq:ONboson}, that there should be small modifiations to the duality conditions, but this is not the problem.}
 Nevertheless, let us try to emulate this structure and see how far we can go.

\subsection{Ansatz}


The duality conditions above imply that the functionals are such that the associated functional actions will have double zeros on the spectrum of the generalized free field solution, so this is what we should aim for. As we've seen in section \ref{sec:review} (and appendix \ref{sec:bases}), obtaining such functionals in $d=1$ is not at all a trivial task. A nice surprise is that in general dimension, it is quite easy to get functionals which (almost) behave in the correct way. The reason is that even in $d=1$ it is simple to obtain functionals with first order zeros. In higher dimensions we can simply take a product of these to obtain the desired double zero structure.

Starting from the functional action \reef{eq:funcactionh}, we set $h_{+-}=0$ and choose $h_{++}(z,\bar z)\equiv h(z,\bar z)$ to be a meromorphic function with possible poles at $z,\bar z=0,1$. This immediately gives $g=\tilde g=e=\bar e=0$ and $f(z,\bar z)=h(z,\bar z)-h(1-z,1-\bar z)$. In this case our ansatz can be written simply as:
\bea
\omega(\Delta,\ell)= \int_{\Gamma} \frac{\ud z}{2\pi i} \int_{\Gamma} \frac{\ud \bar z}{2\pi i}  \, h(z,\bar z) F_{\Delta,\ell}(z,\bar z|\Df) \label{eq:hppsdef}
\eea
The boundedness conditions at infinity discussed in section \ref{sec:constraints}, together with meromorphicity, demand $h(z,\bar z)\underset{z\to \infty}\sim 1/z^2$. In particular this allows us to deform the contours of integration to get
\bea
\omega(\Delta,\ell)=\int_{--}\frac{\ud z \ud \bar z}{\pi^2}\, \left[h(z,\bar z)-h(1-z,1-\bar z)\right] \mathcal I_z\mathcal I_{\bar z}\left[ \frac{G_{\Delta,\ell}(z,\bar z)}{(z \bar z)^\Df}\right]\,,
\eea
as long as possible singularities at $z,\bar z=0$ are integrable, as we'll discuss below. We now use that for even $\ell$:
\bea
\mathcal I_z\mathcal I_{\bar z}\left[ \frac{G_{\Delta,\ell}(z,\bar z)}{(z \bar z)^\Df}\right]=\sin^2\left[\frac{\pi}2\left(\Delta-2\Df\right)\right]\, \frac{G_{\Delta,\ell}(\frac{z}{z-1},\frac{\bar z}{\bar z-1})}{(z \bar z)^\Df},\qquad z<0, \bar z<0
\eea
 After a change of variables $z\to \frac z{z-1}$ we find
\bea
\omega(\Delta,\ell)=\frac{\sin^2\left[\frac{\pi}2\left(\Delta-2\Df\right)\right]}{\pi^2}\, \left[\Omega_h(\Delta|\Df)-\tilde \Omega_h(\Delta|\Df)\right]\label{eq:funcacpq1}
\eea
with
\bea
\label{eq:funcacpq2}
\Omega_h(\Delta,\ell|\Df)=\int_{00} \frac{\ud z\ud \bar z}{z^2 \bar z^2}h\left(\mbox{$\frac{z}{z-1},\frac{\bar z}{\bar z-1}$}\right)\,\left(\frac{(1-z)(1-\bar z)}{z\bar z}\right)^{\Df-2} G_{\Delta,\ell}(z,\bar z)\,, \\
\tilde \Omega_h(\Delta,\ell|\Df)=\int_{00} \frac{\ud z\ud \bar z}{z^2 \bar z^2}h\left(\mbox{$\frac{1}{1-z},\frac{1}{1-\bar z}$}\right)\,\left(\frac{(1-z)(1-\bar z)}{z\bar z}\right)^{\Df-2} G_{\Delta,\ell}(z,\bar z)\,.
\eea
This shows that quite generally the functionals will indeed have double zeros when evaluated on the generalized free spectrum, as desired. In fact this is not exactly so, since these expressions might diverge for sufficiently small values of $\Delta$ due to singularities near $z,\bar z=0,1$. The more precise statement is that the functionals will have finite support on the GFF spectrum.

\subsection{General properties}
There are two sets of functionals we are interested in:
\ba
\alpha_{p,q}:&&\qquad h(z,\bar z)&=\frac 12\left[\frac 1{z^{p+2} \bar z^{q+2}}+(z\leftrightarrow \bar z)\right]\,,\\
\beta_{p,q}:&&\qquad h(z,\bar z)&=\frac 12\left[\frac 1{z^{p+2} (1-\bar z)^{q+2}}+(z\leftrightarrow \bar z)\right]\,.
\ea
We are being slightly inconsistent in notation here, since these functionals will not have as simple duality properties as those in \reef{eq:gffduality}. Nevertheless, they do share some of those properties, so we ask for the reader's indulgence.

We wrote the powers in a funny way because of the constraints of section \ref{sec:constraints}. These imply that for large $z,\bar z$ we must take $p,q \in \mathbb N_{\geq 0}$. Depending on $p,q$, the integrals \reef{eq:funcacpq2} can have divergences as a function of $\Delta,\ell$, arising from the integration regions where $z$ or $\bar z$ approach zero.  These divergences can then cancel partially or completely the double zeros coming from the sine squared prefactor in \reef{eq:funcacpq1}. 

A simple way to understand the precise structure of zeros is to go back to the original functional representation \reef{eq:hppsdef}, which is always valid and finite. Let us consider the $\alpha_{p,q}$ functionals for definiteness. Then we have
\bea
\alpha_{p,q}(\Delta,\ell)= \int_{\Gamma \otimes \Gamma} \frac{\ud z \ud \bar z}{(2\pi i)^2}\,\frac{ G_{\Delta,\ell}(z,\bar z|\Df)}{z^{2+p} \bar z^{2+q}}- \int_{\Gamma \otimes \Gamma} \frac{\ud z \ud \bar z}{(2\pi i)^2}\,\frac{G_{\Delta,\ell}(z,\bar z|\Df)}{(1-z)^{2+p}(1-\bar z)^{2+q}}
\eea
To obtain representation \reef{eq:funcacpq1} we want to close both $z$ and $\bar z$ contours on the left. This is possible as long as the singularities near $z,\bar z=0$ are integrable. Since the conformal blocks have an expansion for small $z$ of the form \reef{eq:lightconeexp} we see that this is possible only if
\bea
\tau>2\Df+2+2\max\{p,q\}\,.\label{eq:condfirstzero}
\eea
Below this point the double zeros will generically become simple zeros. To determine the finite support of the functionals on the GFF spectrum, $\Delta_{n,\ell}=2\Df+2n+\ell$, one simply notes that in this case the blocks have no discontinuity for $z,\bar z<0$, and hence closing the contours on the left merely computes the residues at $z,\bar z=0$. These residues will be non-zero if:
\bea
0\leq n&\leq N,& N&\equiv 1+\min\{p,q\}\\
0\leq \ell&\leq J+2(N-n),& J&\equiv |p-q|\,. \label{eq:finitesupport}
\eea
Note that this residue computation also easily determines the functional action at these points. 

The analysis of the $\beta_{p,q}$ functionals is very similar. In that case, it is easy to see that one can never find a simultaneous non-vanishing residue in both $z,\bar z$ when acting on the GFF spectrum, and hence these functionals are zero on the entire GFF spectrum. However, since single residues are still available the functional action will have {\em double} zeros on the spectrum only if \reef{eq:condfirstzero} holds\,.

To summarize: when acting on GFF operators, the $\beta_{p,q}$ are zero everywhere, whereas the $\alpha_{p,q}$ have finite support in twist and spin, depending on $p,q$. However, when these functionals act on {\em derivatives} of crossing vectors of the GFF spectrum, then both $\alpha_{p,q}$ and $\beta_{p,q}$ have finite support in twist, but not in spin. In this way, we see that those equations in \reef{eq:gffduality} involving derivatives of functional actions can never be satisfied, at least not by taking any finite linear combination of $\alpha_{p,q}$ and $\beta_{p,q}$.  The conclusion is that these functionals, in spite of their nice properties, do not give us exactly what we were looking for. Nevertheless, as we shall see in the following subsections, they still have their uses.

\subsection{Bootstrapping GFFs and completeness}
From the discussion above, we see that the $\alpha_{p,q}$ functionals can be used to bootstrap the generalized free field solution. We write the crossing equation
\bea
F_{0,0}(z,\bar z)+\sum_{n=0}^\infty\sum_{\ell=0,2,\ldots} a_{n,\ell}^{\mbox{\tiny}}\, F_{\Delta_{n,\ell},\ell}(z,\bar z)=0
\eea
and let us imagine we did not know the solution for $a_{n,\ell}$ given by \reef{eq:gffope}. To determine the $a_{n,\ell}^{\mbox{\tiny}}$ we act on this equation with the $\alpha_{p,q}$, where we can take $q\geq p$. Since the functional actions have finite support on the full GFF spectrum, this always leads to equations which involve a finite number of unknown OPE coefficients, namely those allowed by the finite support conditions \reef{eq:finitesupport}. 

These equations can be solved systematically as follows. Let us fix $p$. Then equations with $q=p+2k$ and $q=p+2k+1$ for non-negative integer $k$ will involve the same set of OPE coefficients, whereas for $q=p+2k+2$ two new coefficients make an appearance. Hence the growth in the number of unknowns matches that of the constraints, but we must check initial conditions. On the one hand, the $(p,q)=(0,0)$ equation gives a relation between not two, but three distinct OPE coefficients, $a_{1,0}^{\mbox{\tiny}},a_{0,0}^{\mbox{\tiny}}$ and $a_{0,2}^{\mbox{\tiny}}$. Overall, equations of the form $(0,q)$ leave then one undetermined coefficient which we can choose to be $a_{0,0}^{\mbox{\tiny}}$. On the other hand, once all equations with fixed $p$ have been solved, equations $(p+1,p+1)$ and $(p+1,p+2)$ only involve one new unknown (namely $a_{0,p+2}^{\mbox{\tiny}}$). So it may even seem that we have an overconstrained system, but of course we know a solution exists. The question then is only whether $a_{0,0}^{\mbox{\tiny}}$ eventually becomes fixed to its GFF value or not. While this may seem like a small point, it is important since it determines whether the $p,q$-functional basis is complete or not, i.e. whether it fully captures the constraints of crossing symmetry.

In general, we have not been able to answer this question definitively. To the extent that we were able to investigate, (by writing down and solving equations up to some maximum $p,q$) we do find that $a_{0,0}^{\mbox{\tiny}}$ is never fixed.\footnote{Although interestingly, in some cases  we do find that changing this coefficient sufficiently far away from the GFF value can make some other OPE coefficients become negative, violating unitarity.} 
We can gain further insight into this question by examining in detail a special case. 
Let us set $d=2$ and $\Df=1$. In this case, there is a solution to crossing whose spectrum partially overlaps with the GFF spectrum, namely the $\langle \varepsilon \varepsilon \varepsilon \varepsilon\rangle$ correlator in the 2d Ising model. As reviewed in section \ref{sec:productcorrelators}, this correlator is obtained by taking a (chiral) product of two 1d generalized free fermions, and contains the operator content
\bea
\Delta&=2\Df+2+4n+\ell,& \ell&=0,2,\ldots\,,&\qquad n\in \mathbb N_{\geq 0},\quad \\
\Delta&=\ell,& \ell&=0,2,\ldots
\eea
with $\Df=1$. We now point out that, as is easy to check, the action of $\alpha_{p,q}$ on the conserved currents with $\Delta=\ell$ with $\ell\neq 0$, vanishes for this value of $\Df$. Hence, from the point of view of these functionals, we could modify the crossing equation to:
\bea
F_{0,0}(z,\bar z)+\sum_{\ell=2,4,\ldots}^{\infty} a_\ell F_{\ell,\ell}(z,\bar z)+\sum_{n=0}^\infty\sum_{\ell=0,2,\ldots} a_{n,\ell}^{\mbox{\tiny}}\, F_{\Delta_{n,\ell},\ell}(z,\bar z)=0\label{eq:crossingcurrents}
\eea
It follows then that $a_{0,0}$ must not be fixed by the $\alpha_{p,q}$ equations, since both the GFF and Ising correlators satisfy this equation with different values of $a_{0,0}$. In particular, the Ising correlator corresponds to setting $a_{0,0}=0$, and we have checked that in that case the $\alpha_{p,q}$ equations precisely reproduce the correct OPE coefficients of the Ising solution. For instance we find that the equations set $a_{2k,\ell}\propto a_{0,0}$ for all integer $k$.

But what of the conserved currents? How can then these be determined? Although the action of the $\alpha_{p,q}$ functionals on these is vanishing, this is not the case for the $\beta_{p,q}$. Since these functionals annihilate the GFF spectrum, the $\beta_{p,q}$ equations take the form
\bea
\sum_{\ell=2,4,\ldots} a_{\ell} \beta_{p,q}(\ell,\ell)\,=0\,,
\eea
where
\bea
\beta_{p,q}(\ell,\ell)=\frac 12\left[\int_{\Gamma} \frac{\ud z}{2\pi i} \frac{1}{z^{2+p}} \frac{G_{\ell}(z)}z -(p\leftrightarrow q)\right]\,.
\eea
In particular one can check that
$\beta_{p,q}(\ell,\ell)=0$ for $\ell> 2+\max\{p,q\}$, and hence it is straightforward to solve for the OPE coefficients systematically. In fact, the solution to the full set of equations is obtained by setting
\bea
a_\ell\propto 2 a_{\ell}^{\mbox{\tiny free}}|_{\Df=1}=\frac{4\, \Gamma(\ell)^2}{\Gamma(2\ell-1)}\,,\label{eq:opecurrents}
\eea
since:
\bea
\sum_{\ell=2,4\ldots} a_\ell \beta_{p,q}(\ell,\ell)&\propto \frac 12\left[\int_{\Gamma} \frac{\ud z}{2\pi i} \frac{1}{z^{2+p}} \left(\sum_{\ell=2,4,\ldots}\,a_{\ell}^{\mbox{\tiny free}} \frac{G_{\ell}(z)}{z}\right) -(p\leftrightarrow q)\right]\\
&=\frac 12\left[\int_{\Gamma} \frac{\ud z}{2\pi i} \frac{1}{z^{2+p}} \left(\frac 1{1-z}-1\right) -(p\leftrightarrow q)\right]=0\,.
\eea
These results are consistent with the OPE coefficients of the conserved currents for the Ising correlator as they should be, with the exact results obtained by setting the proportionality constant in \reef{eq:opecurrents} equal to one, cf. equation \reef{eq:isingope}. Again, this proportionality constant could not have been fixed by the $\beta_{p,q}$ equations, since after all the GFF solution exists and there this constant is zero.

The conclusion then seems to be that as far as the $p,q$ functionals are concerned, there are two undetermined degrees of freedom in this case, which we can take to be $a_{0,0}$ and $a_{\ell=2}$. At this point, either we are missing a new solution to crossing, or the $p,q$ basis of functionals is not complete, in that it does not fully capture the constraints of crossing. As it turns out, it is this latter possibility that is correct. To see this is the case, we shall explicitly introduce a new functional which will fix the relation between $a_{0,0}$ and $a_{\ell=2}$. This functional can be chosen from the set of $d=1$ functionals:
\bea
\theta_n\left[\mathcal F\right]:= \int_{\Gamma} \frac{\ud z}{2\pi i}\, \frac{1}{z^{2+n}} \mathcal F(z)\,,\qquad
\tilde \theta_n\left[\mathcal F\right]:= \int_{\Gamma} \frac{\ud z}{2\pi i}\, \frac{1}{(1-z)^{2+n}} \mathcal F(z)\,.
\label{eq:thetafuncs}
\eea
As usual we have in mind acting on the higher dimensional crossing equation by restricting it to the line $z=\bar z$.
Acting with $\theta_0$ on the crossing equation \reef{eq:crossingcurrents} gives:
\bea
a_{0,0}+a_{\ell=2}=2\,.
\eea
In particular this gives the correct result for the GFF ($a_{\ell=2}=0$) and Ising $(a_{0,0}=0)$. At this point we have fixed all possible OPE coefficients, and one can check that the conditions obtained by acting with other $\theta_n$ are  also satisfied.

To summarize, in the special case $d=2$ and $\Df=1$ we can prove definitively that the $p,q$ functionals cannot fix $a_{0,0}$, and that this is a direct consequence of the existence of the Ising solution. The $p,q$ functionals correctly bootstrap these solutions up to two undetermined coefficients. For more general $d$ and $\Df$, we have checked to high orders in $p,q$ that the $\alpha_{p,q},\beta_{p,q}$ functionals do not determine $a_{0,0}$. On the other hand, one can easily check that for any $d$ and $\Df$ the $\theta_0$ equation does correctly determine $a_{0,0}=2$.  The $\theta$ functionals should be thought of as infinite sums of $p,q$ functionals since we can think of them as including a factor $\frac{1}{z-\bar z}$ over which we take a residue, a statement which will be made more rigorous in section \ref{sec:1duplift}. 

Overall these results strongly suggest that the $p,q$ functionals do not form a complete basis, in the sense that they do not fully capture the constraints of crossing symmetry. To complete the basis requires adding at least one extra functional, which we have chosen to be $\theta_0$. Whether this, or even including the full set of $\theta_n$, is enough to obtain a complete basis of functionals, remains to be understood.

\subsection{Bootstrapping contact interactions}
We will now show how the HPPS functionals may be used to bootstrap contact interactions of scalar field in AdS space. It is not our goal to repeat the analysis of \cite{Heemskerk2009}, but rather to show how it can be recovered with our language of functionals. 

The correlation functions we are interested in here arise from considering scalar fields in AdS space perturbed by four point contact interactions, which may have various numbers of derivatives, viz. 
\bea
\int_{\mbox{\tiny AdS}} \Phi^4,\qquad \int_{\mbox{\tiny AdS}} (\nabla\Phi\cdot \nabla \Phi)^2\,, \qquad \ldots
\eea
In the presence of such interactions, the generalized free field correlator gets perturbative corrections which may be written as
\bea
\mathcal G^{(1)}(z,\bar z)=\sum_{n=0}^\infty\sum_{\ell=0}^{L}\left[ a^{(1)}_{n,\ell} G_{\Delta_{n,\ell},\ell}(z,\bar z|\Df)+a^{\mbox{\tiny gff}}_{n,\ell} \gamma_{n,\ell} \partial_{\Delta} G_{\Delta_{n,\ell},\ell}(z,\bar z|\Df)\right]\label{eq:g1}
\eea
We can act with functionals on the crossing equation for this correlator, which will lead to equations on the unknown parameters $a_{n,\ell}^{(1)}$ and $\gamma_{n,\ell}$, which correct OPE coefficients and scaling dimensions respectively. It will be important for us below to note that if $2k$ is the number of derivatives then\footnote{This follows e.g. from thinking about the flat space limit of the amplitude \cite{Penedones2011}.}
\bea
\gamma_{n,\ell} \underset{n\to \infty}{=} O(n^{4k+\frac d2-\frac 32}) \label{eq:gammalargen}
\eea
Unlike the computations in the previous subsection, the crossing equation now involves derivatives of blocks evaluated on the generalized free spectrum. As we pointed out before, when functionals act on these they have finite support in the twist  $\tau\sim n$ but not in the spin $\ell$. So it would seem that acting with functionals would lead to equations involving infinite numbers of variables which would be too difficult to solve. What saves the day is that for contact interactions such as the ones above, the sum over operators only involves those up to a fixed spin $L$, which depends on the number of derivatives of the contact interaction. Hence, for this special set of deformations, we still obtain equations in a finite number of variables which we can then hope to solve for. 

An important point is that $\mathcal G^{(1)}(z,\bar z)$ is not an ordinary crossing symmetric correlator, both because of the fact that it involves derivative of conformal blocks, as well as the fact that the signs of the coefficients above are not a priori positive. This means that in principle the behaviour of this function as we take various limits in $z,\bar z$ does not have to be, and in general is not, the same as an ordinary four-point function. Hence the assumptions that led to the constraints on the functional kernels in section \ref{sec:constraints} are in general not applicable. In particular, for the HPPS functionals we are considering here, the integers $p,q$ will be constrained differently depending on the specific contact interaction we are considering.

To make these two points clear, we will shortly see that in the case of the $\phi^4$ interaction with no derivatives we can apply $\beta_{-1,0}$ to the crossing equation (notice $p<0$, which is disallowed for ordinary correlation functions).  This leads to the equation:
\bea
\gamma_{1,0}=\frac{d(d-4\Df) \Df}{2(1+2\Df)(d-2-2\Df)}\,\gamma_{0,0}\,. \label{eq:beta10}
\eea
This matches the results for the $\phi^4$ interaction deduced in \cite{Heemskerk2009} for $d=2,4$. It is straightforward to apply other functionals and reconstruct the data of the contact interactions in this way, and we have done this in a number of cases.  However, since the main focus of this work is understanding the general structure of functionals we will not do this here, but instead describe how the functional  kernels must be constrained depending on the contact terms we consider. 

\subsubsection{Constraints on kernels}
We can write the functional action as
\bea
\int_{\Gamma^+} \frac{\ud z}{2\pi i}\int_{\Gamma^+} \frac{\ud \bar z}{2\pi i}\left[ f(z,\bar z) \mathcal F(z,\bar z)+f(z,1-\bar z) \mathcal F(z,1-\bar z)\right]
\eea
As usual we must ask whether the action of the functional commutes with infinite sums of crossing symmetric vectors. One condition for this is that the action on the correlator $\mathcal G^{(1)}(z,\bar z)$ should be finite. 
We start from the observation that
\bea
\mathcal G^{(1)}(z,\bar z)&=\frac{\mathcal G^{(1)}(\mbox{$\frac{z}{z-1},\frac{\bar z}{\bar z-1}$})}{(z-1)^\Df(\bar z-1)^{\Df}}+i\pi\,\mbox{sgn}\left[ \mbox{Im} z\right]\,
\mathcal H^{(1)}(z,\bar z)\,,& \qquad \mbox{Im}z \times \mbox{Im} \bar z>&0 \\
\mathcal G^{(1)}(z,\bar z)&=\frac{\mathcal G^{(1)}(\mbox{$\frac{z}{z-1},\frac{\bar z}{\bar z-1}$})}{(z-1)^\Df(\bar z-1)^{\Df}}\,,&\mbox{Im}z \times \mbox{Im} \bar z<&0
\eea
with
\bea
\mathcal H^{(1)}(z,\bar z)\equiv 
\sum_{n,\ell} a_{n,\ell}^{\mbox{\tiny gff}} \gamma_{n,\ell} \frac{ G_{\Delta_{n,\ell},\ell}(\mbox{$\frac z{z-1},\frac{\bar z}{\bar z-1}$})}{(z \bar z)^{\Df}},\label{eq:defH}
\eea
As in section \ref{sec:constraints}, we send $z,\bar z$ to infinity at different rates ($z\sim \Lambda^\alpha, \bar z\sim \Lambda$). Consider first finiteness when acting on the $\mathcal G^{(1)}$ piece. We use crossing to find
\bea
\frac{\mathcal G^{(1)}(\mbox{$\frac{z}{z-1},\frac{\bar z}{\bar z-1}$})}{(z-1)^\Df(\bar z-1)^{\Df}}=\frac{\mathcal G^{(1)}(\mbox{$\frac{1}{1-z},\frac{1}{1-\bar z}$})}{(z-1)^\Df(\bar z-1)^{\Df}}\underset{\Lambda \to \infty}= O(\Lambda^{-\Df(1+\alpha)})\,,
\eea
which follows from the OPE expansion of $\mathcal G^{(1)}$. Hence, we must surely require
\bea
f(\Lambda^\alpha x, \Lambda \bar x)\underset{\Lambda\to \infty}{=} O(\Lambda^{(\Df-1)(1+\alpha)-\epsilon})\qquad \mbox{for some}\quad \epsilon>0 \label{eq:condf1}
\eea
Notice that as long as $\Df>0$ this is weaker than for an ordinary four point function (cf. \reef{eq:constraintskernels}), and can be traced back to the absence of an identity operator for the contact term. Roughly speaking this demands that near infinity we have $f(z,\bar z)\sim (z\bar z)^{\Df-1-\epsilon}$.

Let us turn to $\mathcal H^{(1)}(z,\bar z)$. In this case we cannot use crossing to determine the behaviour at infinity, so we'll have to do a direct computation. We will do our analysis in $d=2$ for simplicity, although the results will be $d$ independent. This is because as functions, the contact term correlators $\mathcal G(z,\bar z)$are independent of $d>1$.
Using the known form of the OPE coefficients together with \reef{eq:gammalargen} we find
\bea
a^{(0)}_{n,\ell}\gamma_{n,\ell}\frac{\Gamma(2\Df+2n)}{\Gamma(\Df+n)^2}\frac{\Gamma(2\Df+2n+2\ell)}{\Gamma(\Df+n+\ell)^2}\, \underset{n\to \infty}{=} O(n^{4\Df+2k-3})
\eea
On the other hand, zooming on the limit where $z,\bar z \to \infty$ with $n/\sqrt{z}, n/\sqrt{\bar z}$ both kept fixed, we have \cite{Fitzpatrick2013}
\bea
G_{\Delta_{n,\ell},\ell}(\mbox{$\frac z{z-1},\frac{\bar z}{\bar z-1}$})\sim \frac{\Gamma(2\Df+2n)}{\Gamma(\Df+n)^2}\frac{\Gamma(2\Df+2n+2\ell)}{\Gamma(\Df+n+\ell)^2} K_0(n/\sqrt{z})K_0(n/\sqrt{\bar z})
\eea
and hence
\bea
\mathcal H^{(1)}(z,\bar z)&\underset{z,\bar z\to \infty}{\sim} z^{-\Df} \bar z^{-\Df} \int^\infty \ud n\, n^{4\Df+2k-3}K_0(n/\sqrt{z})K_0(n/\sqrt{\bar z})\\
& \sim z^{k-1} \left(\frac{z}{\bar z}\right)^{\Df}\, _2F_1(2\Df+k-1,2\Df+k-1,4\Df+2k-2,1-z/\bar z)\,.
\eea
We conclude that finiteness of the functional action requires:
\bea
f(\Lambda^{\alpha} x,\Lambda \bar x)=O(\Lambda^{-1-\alpha k+\Df(1-\alpha)-\epsilon}) \label{eq:condf2}\,.
\eea

It is not hard to see using the kind of arguments of section \ref{sec:constraints} that these constraints are also sufficient for the swapping condition to hold. This constraint is in general stronger than that of \reef{eq:condf1}, namely for $k> 0$ or $k=0$ and $\Df>1/2$. The key point to retain here is that as we increase the number of derivatives in the contact term, we must ask for softer behaviour of the functional kernels at large $z,\bar z$. As we do this, some functionals become disallowed, which means less functional equations, and therefore less constraints on the coefficients $a_{n,\ell}^{(1)}$ and $\gamma_{n,\ell}$. This allows for more solutions, solutions which precisely correspond to the contact terms we are trying to bootstrap.

As an example, when $k=0$ and $\Df=1$ we can use functionals with $p,q \geq -1$, but when $k=2$ we must set $p\geq -1$ and $q\geq 1$ or $p,q\geq 0$. This means that in the latter we lose those functional equations obtained by acting with functionals with $(p,q)=(-1,-1)$ and $(p,q)=(-1,0)$. We have checked that this leads to one extra solution which precisely corresponds to the contact term with $k=2$.

\section{$d=2$ product functionals}
\label{sec:d2prod}

In this section we will study a class of functionals which act on the $d=2$ crossing equation. The form of these functionals follows from the factorized structure of the $d=2$ conformal blocks. Indeed, these are essentially the product of two copies of the $d=1$ conformal blocks \reef{eq:sl2block}, one for each cross-ratio $z,\bar z$. This suggests then acting on each factor independently with a $d=1$ functional. In the next subsection we give more details on these product functionals and their action on the crossing equation. In particular we focus on the case where both factors in the product functionals belong to the $d=1$ bases discussed in section \ref{sec:review} and appendix \ref{sec:bases}, which are dual to generalized free fields. In subsection \ref{sec:productcorrelators} we show that the resulting set of functionals bootstraps a simple set of correlation functions in $d=2$, which includes the energy correlator for the 2d Ising model. In subsections \ref{sec:analyticnumericdimbounds} and \ref{sec:opebound} we show how these functionals can be used to obtain analytic and numeric bounds on 2d CFT data.

\subsection{Functional form and basis decomposition}

We begin by introducing convenient notation:
\bea
k_h(z|\Df)&:=z^{\frac{h}2-\Df}\,_2F_1\left(\frac h2,\frac h2,h;z\right),\\
F_{h}(z|\Df)&:= k_h(z|\Df)-k_h(1-z|\Df),\\
 H_h(z|\Df)&:=k_h(z|\Df)+k_h(1-z|\Df)
\eea
Note that in terms of the 1d notation of section \ref{sec:review} we have
\bea
k_h(z|\Df)=G_{\frac h2}(z|\mbox{$\frac {\Df}2$}),\quad F_h(z|\Df)=F_{-,\frac h2}(z|\mbox{$\frac {\Df}2$}),\quad H_h(z|\Df)=F_{+,\frac h2}(z|\mbox{$\frac {\Df}2$})\,. \label{eq:newnotation}
\eea
For convenience we will also define $\partial F_h=\partial_{\Delta} F_{-,\Delta}|_{\Delta=h/2}$ and similarly for $\partial H_h$.

In two dimensions the conformal blocks take on a factorized form:
\bea
G_{\Delta,\ell}(z,\bar z)=\frac 12 \left[k_\tau(z) k_\rho(\bar z)+\left(z\leftrightarrow \bar z\right)\right]\,, \qquad \tau=\Delta-\ell,\quad \rho=\Delta+\ell\,.\label{eq:2dblock}
\eea
Accordingly the crossing vectors become:
\bea
F_{\Delta,\ell}(z,\bar z|\Df)=\frac 14\left[ F_{\tau}(z|\Df)H_{\rho}(\bar z|\Df)+ F_{\rho}(z|\Df)H_{\tau}(\bar z|\Df)+(z\leftrightarrow \bar z)\right] \label{eq:f2d}
\eea
In section \ref{sec:review} we have seen that in $d=1$ there are functional bases which act nicely on $F_h, H_h$. Hence it is natural to take a tensor product of those functionals acting separately on each such factor in the equation above. This can be achieved by defining functionals $\omega_-\otimes \omega_+$ which act as follows:
\bea
(\omega_-\otimes \omega_+)(\Delta,\ell):=2\,\int_{++} \frac{\ud z \ud \bar z}{\pi^2}\, h_-(z)h_+(\bar z)\,\left[ \mathcal I_z \mathcal I_{\bar z} F_{\Delta,\ell}(z,\bar z)+\mathcal I_z \mathcal I_{\bar z} F_{\Delta,\ell}(z,1-\bar z)\right]
\eea
with $h_{\pm}$ being $d=1$ functional kernels. The functional form was chosen without loss of generality such that we symmetrize under $\bar z\leftrightarrow 1-\bar z$ before acting with the functionals. The above is a $d=2$ functional which in the notation of section \ref{sec:ansatz} corresponds to setting
\bea
h_{++}(z,\bar z)&=h_-(z) h_{+}(\bar z)+h_-(\bar z)h_+(z)\,,\\
h_{+-}(z,\bar z)&=h_{+}(z) h_-(1-\bar z) -h_-(z) h_{+}(1-\bar z)\,,
\eea
or in terms of representation \reef{eq:fgerep}:
\bea
f(z,\bar z)/\pi^2&=-\left[f_{-}(z)f_{+}(\bar z)+f_{-}(\bar z)f_{+}(z)\right]\,,\\
e(z,\bar z)/\pi^2&=\bar e(\bar z,z)/\pi^2=-i\left[g_{-}(z)f_{+}(\bar z)+g_+(z)f_-(\bar z)\right]\,,\\
g(z,\bar z)/\pi^2&=g_{-}(z)g_+(\bar z)+g_-(\bar z)g_+(z)\,,\\
\tilde g(z,\bar z)/\pi^2&=g_-(z)g_+(\bar z)-g_-(\bar z)g_+(z)\,,\label{eq:ftof}
\eea
where the single variable $f_{\pm},g_{\pm}$ are determined in terms of $h_{\pm}(z)$, as explained in appendix \ref{sec:bases1}.\footnote{A small technicality is that it is assumed that, where they appear in equations \reef{eq:ftof}, the $f_{\pm}$ should be evaluated with their argument on the upper half-plane; otherwise they should be multiplied by an extra minus sign).} 
Note that for the moment the above are true for any choice of $d=1$ functionals, whether they belong to GFF-dual bases or not.

In order for the above to be valid $d=2$ functionals, we must ensure that the constraints determined in section \ref{sec:constraints} are satisfied. This in turn implies constraints on the $d=1$ functionals we can use. These constraints turn out to be the same as the constraints that guarantee that the $d=1$ functionals are consistent with the $d=1$ crossing equation for $d=1$ correlators. These are determined as in the higher dimensional case by demanding that the action of the functional commutes with infinite sums of crossing vectors.\footnote{To be precise there are many kinds of such sums, but here we have in mind those which are associated to crossing symmetry of an ordinary, unitary 1d correlation function of identical operators.} They demand \cite{Mazac:2018}:\footnote{In the constraint on $g(z)$ we have taken $\Df\to \Df/2$, consistently with \reef{eq:newnotation}.}
\bea
f(z)&\underset{z\to i\infty}{=} O(z^{-1-\epsilon})\\
g(z)&\underset{z \to 1^-}{=} O[(1-z)^{\Df-1+\epsilon}]\,.
\eea
Assuming these it is easy to see that the constraints on $f,g,e,\tilde e$ summarized in \ref{eq:constraintskernels} also hold, and in fact they do so in the tightest sense possible, meaning that the $\epsilon$ factors in those equations and in the above turn out proportional to each other.

In this way any well-defined $d=1$ functionals induce well-defined  tensor product functionals in $d=2$, and acting with these functionals on the crossing equation leads to a valid functional bootstrap equation:
\bea
\sum_{\Delta,\ell} a_{\Delta,\ell}\, \left(\omega_{-}\otimes \omega_+\right)(\Delta,\ell|\Df)&=0\,. \label{eq:prodequations}
\eea
The functional action itself reduces to a product of $d=1$ actions. Given \reef{eq:newnotation} it will be convenient throughout this section to redefine the $d=1$ functional actions as
\bea
\omega\left(\frac{h}2\bigg|\frac{\Df}2\right)\to \omega(h|\Df)\,. 
\eea
Keeping this is mind, the functional action of a tensor product functional $\omega_-\otimes \omega_+$ is given by:
\bea
(\omega_-\otimes \omega_+)(\Delta,\ell|\Df)=\frac 12\left[\omega_-(\tau|\Df)\omega_+(\rho|\Df)+\omega_-(\rho|\Df)\omega_+(\tau|\Df)\right]
\eea

We can choose any 1d functionals we wish as long as they are consistent with the 1d crossing equation. As an example, we could use the $\theta, \tilde \theta$ functionals introduced in \reef{eq:thetafuncs}. In that case it is easy to see that the tensor product functionals are simply the $d=2$ HPPS functionals of the previous section. Instead, here we are interested in using the $d=1$ functionals reviewed in section \ref{sec:review} and appendix \ref{sec:bases}, namely $\alpha_{\pm},\beta_{\pm}$. For definiteness we will focus on the fermionic bases here and for convenience change notation as $\omega^F_{-,n}\to \omega_n^-$. We will denote schematically the four different kinds of functionals as
\bea
\omega^-_n \otimes \omega^+_m \to \alpha \alpha_{n,m},\quad \alpha \beta_{n,m},\quad \beta \alpha_{n,m}, \quad \beta \beta_{n,m}\,.
\eea
Notice that in this notation the first functional in the product is always of the $-$ type. The tensor product functional actions take the schematic form
\bea
\omega\omega_{n,m}(\Delta,\ell)=\sin^2\left[\frac{\pi}4(\tau-h_n)\right]\sin^2\left[\frac{\pi}4(\rho-h_m)\right]
\,R_{n,m}(\tau,\rho)+\left(\rho \leftrightarrow \tau\right)\,.
\eea
Here the $h_n$ are determined by the 1d generalized free fermion operator dimensions and given by $h_n=2\Df+2+4n$. We see that the functionals have generically quadruply spaced fourth-order zeros for even spins $\ell=2(n-m)$ and $\Delta=2\Df+2+4 m+\ell$ (setting $n\geq m$ without loss of generality). 

Recall that the 1d functionals provide basis decompositions of the form \reef{eq:basesdecomp}
\begin{subequations}
\begin{align}
H_{h}(z)&=\sum_{m=0}^{+\infty} \left[\alpha_m^+(h) H_{h_m}(z)+\beta^+_m(h) \partial H_{h_m}(z)\right]\\
F_{h}(z)&=\sum_{m=0}^{+\infty} \left[\alpha_m^-(h) F_{h_m}(z)+\beta_m^-(h) \partial F_{h_m}(z)\right]\label{eq:fbasis}.
\end{align}
\end{subequations}
Plugging in these expressions into the 2d crossing vector \reef{eq:f2d} we obtain:
\bea
F_{\Delta,\ell}(z,\bar z)=\mathcal F_{\Delta,\ell}(z,\bar z)&+\mathcal F_{\Delta,\ell}(\bar z,z)\\
\mathcal F_{\Delta,\ell}(z,\bar z)=
\frac 14\sum_{m,n=0}^{\infty} \bigg\{&  \left[ \alpha_m^-(\tau)\alpha_n^+(\rho)+\alpha_m^-(\rho)\alpha_n^+(\tau)\right]\,F_{h_m}(z) H_{h_n}(\bar z) \\
+&\left[ \beta_m^-(\tau)\alpha_n^+(\rho)+\beta_m^-(\rho)\alpha_n^+(\tau)\right]\,\partial F_{h_m}(z) H_{h_n}(\bar z)
\\
+&\left[ \alpha_m^-(\tau)\beta_n^+(\rho)+\alpha_m^-(\rho)\beta_n^+(\tau)\right]\,F_{h_m}(z) \partial H_{h_n}(\bar z)
\\
+&\left[ \beta_m^-(\tau)\beta_n^+(\rho)+\beta_m^-(\rho)\beta_n^+(\tau)\right]\,\partial F_{h_m}(z) \partial H_{h_n}(\bar z)
\bigg\}\,.
\label{eq:fbasis2d}
\eea

Hence, we can think of the product functionals as extracting the decomposition coefficients of the crossing vector into a basis of functions made up of products of $F,H$ and their derivatives. In particular the product functionals satisfy orthonormality relations which follow directly from those stated in \reef{eq:ONfermion}. Setting $\Delta_{p,\ell}=2+2\Df+4p+\ell$ we have for instance
\bea
\alpha\alpha_{m,n}(\Delta_{p,\ell},\ell)=\delta_{n,p} \delta_{\ell,2(m-n)}+\delta_{m,p} \delta_{\ell,2(n-m)}\,,\\
\partial_\Delta \alpha\alpha_{m,n}(\Delta_{p,\ell},\ell)=\partial_\ell \alpha\alpha_{m,n}(\Delta_{p,\ell},\ell)=\partial_\Delta \partial_\ell\alpha\alpha_{m,n}(\Delta_{p,\ell},\ell)=0\,. \label{eq:ONprodbasis}
\eea
and similar ones for $\alpha\beta,\beta\alpha$ and $\beta\beta$.
It is interesting to compare these with the duality conditions \reef{eq:gffduality} (for functionals which, we remind the reader, we did not obtain). There we had two sets of functionals which were associated with the GFF spectrum. The GFF spectrum has operators spaced in steps of two. Here instead we have two times as many functionals, but they are dual to a spectrum with operators spaced in steps of four, so in a (ill-defined) sense, the number of degrees of freedom is still the one to be expected.

Relatedly, we may wonder whether the present basis is associated to some simple solution to crossing, as it would be the case for generalized free fields. In $d=1$ one way to see this was to start from the basis decomposition equation \reef{eq:fbasis} and set $h=0$. This eliminates the derivative terms on the righthand side (since $\beta_p^-(0)=0$), and turns the equation into the crossing equation for the 1d GFF correlator. In fact, even for non-zero $h$ the same equation expresses crossing symmetry of (crossing-symmetric sums of) Witten exchange diagrams in AdS$_2$. In the present context, the situation seems to be very different. For instance, the decomposition \reef{eq:fbasis2d} does not even express the general crossing vector $F_{\Delta,\ell}$ in terms of other crossing vectors or their derivatives. In the next subsection we will see how the basis decomposition can nevertheless sometimes be associated to a special class of solutions to crossing, although as far as we able to determine, not in general. 

To summarize, we have shown that there exists a set of product functionals which acts on the 2d crossing equation. In fact although we have focused on taking products of fermion $\times$ fermion in the holomorphic and antiholomorphic sectors, nothing stops us from considering other choices, such as boson$\times$ boson or even mixed bases fermion $\times$ boson or boson $\times$ fermion. Acting with these functionals leads to a set of functional bootstrap equations in~2d:
\bea
\sum_{\Delta,\ell} a_{\Delta,\ell}\, \alpha\alpha_{m,n}(\Delta,\ell|\Df)&=0\,,& \sum_{\Delta,\ell} a_{\Delta,\ell}\, \beta\beta_{m,n}(\Delta,\ell|\Df)&=0\,,\\
\sum_{\Delta,\ell} a_{\Delta,\ell}\, \alpha\beta_{m,n}(\Delta,\ell|\Df)&=0\,,&
\sum_{\Delta,\ell} a_{\Delta,\ell}\, \beta\alpha_{m,n}(\Delta,\ell|\Df)&=0\,,
\label{eq:funcequations2d}
\eea
which must hold for all integer $m,n\geq 0$. These equations are certainly necessary for crossing symmetry to hold. Whether they are also sufficient is a harder question which we will not resolve in this work. We will comment on the missing ingredients to proving this in the discussion section.

\subsection{Bootstrapping the $\langle \varepsilon\varepsilon\varepsilon\varepsilon\rangle$ correlator}
\label{sec:productcorrelators}
In the 2d Ising model, the correlator of four energy operators $\varepsilon$ takes on a simple factorized form
\bea
\langle \varepsilon\varepsilon\varepsilon\varepsilon\rangle=\frac{\mathcal G(z,\bar z)}{x_{13}^{2\Df}x_{24}^{2\Df}}, \qquad \mathcal G(z,\bar z)=\left(\frac{1}{z^{\Df}}+\frac{1}{(1-z)^\Df}-1\right)\left(\frac{1}{\bar z^{\Df}}+\frac{1}{(1-\bar z)^\Df}-1\right)\,,
\eea
with $\Df=1$. The correlator is a product of two $d=1$ generalized free fermion solutions, and is part of an infinite family of solutions to crossing with $\Df$ an odd integer (had we taken the product of two 1d generalized free boson solutions, we should have instead chosen $\Df$ even). This restriction to integer $\Df$ arises not from the crossing equation itself, but from demanding that only integer (even) spins appear in the conformal block decomposition of the correlator. To see this, we begin by noting:
\bea
\frac{1}{(1-z)^\Df}-1=\sum_{n=0}^{+\infty} a_{h_n}^{\mbox{\tiny free}} \frac{k_{h_n}(z)}{z^{\Df}}
\eea
where $h_n=2+2\Df+4n$ and the coefficients $a_{h_n}^{\mbox{\tiny free}}$ are simply related to the $a_{\Delta}^{\mbox{\tiny gff}}$ introduced in section \ref{sec:review}:
\bea
a_{h}^{\mbox{\tiny free}}=\frac{2 \Gamma\left(\frac{h}2\right)^2}{\Gamma(h-1)\Gamma(\Df)^2}\, \frac{\Gamma\left(\frac{h+2\Df-2}2\right)}{\Gamma\left(\frac{h+2\Df+2}2\right)} \label{eq:opetwist}\,.
\eea
Using expression \reef{eq:2dblock} for the 2d conformal blocks, the correlator becomes:
\bea
(z \bar z)^{\Df}\mathcal G(z,\bar z)=1+\sum_{k=0}^\infty2 a_{2\ell}^{\mbox{\tiny free}}  G_{\ell,\ell}(z,\bar z)\bigg|_{\ell=1+\Df+2k}+\sum_{n,m=0}^{+\infty}a_{h_n}^{\mbox{\tiny free}}a_{h_m}^{\mbox{\tiny free}} G_{\frac{h_n+h_m}2,\frac{|h_n-h_m|}2}(z,\bar z)\,.
\eea
We see that the spectrum contains twist zero operators as well as operators with dimension
\bea
\Delta_{n,\ell}^{\mbox{\tiny prod}}=2+2\Df+4n+\ell,\qquad \mbox{with}\quad \ell \quad \mbox{even}\,,\qquad n\in \mathbb N_{\geq 0}\,,
\eea
and corresponding OPE coefficients
\bea
a^{\mbox{\tiny prod}}_{n,\ell}&:=a_{\Delta_{n,\ell}^{\mbox{\tiny prod}},\ell}=\left\{
\begin{array}{cc}
\left(a_{2+2\Df+4n}^{\mbox{\tiny free}}\right)^2,& \ell=0 \\
2\,a_{2+2\Df+4n}^{\mbox{\tiny free}}\,a_{2+2\Df+4m}^{\mbox{\tiny free}},& \ell=2|n-m|.
\end{array}
\right.\,,\\
a_{\ell,\ell}&=2 a_{2\ell}^{\mbox{\tiny free}}\, \qquad \ell=1+\Df+2k, \qquad k\in \mathbb N_{\geq 0}\,.
 \label{eq:isingope}
\eea
The twist zero states will only have even integer spin if $\Df$ is an odd integer. It is interesting to note that regarding the functional bootstrap equations \reef{eq:funcequations2d}, the constraint of integrality of spin is somewhat artificial. The above solutions do satisfy those equations if we allow for a moment arbitrary real spin. The upshot is that while for generic $\Df$ the solutions to crossing strictly speaking do not exist, the dual functional bases still do and in fact make perfect sense, as do the set of associated bootstrap equations.

It is interesting to see how these product solutions actually solve the functional equations. We write the crossing equation as
\bea
-F_{0,0}(z,\bar z)-\sum_{k=0}^{+\infty}a_{\ell,\ell} F_{\ell,\ell}(z,\bar z)\bigg|_{\ell=1+\Df+2k}=\sum_{n,\ell}^{+\infty}a^{\mbox{\tiny prod}}_{n,\ell} F_{\Delta_{n,\ell}^{\mbox{\tiny prod}},\ell}(z,\bar z)\,,\label{eq:crossingprod}
\eea
and attempt to recover the correct OPE coefficients \reef{eq:isingope} using the functional equations. To begin with we note that the action of functionals on the righthand side is particularly simple: the $\alpha\alpha_{p,q}$ functionals pick out the term with $\ell=2|p-q|$ and $n=\mbox{min}\{p,q\}$ (as follows from \reef{eq:ONprodbasis}), whereas all other functionals completely annihilate it. The action of the $\beta \alpha$ and $\beta \beta$ is even simpler, since they also trivially annihilate the left-hand side.\footnote{The difference between $\alpha \beta$ versus $\beta \alpha$ for instance arises because $\beta^-_p(0)=\beta^-_p(2+2\Df+4k)=0=\beta^+_p(2+2\Df+4k)$, but $\beta^+_p(0)\neq 0$.} Acting first with $\alpha \beta$ we find 
\bea
-2\alpha_q^-(0)\beta_p^+(0)-\sum_{k=0}^{+\infty}a_{\ell,\ell} \left[\alpha_q^-(0)\beta_p^+(2\ell)+\alpha_q^-(2\ell)\beta_p^+(0)\right]\bigg|_{\ell=1+\Df+2k}=0\\
\Leftrightarrow \beta_p^+(0)\left(2 a_{h_q}^{\mbox{\tiny free}}-a_{\ell,\ell}\right)\bigg|_{\ell=1+\Df+2q}=0
\eea
which agrees with \reef{eq:isingope} since $h_q=2+2\Df+4q$. Here we have used the duality properties \reef{eq:ONfermion} as well as the identity action \reef{eq:idaction}. We should thus think of the $\beta \alpha$ functionals as bootstrapping the twist zero Regge trajectory. Acting now with the $\alpha\alpha$ functionals fixes the higher twist sector in terms of this low twist data:
\bea
(1+\delta_{p,q})\,a_{n,\ell}^{\mbox{\tiny prod}}&=-2\alpha_p^+(0)\alpha_{q}^-(0)-\sum_{k=0}^{+\infty}2 a_{h_k}^{\mbox{\tiny free}}\left[\alpha_p^+(0)\alpha_q^-(2\ell)+\alpha_p^+(2\ell)\alpha_q^-(0)\right]\bigg|_{\ell=1+\Df+2k}\\
\Leftrightarrow a_{n,\ell}^{\mbox{\tiny prod}}&=\frac{2}{1+\delta_{p,q}} a_{h_p}^{\mbox{\tiny free}} a_{h_q}^{\mbox{\tiny free}}\,, \qquad n=\mbox{min}\{p,q\},\qquad \ell=2|p-q|\,.
\eea
A different perspective on this solution of the bootstrap equations comes from the basis decomposition \reef{eq:fbasis2d}. As we've pointed out, and unlike the analogous equation \reef{eq:basesdecomp} in 1d, this equation does not relate crossing vectors on both sides, so there is no obvious way to recover from it something resembling a crossing-symmetry equation. This is the case for instance if we take that decomposition and set $\Delta=\ell=0$. However, if instead we consider the full twist zero trajectory i.e. the left-hand side of equation \reef{eq:crossingprod} and use then \reef{eq:fbasis2d}, then it does lead to the righthand side of the former equation on the nose. Indeed, this is a completely equivalent way of bootstrapping the tensor product solution. This suggests that basis decomposition equations might take a simpler form if we use them to express whole Regge trajectories rather than individual operators.

\subsection{Analytic and numeric dimension bounds}
\label{sec:analyticnumericdimbounds}
The functional bootstrap equations \reef{eq:funcequations2d} constitute a set of sum rules that the OPE density $a_{\Delta,\ell}$ must satisfy. In this subsection and the next we explore how these can be used derive bounds on the CFT data, by exploiting positivity of the OPE density together with positivity properties of the functional actions.

A typical bootstrap question is to ask whether there is an upper bound on the scaling dimension of the first non-trivial scalar operator \cite{Rattazzi:2008pe}. In the functional language we ask what is the largest allowed $\Delta_0$ such that the functional bootstrap equations can still be satisfied,
\bea
\omega(0,0)+\sum_{\Delta\geq \Delta_0} a_{\Delta,0}\, \omega(\Delta,0)+\sum_{\Delta\geq \ell,\ \ell=2,4,\ldots} a_{\Delta,\ell} \omega(\Delta,\ell)=0\label{eq:funcbound}
\eea
where $\omega$ can stand for any functional in the product basis. In practice we can tackle such problems by taking $\omega$ to be a finite linear combination of basis elements and demanding
\bea
\omega(0,0)>0\,,\qquad \omega(\Delta,0)\underset{\Delta\geq \Delta_0}{\geq} 0\,,\qquad \omega(\Delta,\ell)\underset{\Delta\geq \ell}{\geq} 0 \quad \ell=2,4,\ldots\label{eq:funcconds}
\eea
which together with positivity of the $a_{\Delta,\ell}$ would contradict \reef{eq:funcbound}.

\begin{figure}%
\begin{center}
\begin{tabular}{cc}
\includegraphics[width=7cm]{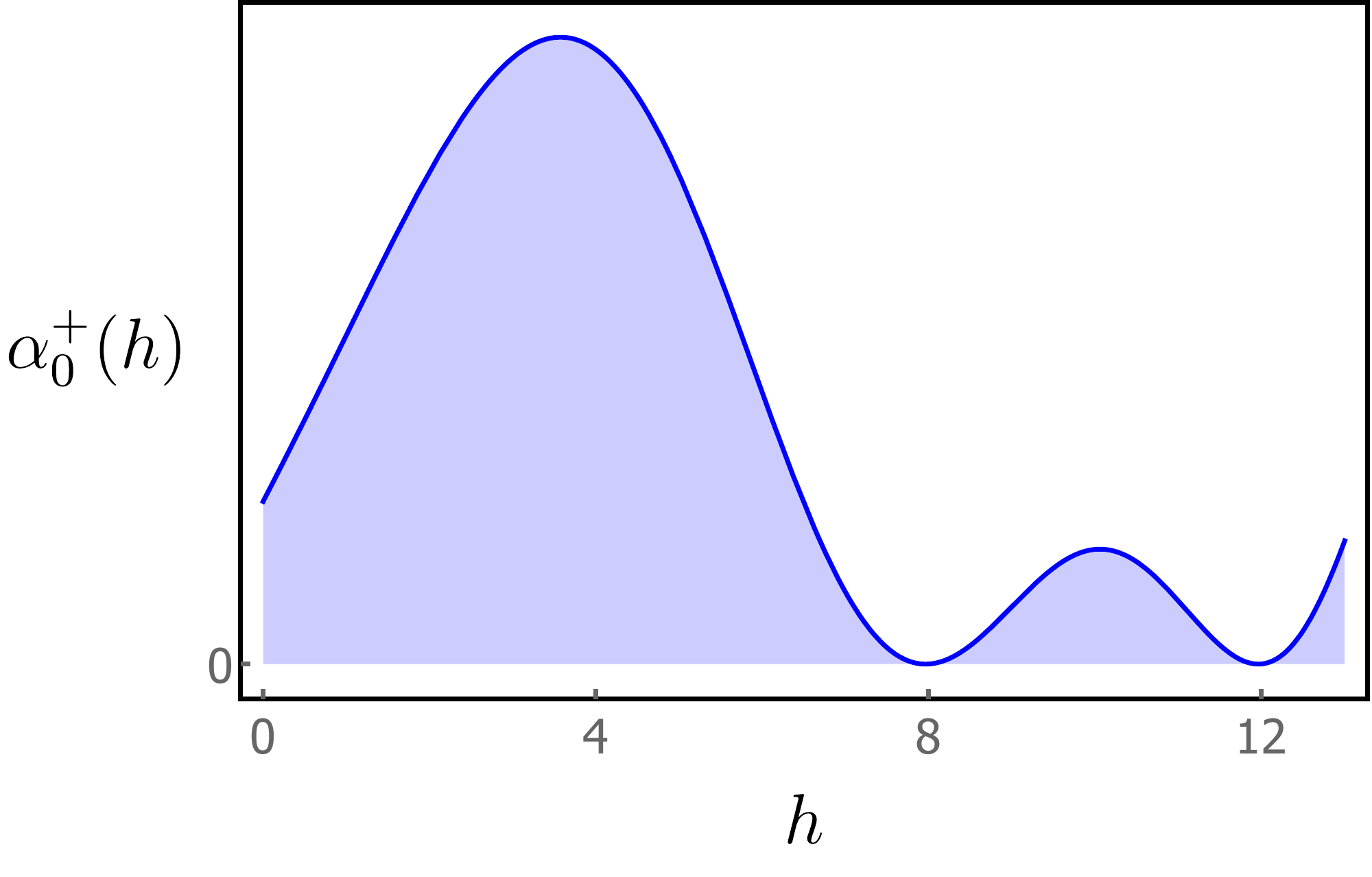}%
&
\includegraphics[width=7cm]{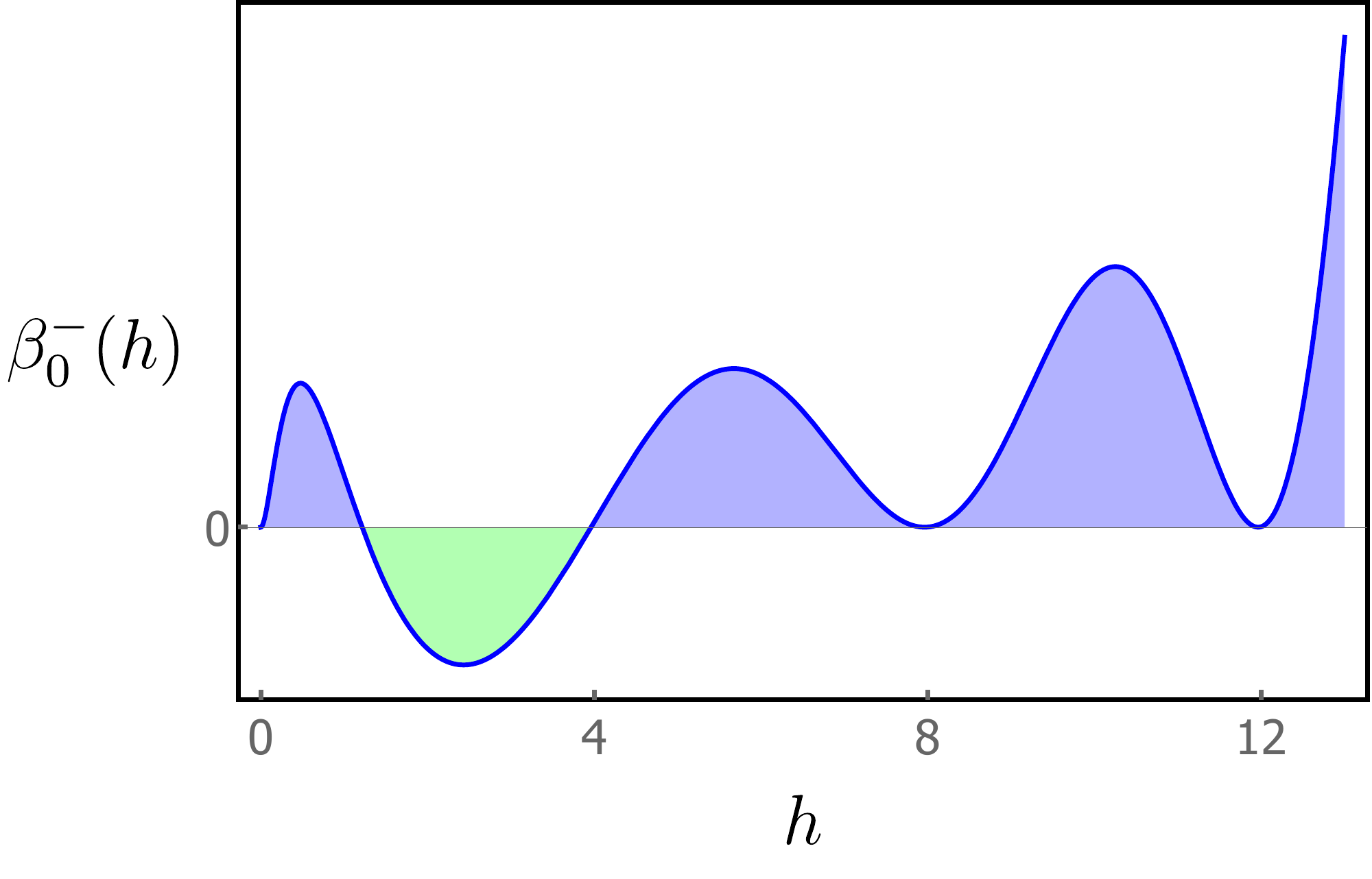}%
\end{tabular}
\caption{Schematic plots of the functional actions $\alpha_0^+(h)$ and $\beta_0^-(h)$ for $\Df=1$. Since we only care about their positivity properties, we have rescaled them by non-negative functions of $h$ for clarity. For $\Df<1$ the curves look similar replacing $4,8,\ldots$ by $2+2\Df, 4+2\Df,\ldots$.}%
\label{fig:funcs0}%
\end{center}
\end{figure}

We begin by showing that simply taking $\omega=\beta\alpha_{0,0}$ gives an upper bound $\Delta_0\leq 2+2\Df$ when $\Df\leq 1$.\footnote{More rigorously, strict positivity of the functional action on the identity demands the addition of $\alpha\alpha_{00}$ with an infinitesimal coefficient.} When $\Df=1$ this bound is optimal as it is saturated by the $\langle \varepsilon\varepsilon\varepsilon\varepsilon\rangle$ correlator in the 2d Ising model which we discussed in the previous subsection. The functional action is
\bea
\beta\alpha_{00}(\Delta,\ell)=\beta_0^-(\tau)\alpha_0^+(\tau+2\ell)+\beta_0^-(\tau+2\ell)\alpha_0^+(\tau) \label{eq:exactfunc}
\eea
In figure \ref{fig:funcs0} we plot the functionals $\alpha_0^+, \beta_0^-$ for $\Df=1$. 
The figure changes little for $\Df\leq 1$. By inspection, when $\Df\leq 1$ the functional $\alpha_0^+$ is non-negative when its argument is. As for $\beta_0^-$, it is also definitely positive when its argument is larger than $2+2\Df$. It follows that the complete action is positive if $\tau\geq 2+2\Df$ independently of spin. However to get a valid bound conditions \reef{eq:funcconds} tell us that we need to ensure  that for $\ell\geq 2$ the functional action is actually positive for all $\tau\geq 0$, i.e. for all operators consistent with unitarity for those spins. 

\begin{figure}%
\begin{center}
\begin{tabular}{cc}
\includegraphics[width=7cm]{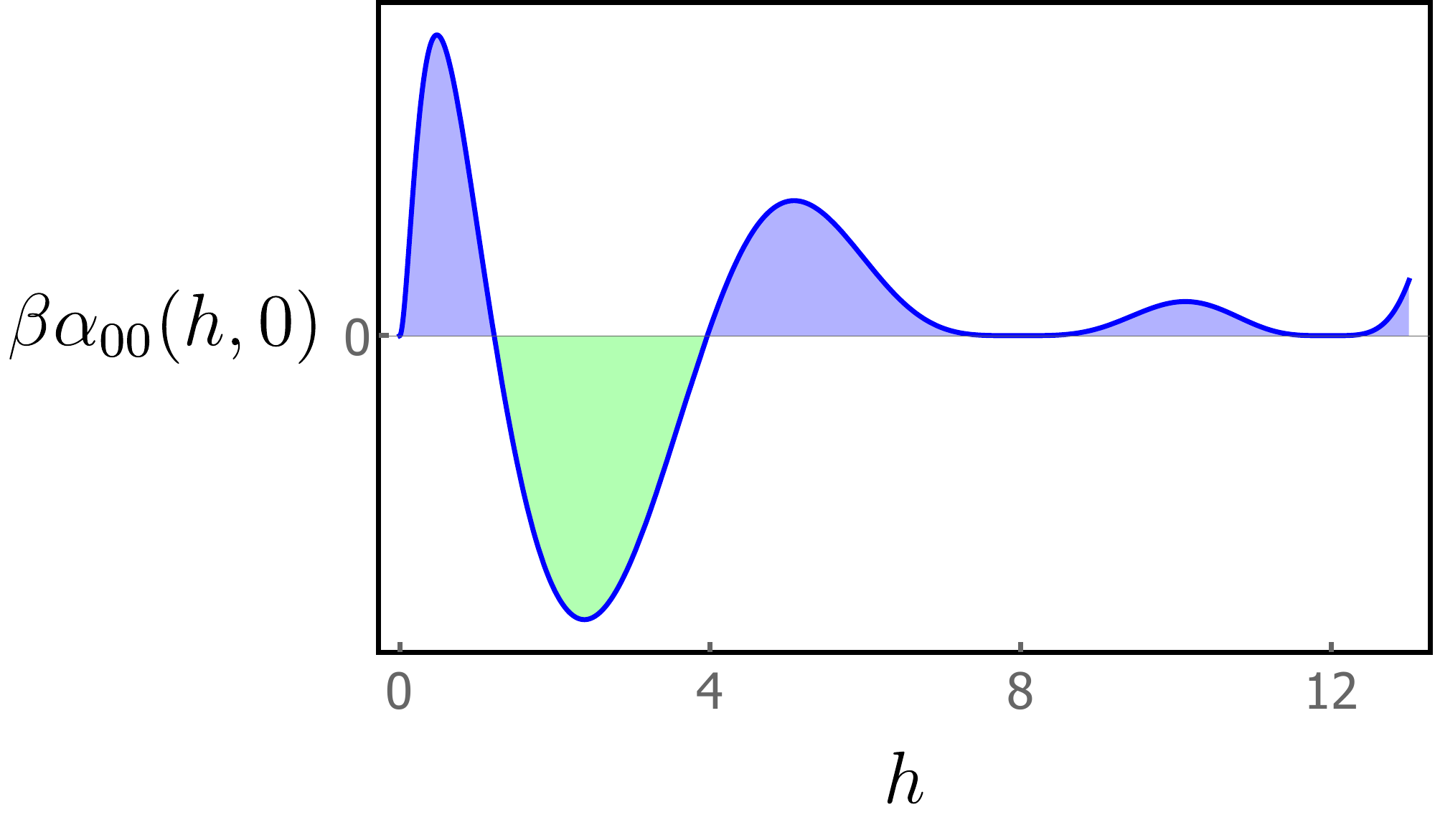}%
&
\includegraphics[width=7cm]{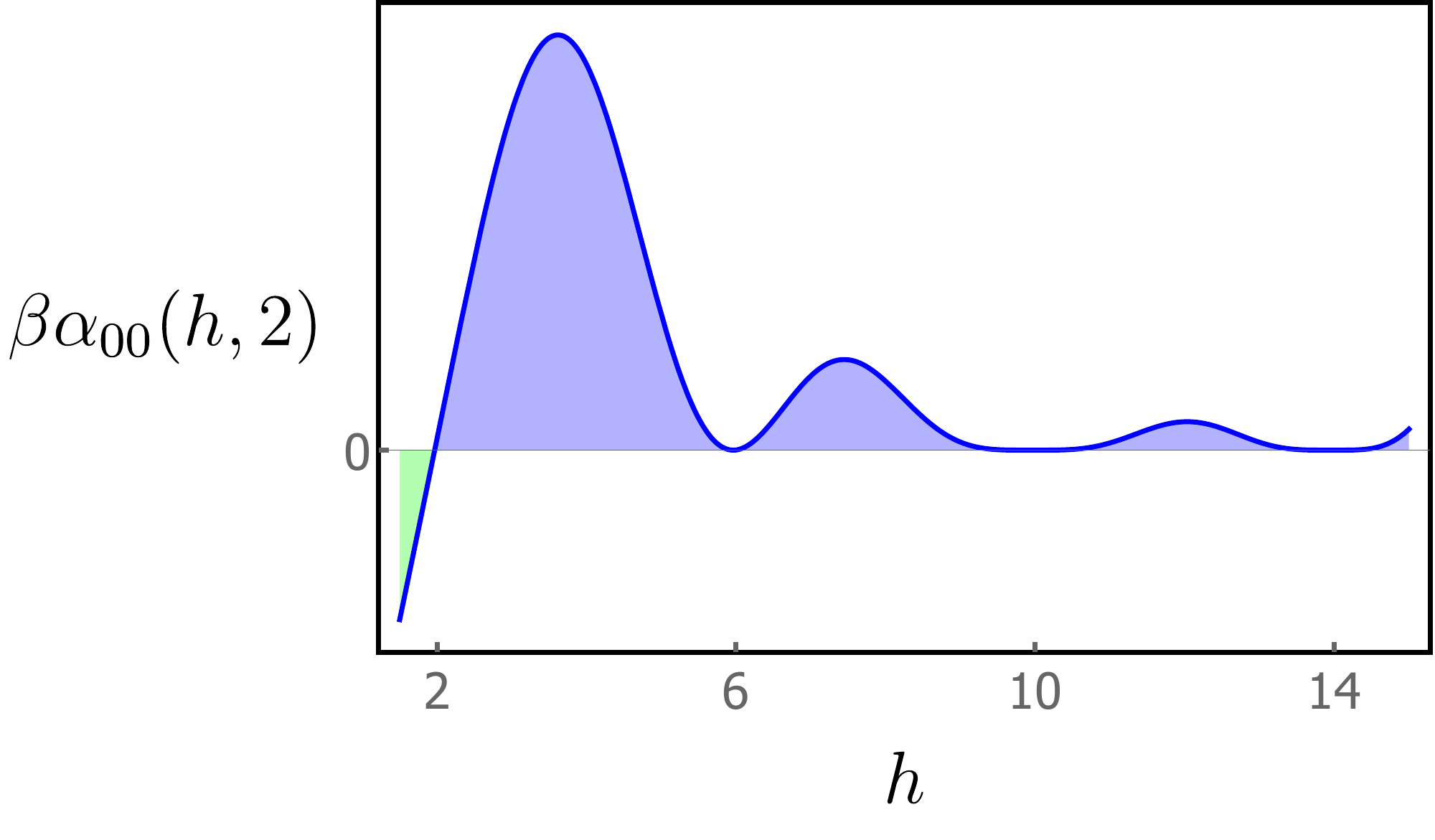}%
\\
\includegraphics[width=7cm]{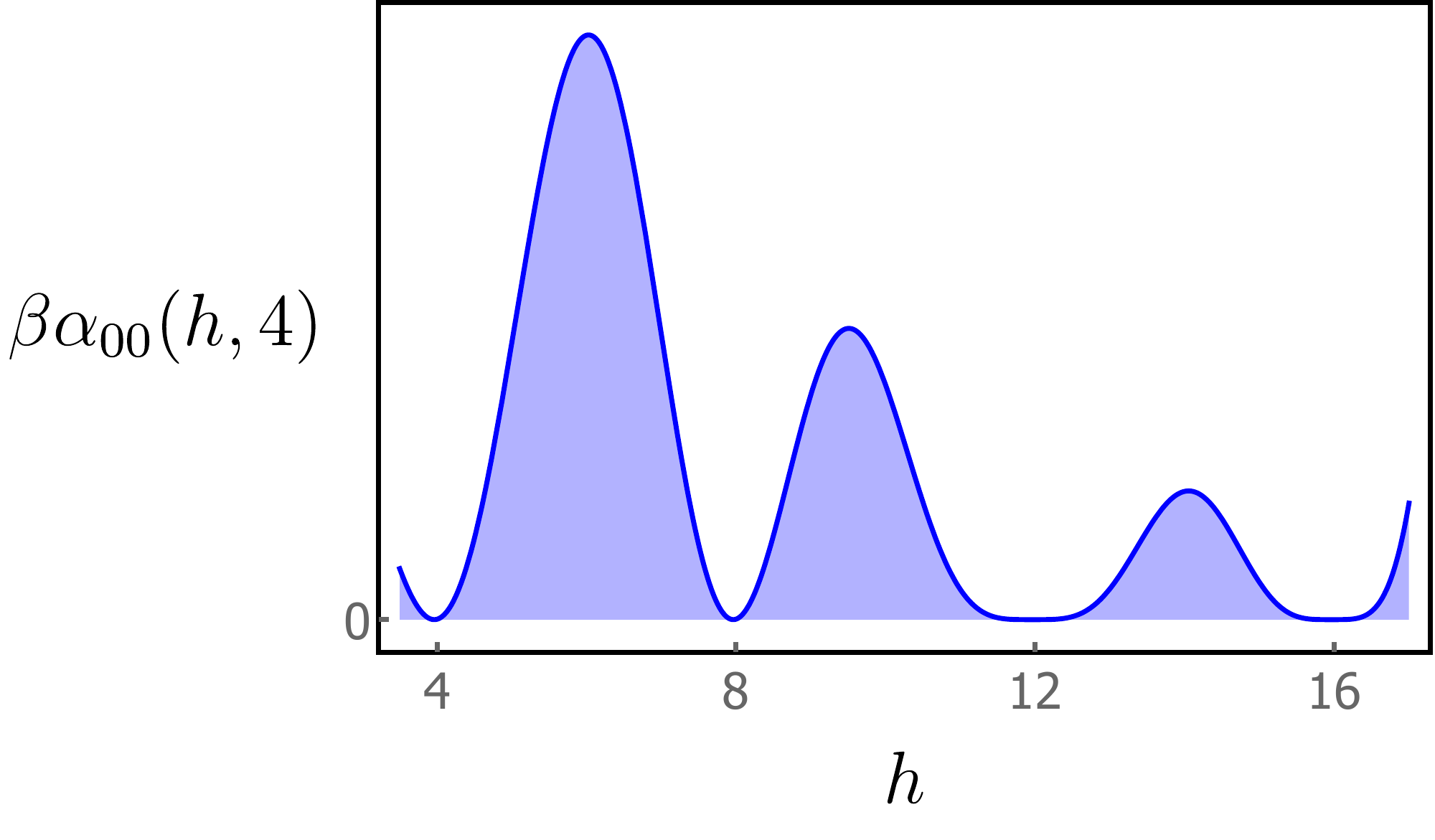}%
&
\includegraphics[width=7cm]{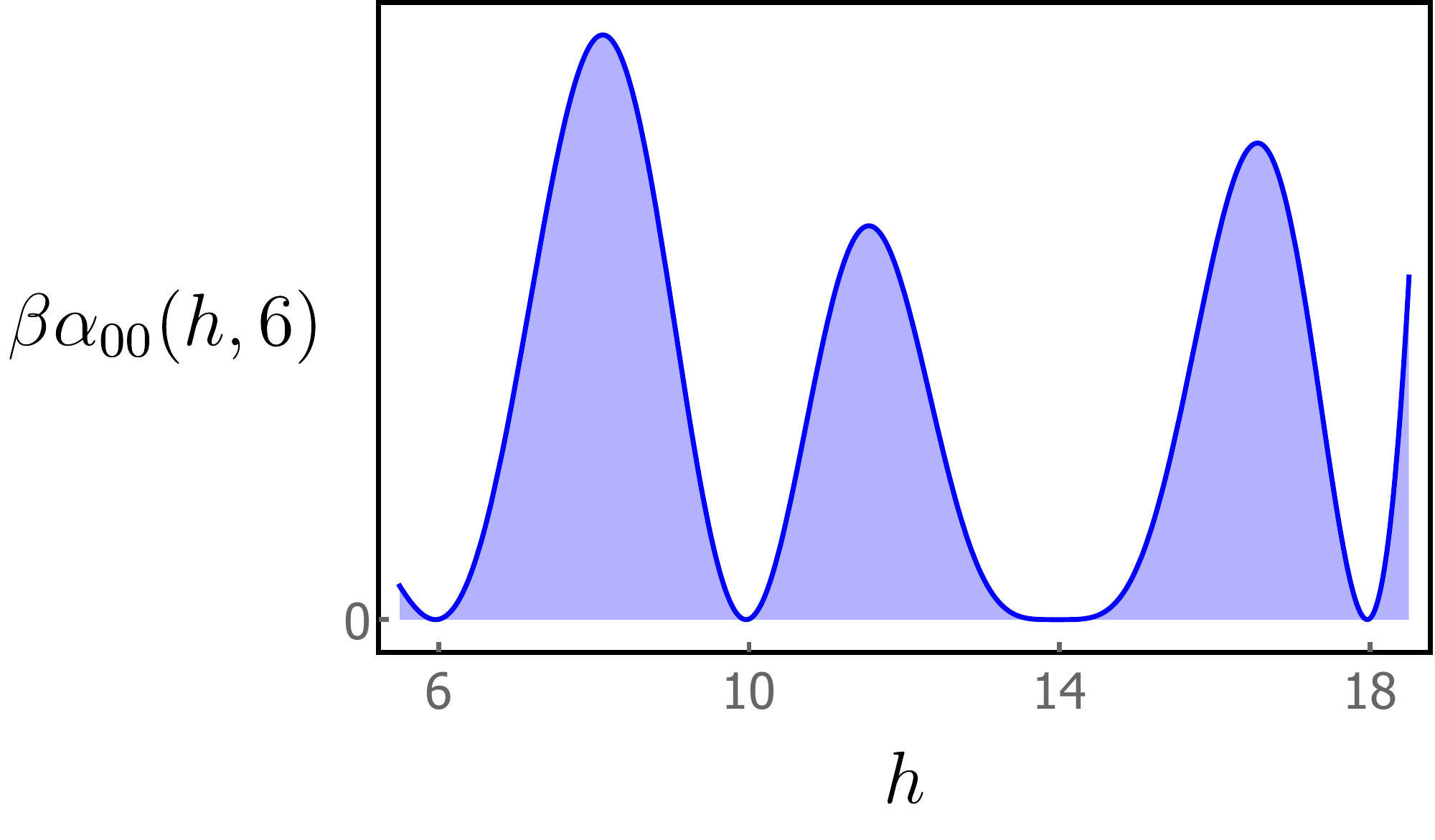}%
\end{tabular}
\caption{Functional actions $\beta\alpha_{00}(\Delta,\ell)$ for various spins and $\Df=1$. For $\ell=0$ the functional becomes positive when $\Delta>4$. Similarly there is a first order zero at $\Delta=\ell=2$. These imply that for $\Df=1$ there must be at least one scalar operator with dimension smaller or equal to four, and higher than $\sim 1.215$.}%
\label{fig:positivity}%
\end{center}
\end{figure}

Let us therefore examine the action \reef{eq:exactfunc} for small $\tau$ and $\ell\geq 2$. As $\tau$ approaches zero the first term vanishes, since $\beta_0^-(0)=0$. The second term will be positive only if $2\ell\geq 2+2\Df$. The strongest constraint thus comes from $\ell=2$, which implies $\Df\leq 1$. As we turn on the twist, the first term in \reef{eq:exactfunc} starts out negative ($\beta_0^-(\tau)$ is negative for small enough $\tau$), and we must check whether it does not swamp the contribution of the second term. By inspection, we find that indeed it does not: figure \ref{fig:positivity} shows the functional action computed numerically for several spins, which demonstrates positivity explicitly. We conclude that the functional $\beta\alpha_{00}$ provides a valid upper bound $2+2\Df$ for $\Df\leq 1$, a bound which is optimal for $\Df=1$.

\begin{figure}%
\begin{center}
\includegraphics[width=12cm]{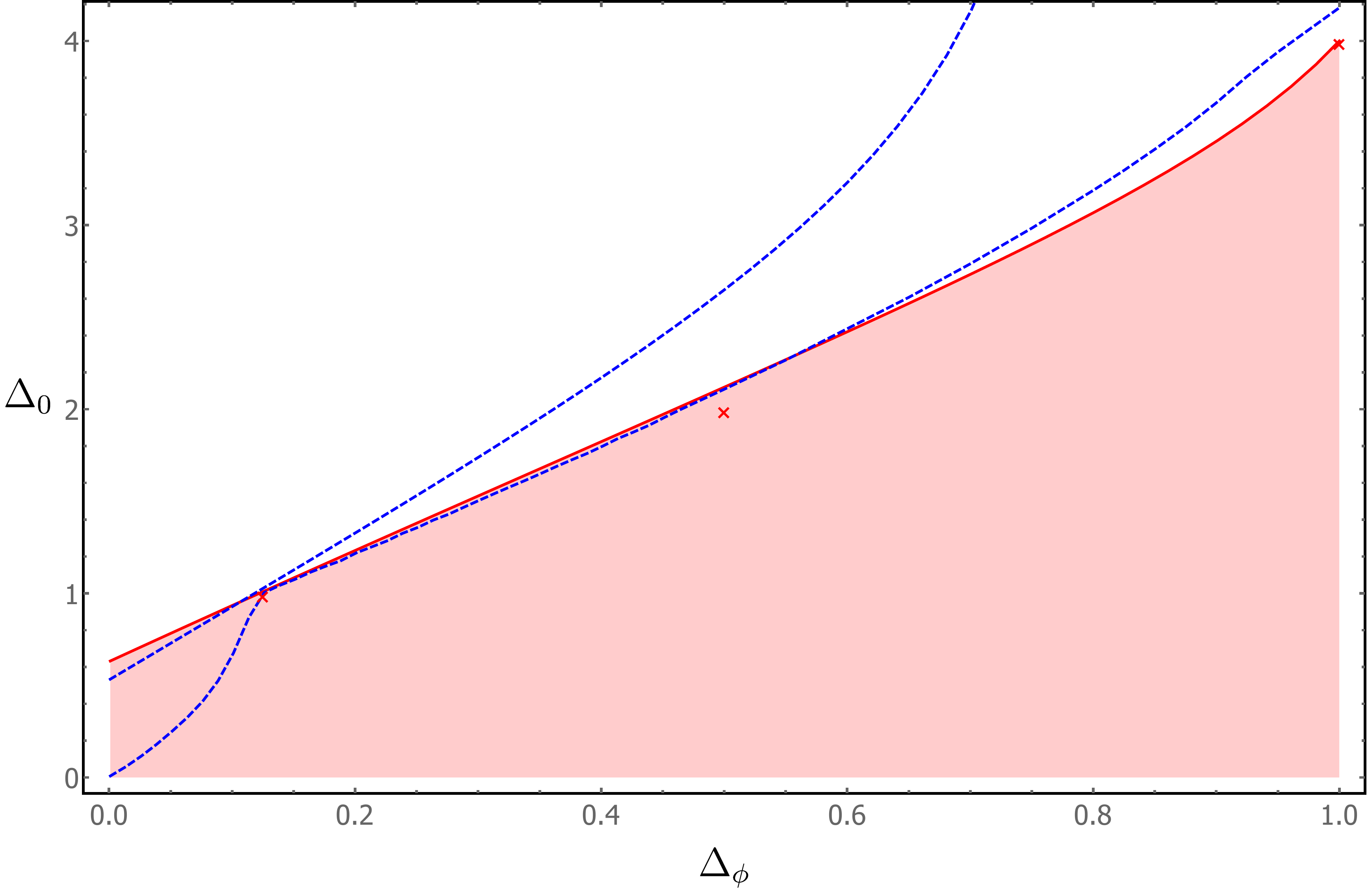}%
\caption{Scaling dimension bound. The thick red curve is the best bound with a basis of functionals $\alpha\alpha_{00}$, $\alpha\beta_{00}$, $\beta\alpha_{00}$ and $\beta\beta_{00}$. The blue dashed curves represent bounds obtained with the ordinary derivative basis, with 4 and 21 components. The crosses at (1/8,1) and (1,4) indicate the 2d Ising $\langle \sigma\sigma\sigma\sigma\rangle$ and $\langle \varepsilon\varepsilon\varepsilon\varepsilon\rangle$ correlators and the one at (1/2,2) the $c\to 1$ limit of minimal model's $\phi_{1,2}$ correlator. These are all points which we expect will saturate an optimal dimension bound.}%
\label{fig:bound}%
\end{center}
\end{figure}

Allowing for more general linear combinations of functionals can lead to stronger bounds, although in general one cannot obtain these analytically. Typically such bounds are obtained by considering functionals which are linear combinations of derivatives acting at particular points. Recently it has been argued that in $d=1$, the functional basis reviewed in section \ref{sec:review} vastly improves on the derivative basis \cite{Paulos:2019fkw}. Is this also true for the product basis in $d=2$?
It seems quite likely this can be the case, since after all for $\Df=1$ we get an exact optimal bound on the nose with a single functional component. As a cursory exploration of this question, we consider a simple basis of functionals made up of $\alpha\alpha_{00}$, $\alpha\beta_{00}$, $\beta\alpha_{00}$ and $\beta\beta_{00}$. We have seen that $\beta\alpha_{00}$ alone already determines a valid bound $\Delta_0\leq 2+2\Df$ for $\Df\leq 1$. In figure \ref{fig:bound} we show the best upper bound obtainable with this larger set of functionals in the same region.
We see that for the same number of components the functional basis compares favorably with the derivative basis for $\Df\gtrsim 1/8$ but eventually gives worse results for small enough $\Df$. In fact, for $\Df\gtrsim 1/8$ the results obtained with the functionals are in fact comparable with a derivative bound using a much greater number of components. Of course, independently of the number of derivatives, the functional basis is bound\footnote{No pun intended!} to do better as we approach $\Df=1$. The fact that the bounds are so close to those of the derivative basis even with so few components also strongly suggests that the functional basis is complete (since the derivative basis certainly is). Overall we find these results promising, although further work is necessary, in particular to determine how the numerical bounds converge as the size of the functional bases increase.

\subsection{Upper bound on the OPE density}
\label{sec:opebound}
We will now show that the tensor product functionals lead to an upper bound on the OPE density. The discussion mimics the one for the $d=1$ case \cite{Mazac2019a}. The idea is to consider the $\alpha\alpha_{p,q}$ functionals, or rather small modifications of these. These functionals have a small negative region at small twist and a positive bump centered around $\Delta^{\mbox{\tiny prod}}_{n,\ell}$ with $n$, $\ell$ fixed in terms of $p,q$. The corresponding functional bootstrap equation then bounds contributions from the OPE density in the bump in terms of that in the negative region at small twist. In the large $p,q$ limit contributions from the negative region are suppressed for all operators except the identity, whose OPE coefficient is trivially known, and we obtain a simple bound.

We begin by introducing a small modification of the $\alpha$ functionals:
\bea
\tilde \alpha^{\pm}_{m}\equiv \alpha_m^{\pm}+c_m^{\pm} \beta_m^{\pm}\,.
\eea
The coefficients $c_m^{\pm}$ are chosen such that the corresponding kernels $f_{\tilde \alpha^{\pm}_m}$ (defined in appendix \ref{sec:bases}) have softer behaviour at large $z$,
\bea
f_{\tilde \alpha^{+}_m} \underset{z\to \infty}\sim O(z^{-5})\\
f_{\tilde \alpha^{-}_m} \underset{z\to \infty}\sim O(z^{-4}),
\eea
as compared to $z^{-3}$ and $z^{-2}$ fall-offs respectively. Equivalently, these particular functional combinations correspond to functional actions $\tilde \alpha_m^{\pm}(h)$ which have softer behaviour for large $h$.
\footnote{The functional actions at large $\Delta$ involve the integral 
$$
\int_0^1 \ud z (1-z)^{\Df-2} f(\mbox{$\frac 1{1-z}$}) k_h(z|\Df)\,.
$$
For large $h$ the integral is dominated by the $z\sim 1$ region and hence controlled by the large $z$ behaviour of $f(z)$.
}
The point of this definition is that it leads to the following properties:
\bea
\tilde \alpha^+_m(\tau)&\geq 0& \qquad &\text{for all}&\quad \tau&\geq 0\,, \\
\tilde \alpha^-_m(\tau)&\geq 0 &\qquad &\text{for all}&\quad \tau &\geq \tau_0(\Df)\,.
\eea
In particular, in the above we can choose $\tau_0$ independent of $m$. Furthermore, we find $\tau_0< 2\Df$ in all cases we've checked.

We set $n\geq m$. Acting with the functional $\tilde \alpha_m^-\otimes \tilde \alpha_n^+$ on the crossing equation we obtain the exact sum rule:
\bea
\sum_{\Delta,\ell} a_{\Delta,\ell}\left[\tilde \alpha_m^-(\tau)\,\tilde \alpha_n^+(\rho)+\tilde \alpha_m^-(\rho )\,\tilde \alpha_n^+(\tau)\right]=0\,.
\eea
The $\tilde \alpha_m(h)$ have bumps centered around $h=h_m=2+2\Df+4m$. Using the positivity properties mentioned above, the sum rule implies
\bea
\sum_{\substack{|\Delta-\Delta_{m,\ell}^{\mbox{\tiny prod}}|\leq 2,\\\ell=2(n-m)}} (1+\delta_{\ell,0})a_{\Delta,\ell} \tilde \alpha_m^-(\tau) \tilde \alpha_n^+(\rho)\leq -\!\!\!\sum_{\substack{0\leq \tau\leq \tau_0,\\\ell=0,2,\ldots}} a_{\Delta,\ell} \left[\tilde \alpha_m^-(\tau) \tilde \alpha_n^+(\rho)+\tilde \alpha_m^-(\rho) \tilde \alpha_n^+(\tau)\right] \label{eq:prebound}
\eea
This places an upper bound on the OPE density for spin $\ell=2(n-m)$, inside the bin in $\Delta$ space centered around $\Delta_{m,\ell}^{\mbox{\tiny prod}}\equiv 2+2\Df+2m+\ell$. The bound is determined in terms of the contributions of small twist operators. Note that we can guarantee optimality of the bound for $\Df$ an odd integer, since in this case the tensor product solutions of section \ref{sec:productcorrelators} saturate the above inequality for all $m,n$.

The bound takes a simpler form when we consider the large $\Delta$ limit at fixed spin, i.e. $m,n \to \infty$ with $n-m$ fixed. We examine the functional actions in this limit in appendix \ref{sec:asymptotics}. One result is that in this limit the contributions at fixed $\tau>0$ are suppressed relative to those with $\tau=0$ since:
\bea
\tilde \alpha^+_n(0)&\underset{n\to \infty}{\sim} a_{h_n}^{\mbox{\tiny free}},&\qquad \tilde \alpha^+_n(h)&\underset{n\to \infty}{\sim} a_{h_n}^{\mbox{\tiny free}} \times O(h_n^{-\mbox{\tiny min}\{h,2\}})\\
\tilde \alpha^-_m(0)&\underset{m\to \infty}{\sim}-a_{h_m}^{\mbox{\tiny free}},&\qquad \tilde \alpha^-_m(h)&\underset{m\to \infty}{\sim} a_{h_m}^{\mbox{\tiny free}}\times  O(h_m^{-\mbox{\tiny min}\{h,3\}})\,.
\eea
The functional actions $\tilde \alpha_{m}^{\pm}(h)$ also simplify in the region where both $h,h_m$ are large but $h-h_m$ is fixed:
\bea
\tilde \alpha_m^{\pm}(h)\underset{h,h_m\to \infty}=\left(\frac{a_{h_m}^{\mbox{\tiny free}}}{a_h^{\mbox{\tiny free}}}\right) \left(\frac{4}{\pi} \frac{\sin\left[\frac \pi 4 (h-h_m)\right]}{h-h_m}\right)^2
\eea
Let us examine the contribution of the twist zero operators with $\ell\geq 2$ appearing on the righthand side of \reef{eq:prebound}:
\begin{multline}
\limsup_{n,m\to \infty 
}\, -\!\!\!\sum_{\ell=2,4,\ldots} a_{\ell,\ell} \left[\tilde \alpha_m^-(0) \tilde \alpha_n^+(2\ell)+\tilde \alpha_m^-(2\ell) \tilde \alpha_n^+(0)\right]=\\
=\limsup_{n,m\to \infty 
} a_{h_m}^{\mbox{\tiny free}}a_{h_n}^{\mbox{\tiny free}} \,\sum_{\ell=2,4,\ldots}\left(\frac{a_{\ell,\ell}}{a_{2\ell}^{\mbox{\tiny free}}}\right)\left[\left(\frac{2}{\pi} \frac{\sin\left[\frac \pi 2 (\ell-\ell_n)\right]}{\ell-\ell_n}\right)^2-\left(\frac{2}{\pi} \frac{\sin\left[\frac \pi 2 (\ell-\ell_m)\right]}{\ell-\ell_m}\right)^2\right]
\end{multline}
with $\ell_p=1+\Df+2p$. It now follows that, as long as the fast, exponential dependence of $a_{\ell,\ell}$ on spin matches that of $a_{2\ell}^{\mbox{\tiny free}}$, then this limit vanishes. More precisely, the result holds as long as at large spin we can bound $a_{\ell,\ell}$ from above by $a_{2\ell}^{\mbox{\tiny free}}$ times some power of $\ell$. Notice that this is consistent with our conjecture in section \ref{sec:constraints} regarding the double lightcone behaviour of 2d CFT, which will be satisfied if $a_{\ell,\ell}\leq C a_{2\ell}^{\mbox{\tiny free}}$ for some constant $C$ (which must be larger or equal than two, given result \reef{eq:isingope}).

With this assumption, we can finally write the simplified version of the bound \reef{eq:prebound}:
\bea
\limsup_{n,m\to \infty
} \sum_{\substack{|\Delta-\Delta_{m,\ell}^{\mbox{\tiny prod}}|\leq 2,\\\ell=2(n-m)}} \left(\frac{a_{\Delta,\ell}}{a_{\Delta,\ell}^{\mbox{\tiny prod}}}\right)\, \left( \frac{4}\pi \frac{\sin\left[\frac \pi 4 (\Delta-\Delta_{m,\ell}^{\mbox{\tiny prod}})\right]}{\Delta-\Delta_{m,\ell}^{\mbox{\tiny prod}}}\right)^4\leq 1\,.\label{eq:opebound}
\eea
where
\bea
a_{\Delta,\ell}^{\mbox{\tiny prod}}=\frac{2}{1+\delta_{\ell,0}}\,a_{\tau}^{\mbox{\tiny free}}a_{\rho}^{\mbox{\tiny free}}
\eea
captures the OPE density of the higher twist operators in the product solutions of section~\ref{sec:productcorrelators}.

A few comments are in order. On the left-hand side we are free to extend the summation limits in $\Delta$, and we may also add up the contributions of other spins, as long as these extensions do not scale with $m,n$. We can alternatively shrink the summation region. In particular we can find an upper bound on individual OPE coefficients
\bea
\limsup_{\Delta \to \infty} a_{\Delta,\ell} \leq C a_{\Delta,\ell}^{\mbox{\tiny prod}},\qquad 1\leq C\leq \left(\frac \pi 2\right)^4
\eea
The bound is strongest when $\Delta=\Delta_{m,\ell}^{\mbox{\tiny prod}}, \ell=2(n-m)$ for some $n,m$, and if $\Df$ is such that a tensor product solution exists it is also optimal.\footnote{Existence of solution here means in particular that the spins of all operators should be even integers.} In particular this shows that the strongest possible bound is not in general saturated by generalized free fields. For these, $\Delta_{n,\ell}=2\Df+2n+\ell$ so that in each bin we sum over three states, but the OPE coefficients are smaller so that the bound \reef{eq:opebound} is indeed satisfied.

\section{Discussion}
\label{sec:discussion}
In this work we have begun the construction of interesting classes of functionals which act on the crossing equation of CFTs in general dimension. 

We have proposed a general functional ansatz given by equations \reef{eq:fgerep}, \reef{eq:fgeprops1} and \reef{eq:fgeprops2}. In order for these to be valid crossing-compatible functionals, the kernels should satisfy the boundary conditions summarized in equation \reef{eq:constraintskernels}. Some of these conditions relied on an understanding of the double lightcone limit of correlation functions, an understanding which remains incomplete. This did not affect any of the other results in this work, and given our general analysis it would be straightforward to modify these constraints in light of any new information concerning this limit.

Assuming our analysis is correct, it is interesting to note that there can be discontinuities in the boundary conditions, either at $d=2$ due to the absence of a twist gap in that case, or for fixed dimension as we cross a certain threshold in the external dimension $\Df$. This suggests that quantities such as bounds on scaling dimensions or OPE coefficients might be discontinuous as we vary these parameters. Such discontinuous behaviour is indeed manifest in some contexts, such as the lightcone bootstrap \cite{Fitzpatrick2013,Komargodski2013,Fitzpatrick2014} in the behaviour of the large spin spectrum as a function of $d$, but to our knowledge such discontinuities have never been observed in any numerical bounds. Presumably this is because they are sensitive to the large spin behaviour which is hard to access numerically.

We have introduced a simple set of functionals which we have called the HPPS class. The functional actions for these have finite support on the generalized free field spectrum which can be easily computed.  Unfortunately, if we attempt to bootstrap away from this solution, these functionals will mix up anomalous dimensions of all spins for low enough twist, so they are only useful for analytic computations if such perturbations are bounded in spin. In this sense they are like a poor man's version of the Polyakov bootstrap functionals. Nevertheless they can still be useful. For instance it would be interesting to combine them with other methods which determine the part of the spectrum which is analytic in spin \cite{Caron-Huot:2017vep}.

The HPPS functionals are simple to compute, have nice positivity properties for large enough twist, and automatically incorporate the expected structure of crossing-symmetric solutions with generalized-free field type asymptotics. It is therefore tempting to try to use these functionals for the numerical bootstrap, perhaps combined with $d=1$ functionals. An obstacle seems to be that the functional actions will typically have negative regions for small enough twist and for all spins, whereas in typical applications we would want to demand positivity of functionals above unitarity for all spins above some small value. Adding $d=1$ functionals into the mix does not seem to help, since (experimentally) the functional actions for these care about dimension, not twist, and hence could make positive at most a finite number of spins. It would be very interesting to explore this and related problems in more detail.

In $d=1$, up to now all available evidence seems to show that interesting sets of functionals fall into the GFF class: that is, they are dual to solutions to crossing whose spectra eventually asymptotes to that of a generalized free field. Whether this is really true or not in $d=1$, we now know that in higher dimensions it definitely isn't. In this work we have considered product functionals in $d=2$ which have a very different structure, but which nevertheless seem to provide a perfectly good basis of functionals. In particular they are dual to a solution to crossing which saturates a bound. On the other hand, in the same dimension it can be observed numerically that bounds on leading operators with spin $\ell\geq 2$ are saturated
by generalized free fields. The associated basis of functionals should allow us to bootstrap small deformations away from that solution, and hence should be directly linked to the Polyakov bootstrap. For this to be possible, the associated functional actions should have doubly-spaced double zeros on the generalized free field spectrum. Together, these results suggests that in general one should consider different bases of functionals depending on the problem under consideration. In particular some should be better suited for numerical applications than others. In fact this is already manifest in $d=1$, where choosing between bosonic or fermionic functionals can lead to dramatic differences in convergence of numerical bounds \cite{Paulos:2019fkw}.

It remains to be shown that the product basis is complete. To prove this we would need to show that it is possible to plug in the basis decomposition \reef{eq:fbasis2d} into the 2d crossing equation and commute the series over basis elements with the one over the spectrum. To show this we must have sufficient control over the tails of those series, and in practice this requires establishing upper bounds on the OPE density $a_{\Delta,\ell}$ both at large dimension $\Delta$ and fixed spin $\ell$, as well as large spin and fixed twist. We have established the former but not the latter. In fact, even this is not quite right, since for the bound at large $\Delta$ we obtained a simplified answer only after assuming a bound on OPE coefficients at twist zero. Perhaps such a bound can be found by taking clever combinations of the product basis functionals. We have made some attempts in this direction without success. Here we would just like to point out that the meaning of such a bound would be somewhat puzzling, since it would be obtained by acting with functionals which were themselves constrained by assuming a bound on the double lightcone behaviour of correlation functions, and this is the very same bound we are trying to determine. So it seems the most we could aim for is self-consistency.

There are a few difficulties with the notion of functionals which would rigorously define the Polyakov bootstrap. Naively, based on the results in $d=1$, one might expect there to be functionals $\alpha_{n,\ell} ,\beta_{n,\ell}$ which would allow us to bootstrap arbitrary Witten exchange diagrams, i.e. that the corresponding crossing equation can be written as
\bea
F_{\Delta,J}=\sum_{n,\ell}\left[ \alpha_{n,\ell}(\Delta,J) F_{\Delta_{n,\ell},\ell}+\beta_{n,\ell}(\Delta,J)\partial_n F_{\Delta_{n,\ell},\ell}\right]\,.
\eea
However, it seems unlikely such equations can be true with the coefficient functions being given by functional actions. This is because as we vary $J$ the Regge (large $z,\bar z$ on the same upper half-plane) behaviour of Witten exchanges becomes worse. Therefore, just as for our HPPS functionals and the bootstrapping of contact interactions, we should expect that we would need to constrain functionals more and more to bootstrap higher and higher spin $J$ exchanges. Note that in general it would not be possible to simply add contact interactions to improve the Regge behaviour of exchange diagrams, basically because even for large $z,\bar z$ we would still have a full function of $z/\bar z$ that has to be cancelled. Another reason is that if decompositions as the above were true then, just as in $d=1$ \cite{Mazac2019a}, they would indicate that in perturbation theory one could modify the generalized free field result to arbitrary orders without introducing new operators. However, we expect that for $d>1$ and beyond leading order generically multi-twist operators will have to appear \cite{Simmons-Duffin:2016wlq}.

An interesting possibility which is corroborated by our tensor product basis is to relax decompositions as the one above and include extra basis elements, such as derivatives with respect to spin of the conformal crossing vectors. These spurious elements could then drop out after summing over whole Regge trajectories. Another hint that this is correct is that such sums can have softer Regge behaviour than any individual piece. So perhaps it is really whole Regge trajectories which should satisfy basis decomposition rules such as the one written above, rather than individual crossing vectors. It would be very interesting to make this precise.

\vspace{1 cm}
\section*{Acknowledgments}
We would like to acknowledge A. Kaviraj, D. Maz\'a\v{c}, S. Rychkov, and E. Trevisani for discussions at various points of this collaboration.  MFP acknowledges the Perimeter Institute and Simons Collaboration on the Non-Perturbative Bootstrap for funding and hospitality during the annual Bootstrap conference, where part of this work was completed.

\newpage
\appendix

\section{Bases of 1d functionals}
\label{sec:bases}
In this appendix we show how to construct bases of 1d functionals which are dual to the generalized free fermion or boson solutions to crossing. Note that the full bases of $\omega_-$ type functionals were previously constructed in \cite{Mazac2019a} based on earlier results \cite{Mazac:2018,Mazac:2016qev}. The $\omega_+$ functional bases however have not been fully constructed explicitly before, with the exception of the fermionic $\beta_0^+$ functional in \cite{Hartman2019b}.

\subsection{Fundamental free equation}
\label{sec:bases1}
We begin by recalling the functional action definition:
\bea
\omega_{\pm}(\Delta|\Df)&=\int_{1}^{\infty}\frac{\ud z}{\pi} h_\pm(z) \mathcal I_z F_{\pm,\Delta}(z|\Df)\,.
\eea
We take $h(z)$ to be real analytic for $z>1$ and holomorphic away from the real axis. Hence we can write the above as a contour integral wrapping the $z>1$ half-line. Deforming the contour we get
\bea
\omega_{\pm}(\Delta|\Df)&=\int_{\Gamma_r} \frac{\ud z}{2 \pi i} h_{\pm}(z) F_{\pm,\Delta}(z|\Df)\\
&=
\frac 12 \int_{\Gamma^+}\,\ud z\, f_{\pm}(z) F_{\pm,\Delta}(z|\Df)+\int_{\frac 12}^1 \ud z\, g_{\pm}(z) F_{\pm,\Delta}(z|\Df)
\eea
where the contours $\Gamma_r, \Gamma^+$ are explained in figure \ref{fig:contours} and
\bea
f_{\pm}(z)&=\frac{h_{\pm}(z)\pm h_{\pm}(1-z)}{i\pi},&\qquad \mbox{Im}\, z&>0,\\
g_{\pm}(z)&=-\frac{1}{\pi}\,\mathcal I_z h_{\pm}(z),& z& \in (0,1)\,.
\eea
Real analyticity of $h(z)$ for $z>1$ implies $f_{\pm}(z)=\mp f_{\pm}^*(1-z^*)$. We set $f_{\pm}(z^*)=f_{\pm}^*(z)$ to get $f_{\pm}(z)=\mp f_{\pm}(1-z)$. 
Note that the above imply the gluing condition:
\bea
\mathcal R_z f_\pm(z)=-g_{\pm}(z)\pm g_{\pm}(1-z),\qquad z\in (0,1)\label{eq:gluing}
\eea
with %
\bea
\mathcal R_z f(z):=\lim_{\epsilon\to 0^+} \frac{f(z+i\epsilon)+f(z-i\epsilon)}2\,.
\eea
Now let us go back to the functional action. Doing some contour manipulations we obtain:
\bea
\omega_\pm(\Delta|\Df)&=
\int_1^\infty \frac{\ud z}{\pi} h_{\pm}(z) \mathcal I_z F_{\pm,\Delta}(z|\Df)\nonumber\\
&=-\int_{-\infty}^0 \frac{\ud z}{\pi} \, \mathcal I_z\left\{ \left[\pm h_\pm(1-z)+ h_\pm(z)\right] G_\Delta(z|\Df)\right\}+\int_0^1 \ud z \, g(z)\, G_\Delta(z|\Df)\,.
\eea
The advantage of this representation is that it involves the discontinuities of the conformal blocks for negative argument. These are very simple, since
\bea
\lim_{\epsilon\to 0^+}\,G_\Delta(z+i\epsilon|\Df)=e^{i\pi (\Delta-2\Df)} \frac{G_\Delta\left(\frac{z}{z-1}\right|\Df)}{(1-z)^{2\Df}}, \qquad z<0.
\eea
In particular they are oscillating functions of $\Delta-2\Df$. In order to satisfy the orthonormality conditions \reef{eq:ONboson}, \reef{eq:ONfermion} we want to aim for functional actions which have double zeros for scaling dimensions $\Delta$ in the generalized free spectrum, namely $\Delta_n^B=2\Df+2n$ and $\Delta_n^F=2\Df+2n+1$ for bosons and fermions respectively. A strategy for getting this structure is to get rid of the $\sin \pi(\Delta-2\Df)$ factor and then make use of the trigonometric identity $\cos(2x)=2\sin^2(x)-1$. Accordingly we first demand

\bea
\mathcal R_z [\pm h_{\pm}(z)+h_{\pm}(1-z)]=0, \qquad z<0,\label{eq:recondh}
\eea
or equivalently,
\bea
\mathcal I_z h_{\pm}(z)=\mathcal R_z f_{\pm}(z)=f_{\pm}(z).\qquad z<0\,.
\eea
If we now impose
\bea
g_{\pm}(z)=\eta\, (1-z)^{2\Df-2} f_{\pm}\left(\frac{z}{z-1}\right),\qquad z\in(0,1) \label{eq:gfromf}\,,
\eea
the functional action takes the desired form:
\bea
\omega_{\pm}(\Delta)=\left[1-\eta \cos \pi(\Delta-2\Df)\right] \int_0^1 \ud z\,g_{\pm}(z) G_{\Delta}(z|\Df)
\label{eq:funcact}\,.
\eea
Here $\eta$ is chosen equal to plus (minus) one for bosons (fermions). 

We are not quite done yet, since $f_{\pm}, g_{\pm}$ are not arbitrary, being linked together by the gluing condition. Plugging \reef{eq:gfromf} into the \reef{eq:gluing} we get the fundamental free equation:
\bea
\eta \,\mbox{Re}\, f_{\pm}(z)
= \pm (1-z)^{2\Df-2} f_{\pm}\left(\frac{1}{1-z}\right)-z^{2\Df-2} f_{\pm}\left(\frac 1z\right).
\eea
This must now be solved for $f_\pm(z)=\mp f_{\pm}(1-z)$ real for $z>1$ and analytic on the upper half-plane. 

\subsection{General solution}
We want to construct solutions of the fundamental free equation corresponding to functionals satisfying the duality conditions
\bea
\alpha_{\pm ,n}^{F}(\Delta_m^F)&=\delta_{nm},&\partial \alpha_{\pm,n}^{F}(\Delta_m^F)&=0,\\
\beta_{\pm ,n}^{F}(\Delta_m^F)&=0,&\partial \beta_{\pm,n}^{F}(\Delta_m^F)&=\delta_{nm}
\eea
for the fermionic basis, and
\bea
\alpha_{\pm ,n}^{B}(\Delta_m^B)&=\delta_{nm},&\partial \alpha_{\pm,n}^{F}(\Delta_m^B)&=-c_{\pm,n} \delta_{m0},\\
\beta_{\pm ,n}^{B}(\Delta_m^B)&=0,&\partial \beta_{\pm,n}^{B}(\Delta_m^B)&=\delta_{nm}-d_{\pm,n} \delta_{m0}
\eea
for the bosonic one.
By constructing several solutions of the fundamental free equation for particular $\Df$ it is possible to guess the general solutions. It is important to point out that in order for the functionals to be well defined on infinite sums of crossing vectors, we must demand that the kernels $f(z)\underset{z\to \infty}{\sim} O(z^{-1-\epsilon})$ for some $\epsilon>0$ \cite{Rychkov:2017tpc}, \cite{Mazac:2018}. In practice we are able to find solutions with $\epsilon=1$ for $f_-(z)$ and $\epsilon=2$ for $f_+(z)$. 

To describe the general solution we begin by constructing ``shifted'' functionals
\bea
s\beta_{\pm,n}^F:\qquad f_{\pm}(z)=[z(z-1)]^{-2n} N_{\pm}(z|\Df+3n)\\
s\alpha_{\pm,n}^F:\qquad f_{\pm}(z)=[z(z-1)]^{-2n} L_{\pm}(z|\Df+3n)
\eea
where
\bea
N_+(z|\Df)&=-\frac{\Gamma(3+\Df)\Gamma\left(\frac 52+2\Df\right)}{2^{1+2\Df}\pi\Gamma\left(\frac 32+\Df\right)}\, \frac{_3\tilde F_2\left(-\frac 12,\frac 32,\frac 52+2\Df; \frac 32+\Df,\frac 52+\Df;-\frac{1}{4 w}\right)}{w^{\frac 32}}\\
L_+(z|\Df)&=-\frac{\Gamma(3+\Df)\Gamma\left(\frac {11}2+2\Df\right)}{2^{4+2\Df}\pi\Gamma\left(\frac 52+\Df\right)}\, \frac{_3\tilde F_2\left(\frac 12,\frac 12,\frac {11}2+2\Df; \frac 72+\Df,\frac 72+\Df;-\frac{1}{4 w}\right)}{w^{\frac 52}}\\
N_-(z|\Df)&=-\kappa(\Df)
\frac{2z-1}{w^{3/2}}\left[\, _3\tilde{F}_2\left(-\frac{1}{2},\frac{3}{2},2
   \Df+\frac{3}{2};\Df+1,\Df+2;-\frac{1}{4 w}\right)+\right.\\
	&\,\;\;\left.+\frac{9}{16 w} \,
   _3\tilde{F}_2\left(\frac{1}{2},\frac{5}{2},2 \Df+\frac{5}{2};\Df+2,\Df+3;-\frac{1}{4w}\right)\right],\\
L_-(z|\Df)&=\kappa(\Df)\frac{2(z-2)(z+1)}{(2z-1)w^{3/2}}
\left[
{}_3\tilde{F}_2\left(-\frac{1}{2},-\frac{1}{2},2\Df +\frac{3}{2};\Df +2,\Df +2;-\frac{1}{4 w}\right)+\right.\\
&+\frac{(2 \Df +3) (2 \Df +5)}{16 w} {}_3\tilde{F}_2\left(\frac{1}{2},\frac{1}{2},2 \Df +\frac{5}{2};\Df +3,\Df +3;-\frac{1}{4 w}\right)-\\
&-\frac{3 (4 \Df +5)}{256 w^2}\left.{}_3\tilde{F}_2\left(\frac{3}{2},\frac{3}{2},2 \Df +\frac{7}{2};\Df +4,\Df +4;-\frac{1}{4 w}\right)\right]+\frac{N_-(z|\Df)}{2(1+\Df)}.
\eea
Here ${}_3\tilde F_2$ stands for the regularized hypergeometric function, $w=z(z-1)$ and the normalization factor is given by
\be
\kappa(\Df) = \frac{\Gamma(4\Df+4)}{2^{8\Df+5}\Gamma(\Df+1)^2}\,.
\ee
The normalizations were picked such that near $z=1$ we have
\bea
N_{\pm}(z)&\underset{z\to 1^+}{\sim} -\frac{2}{\pi^2}\frac 1{(z-1)^2}\\
L_{\pm}(z)&\underset{z\to 1^+}{\sim} \frac{2}{\pi^2}\frac{\log(z-1)+c}{(z-1)^2},\qquad c\quad \mbox{constant}\,.
\eea
The shifted functionals provide a complete basis of $\eta=-1$ functionals. They satisfy:
\bea
s\alpha_{\pm ,n}^{F}(\Delta_m^F)&=0,&\partial s\alpha_{\pm,n}^{F}(\Delta_m^F)&=0,\qquad m>n\,,\\
s\beta_{\pm ,n}^{F}(\Delta_m^F)&=0,&\partial s\beta_{\pm,n}^{F}(\Delta_m^F)&=0,\qquad m>n\,.
\eea
Hence for each $n$ one can perform a (finite) Gram-Schmidt orthonormalization procedure to obtain the $\alpha_{\pm,n},\beta_{\pm,n}$. For general $\Df$ the solution to this orthonormalisation step is not currently known in closed form for all $n$, i.e. it has to be done case by case. One can write down a closed form result for special values of $\Df$, as we show in the next section.

Finally let us describe the bosonic functionals. In this case we set
\bea
s\beta_{\pm,n}^B:\qquad f_{\pm}(z)=[z(z-1)]^{1-2n} N_{\pm}(z|\Df-3/2+3n)\\
s\alpha_{\pm,n}^B:\qquad f_{\pm}(z)=[z(z-1)]^{1-2n} L_{\pm}(z|\Df-3/2+3n)
\eea
Demanding that the kernels fall off faster than $1/z$ at infinity we see that only $n\geq 1$ is allowed for the $s\beta$ functionals ($L_+,L_-$ fall off as $1/z^4$ and $1/z^5$ respectively). Again an orthonormalisation procedure may be applied if so wished, by imposing the duality conditions \reef{eq:ONboson},\reef{eq:ONfermion}.

\subsection{Special cases}
For special values of $\Df$ it is possible to find closed form solutions for the orthonormal kernels. These solutions are written in terms of the building blocks
\bea
p\beta_{-,m}^{B,F}:&\quad& f_-(z)&=\frac{2}{\pi^2} \frac{\Gamma(2+2m)^2}{\Gamma(3+4m)}\,\left(\frac{P_{2m+1}\left(\frac{z-2}z\right)}{z^{2-2\Df}}+\eta \frac{P_{2m+1}\left(\frac{1+z}{z-1}\right)}{(z-1)^{2-2\Df}}\right)\\
p\beta_{+,m}^{B,F}:&\quad& f_+(z)&=\frac{2}{\pi^2} \frac{\Gamma(1+2m)^2}{\Gamma(1+4m)}\,\left(\frac{P_{2m}\left(\frac{z-2}z\right)}{z^{2-2\Df}}+\eta \frac{P_{2m}\left(\frac{1+z}{z-1}\right)}{(z-1)^{2-2\Df}}\right)\\
p\alpha_{-,m}^{B,F}:& \quad& f_{-}(z)&=\frac 12 \partial_m f_{\pm,p\beta_{-,m}}(z)\\
&&&-\eta\frac{2}{\pi^2} \frac{\Gamma(2+2m)^2}{\Gamma(3+4m)}\frac{\Gamma(2+2m)^2}{\Gamma(4+4m)}G_{2+2m}(1/z)\\
p\alpha_{+,m}^{B,F}:& \quad& f_{+}(z)&=\frac 12 \partial_m f_{\pm,p\beta_{+,m}}(z)\\
&&&+\eta\frac{2}{\pi^2}\frac{\Gamma(1+2m)^2}{\Gamma(1+4m)} \frac{\Gamma(2+2m)^2}{\Gamma(4+4m)}G_{1+2m}(1/z)\,,
\eea
where as before $\eta=1,-1$ for the bosonic/fermionic case respectively. One can check these prefunctionals satisfy the fundamental free equation for all integer $m$ when:
\bea
(-)&\quad \mbox{case}:&\qquad \Df&\in \frac 12+\mathbb Z& \qquad \mbox{and}\quad \eta&=-1\\
(-)&\quad \mbox{case}:&\qquad \Df&\in  \mathbb Z& \qquad \mbox{and}\quad \eta&=+1\\
(+)&\quad \mbox{case}:&\qquad \Df&\in \frac 12+\mathbb Z& \qquad \mbox{and}\quad \eta&=+1\\
(+)&\quad \mbox{case}:&\qquad \Df&\in  \mathbb Z& \qquad \mbox{and}\quad \eta&=-1
\eea
However, they are not good functionals because they do not satisfy correct fall off conditions for large $z$. For these special cases, the general orthonormal solutions can be obtained by first constructing the $\beta$ functionals:
\bea
\beta_{-,m}^F&=p\beta_{-,\Df-\frac 12+m}^F+\mbox{lower}\\
\beta_{-,m}^B&=p\beta_{-,\Df-1+m}^B+\mbox{lower}\\
\beta_{+,m}^F&=p\beta_{-,\Df+m}^F+\mbox{lower}\\
\beta_{+,m}^B&=p\beta_{-,\Df+\frac 12+m}^B+\mbox{lower}\,,
\eea
with ``lower'' meaning a finite set $p\beta$ and $p\alpha$ of lower $m$, with coefficients chosen as to guarantee that the functional kernels have fall-off faster than $1/z$ at infinity. The $\alpha_m$ functionals are then simply obtained by replacing $p\beta_m$ above with $p\alpha_m$ and differentiating those $m$ dependent coefficients, call them $c_i(m)$, with respect to $m$. For instance for $\Df=1$:
\bea
\beta_{+,m}^F=p\beta^F_{-,m+1}-c_{+,1}(m) p\alpha_{-,0}^F\\
\alpha_{+,m}^F=p\beta^F_{-,m+1}-d_{+,1}(m) p\alpha_{+,0}^F
\eea
with
\bea
c_{+,1}(m)=\frac 43(3+5m+2m^2)\,\frac{\Gamma(1+2+2m)^2}{\Gamma(1+4+4m)}\,,\qquad d_{+,1}(m)=\frac 12 \partial_m c_{+,1}(m)\,.
\eea

\subsection{Asymptotic expansions}
\label{sec:asymptotics}
For the special cases mentioned in the previous subsection it is possible to compute the functional actions very explicitly, as explained in the appendices of \cite{Paulos:2019fkw,Mazac2019a}. In particular, appendix A.2 of the latter explains how asymptotic expansions for the functional actions may be derived in various regimes. We write the functional actions in the following way:
\bea
\omega_n^{B,F}(\Delta|\Df)=\frac{4 \sin^2\left[\frac \pi 2(\Delta-\Delta_n^{B,F})\right]}{\pi^2}\, \left(\frac{a_{\Delta_n^{B,F}}^{\mbox{\tiny free}}}{a_{\Delta}^{\mbox{\tiny free}}}\right)\, R^{B,F}_{\omega}(\Delta,\Delta_n^{B,F}|\Df)+E_\omega^{B,F}(\Delta,\Delta_n^{B,F}|\Df)
\eea
In the limit of large $\Delta$ and $h$ with $\Delta/h$ fixed we find that $E_\omega(\Delta,h|\Df)$ is exponentially suppressed, while
\bea
R_{\beta_{-}}^{B,F}(\Delta,h|\Df) &\underset{\Delta,h\to \infty}{\sim} \frac{4 h^2 \Delta}{(\Delta^4-h^4)}\,,&
\qquad R_{\tilde \alpha_{-}}^{B,F}(\Delta,h|\Df) &\underset{\Delta,h\to \infty}{\sim} \frac{16 h^{5}\Delta}{(\Delta^4-h^4)^2}\\
R_{\beta_{+}}^{B,F}(\Delta,h|\Df)& \underset{\Delta,h\to \infty}{\sim} \frac{4 h^{4}/\Delta}{(\Delta^4-h^4)}\,,&\qquad R_{\tilde \alpha_{+}}^{B,F}(\Delta,h|\Df)& \underset{\Delta,h\to \infty}{\sim} \frac{16 h^{7}/\Delta}{(\Delta^4-h^4)^2}\,.
\eea
On the other hand, with large $h$ holding $\Delta$ fixed we find 
\bea
R_{\beta_{-}}^{B,F}(\Delta,h|\Df)& \underset{h\to \infty}{\sim}- \frac{4 \Delta}{h^2}\,M^{B,F}_{-}(\Delta|\Df)\,,&
\qquad R_{\tilde \alpha_{-}}^{B,F}(\Delta,h|\Df) &\underset{h\to \infty}{\sim} \frac{16 \Delta}{h^3}\,M^{B,F}_{-}(\Delta|\Df)\\
R_{\beta_{+}}^{B,F}(\Delta,h|\Df)& \underset{h\to \infty}{\sim}- \frac{4}{\Delta}\,M^{B,F}_{+}(\Delta|\Df)\,,&
\qquad R_{\tilde \alpha_{+}}^{B,F}(\Delta,h|\Df) &\underset{h\to \infty}{\sim} \frac{16}{h \Delta }\,M^{B,F}_{+}(\Delta|\Df)
\eea
and the universal behaviour
\bea
E_{\beta_\pm}^{B,F}(\Delta,h|\Df)&=\mp\left(\frac{\Gamma(2\Df)}{\Gamma(2\Df-\Delta)}\right)^2\, \frac{a_h^{\mbox{\tiny free}}}{h^{2\Delta}}\,\tan\left[\frac \pi 2(\Delta-h)\right]\,,\\
E_{\tilde \alpha_\pm}^{B,F}(\Delta,h|\Df)&=\pm \left(\frac{\Gamma(2\Df)}{\Gamma(2\Df-\Delta)}\right)^2\, \frac{a_h^{\mbox{\tiny free}}}{h^{2\Delta}}\,. 
\eea
Furthermore we find that the functions $M_{\pm}^{B,F}(\Delta|\Df)$ satisfy
\bea
M_{\pm}^{B,F}(\Delta|\Df) &\underset{\Delta\to \infty}\sim 1\\
M_{\pm}^{B,F}(\Delta|\Df)&\geq 0 \qquad \mbox{for}\quad \Delta\geq \Delta_{\mbox{\tiny pos}}\Df)
\eea
where $\Delta_{\mbox{\tiny pos}}(\Df)$ is never above $2\Df$.

\section{Derivation of constraints on functional kernels}
Here we provide more details on how to derive constraints on the functional kernels, following the general logic described in section \ref{sec:constraints}.
\label{app:constraints}
\subsection{The $g(z,\bar z)$ and $\tilde g(z,\bar z)$ kernels}
 We begin with the simplest case to examine, which is the piece of the functional action~\reef{eq:fgerep} which depends on $g(z,\bar z)$. We want to impose:
\bea
\int_{\frac 12}^1 \ud z \int_{\frac 12}^1 \ud \bar z\, |g(z,\bar z)|\sum_{\Delta,\ell} a_{\Delta,\ell} \bigg|F_{\Delta,\ell}(z,\bar z)\bigg|<\infty\,.
\eea
We begin by noting that
\bea
\int_{\frac 12}^1 \ud z \int_{\frac 12}^1 \ud \bar z\, |g(z,\bar z)|\sum_{\Delta,\ell} a_{\Delta,\ell} \bigg|F_{\Delta,\ell}(z,\bar z)\bigg|\leq 2 \int_{\frac 12}^1 \ud z \int_{\frac 12}^1 \ud \bar z\, |g(z,\bar z)| \mathcal G(1-z,1-\bar z)
\eea
where $\mathcal G(z,\bar z)$ is the crossing symmetric function described previously (in particular $\mathcal G(z,\bar z)$ is positive inside the integration region).
Since $\mathcal G(z,\bar z)$ is smooth away from the OPE limits and we also assume smoothness of the functional kernels along the integration contours, to check convergence it is sufficient to examine the asymptotic regions. 

Although the general logic has already been described in subsection \ref{sec:constraints}, it may be worth repeating it here in a slightly different manner. In the case under consideration the dangerous region is when $z$ and or $\bar z$ approach unity.
We can probe this region by scaling $z, \bar z$ towards one with generically different powers:
\bea
&\int_\frac 12^1 \ud z \int_\frac 12^1 \ud \bar z |g(z,\bar z)| \mathcal G(1-z,1-\bar z)=\int_2^\infty  \frac{\ud z}{z^2} \int_2^\infty \frac{\ud \bar z}{\bar z^2} |g(\mbox{$\frac{z-1}{z},\frac{\bar z-1}{\bar z}$})| \mathcal G\left(\mbox{$\frac{1}z,\frac{1}{\bar z}$}\right)\\
&\geq \Lambda^{-1-\alpha} \int_{c}^{1} \ud x \int_{c}^{1} \ud \bar x  \left|g\left(\mbox{$1-\frac{1}{\Lambda^\alpha x},1-\frac{1}{\Lambda \bar x}$}\right)\right| \mathcal G \left(\mbox{$\frac{1}{\Lambda^\alpha x},\frac{1}{\Lambda \bar x}$}\right)\,. \label{eq:lowerboundint}
\eea
The last equation holds for sufficiently large $\Lambda$, with $0<c<1$ and $0\leq \alpha\leq 1$, and gives a lower bound for the value of the integral. For large $\Lambda$ the Euclidean OPE limit gives 
\bea
\mathcal G \left(\mbox{$\frac{1}{\Lambda^\alpha x},\frac{1}{\Lambda \bar x}$}\right)\underset{\Lambda\to \infty}\sim \frac{\Lambda^{(1+\alpha)\Df}}{x^\Df \bar x^{\Df}}
\eea
It is natural to demand then that
\bea
g\left(\mbox{$1-\frac{1}{\Lambda^\alpha x},1-\frac{1}{\Lambda \bar x}$}\right)\underset{\Lambda\to \infty}{=}O\left(\Lambda^{-(1+\alpha)(\Df-1)-\epsilon}\right) \qquad \mbox{for all} \quad \alpha\in [0,1]\,,
\eea
and some $\epsilon>0$. More precisely, the above argument shows that it is certainly necessary for convergence that $\epsilon\geq 0$; the general argument given at the beginning of the subsection tells us that strengthening this to $\epsilon>0$ is then also sufficient.

The analysis for $\tilde g(z,\bar z)$ is similar:
\bea
\int_{\frac 12}^1 \ud z \int_{\frac 12}^1 \ud \bar z\, |\tilde g(z,\bar z)|\sum_{\Delta,\ell} a_{\Delta,\ell} \bigg|F_{\Delta,\ell}(z,1-\bar z)\bigg|
\leq 
2 \int_{\frac 12}^1 \ud z \int_{\frac 12}^1 \ud \bar z\, |\tilde g(z,\bar z)| \mathcal G(z,1-\bar z)
\eea
Again we must study the approach to the boundary of integration where $z,\bar z \to 1$ at generically different rates. Using the results on the double lightcone behaviour derived in subsection \ref{sec:lightcone} we find:
\bea
\tilde g\left(\mbox{$1-\frac{1}{\Lambda^\alpha x},1-\frac{1}{\Lambda \bar x}$}\right)\underset{\Lambda\to \infty}{=}\left\{\begin{array}{lcll}
O\left(\Lambda^{(1+\alpha)(1-\Df)-\epsilon}\right)\,,& d=2\,,& \alpha\in[0,1]\,,& \Df\geq 0\\
O\left(\Lambda^{1+\alpha-\Df-\epsilon}\right)\,,& d>2\,,& \alpha\in[0,1)\,,&\Df\geq \frac{d-2}2\\
O\left(\Lambda^{2-\Df-\epsilon}\right)\,,& d>2\,,& \alpha=1\,,& \Df\leq d-2\\
O\left(\Lambda^{d-2\Df-\epsilon}\right)\,,& d>2\,,& \alpha=1\,,& \Df\geq d-2
\end{array}
\right.
\label{eq:constgtilde2}
\eea

\subsection{The $f(z,\bar z)$ kernel}

Let us now analyse the term in the functional action involving $f(z,\bar z)$, again demanding absolute convergence:
\bea
\int_{\Gamma^+} \ud z \int_{\Gamma^+}\ud \bar z |f(z,\bar z)| \sum_{\Delta,\ell} a_{\Delta,\ell} \bigg| F_{\Delta,\ell}(z,\bar z)\bigg|<\infty
\eea
where we have dropped some irrelevant factors. The analysis of the term involving $f(z,1-\bar z)$ is similar. On the contour of integration we have $1-z=z^*$ and hence
\bea
\bigg| F_{\Delta,\ell}(z,\bar z)\bigg|\leq 2\frac{\left| G_{\Delta,\ell}(z,\bar z)\right|}{|z \bar z|^{\Df}}\,, \qquad \mbox{for}\quad z,\bar z \in \frac 12+i \mathbb R\,.
\eea
At this point it is useful to use the radial coordinates $\rho(z),\bar \rho(\bar z)$ introduced in \cite{Hogervorst2013}, with
\footnote{The variable $\rho$ here should not be confused with the conformal spin $\rho=\Delta+\ell$ introduced previously.}
\bea
\rho(z)=\frac{z}{(1+\sqrt{1-z})^2},\qquad z=\frac{4\rho}{(1+\rho)^2}\,.
\eea
With slight abuses of notation we have
\bea
|G_{\Delta,\ell}(z,\bar z)|&=|G_{\Delta,\ell}(\rho,\bar \rho)|\leq G_{\Delta,\ell}(|\rho|,|\bar \rho|)=G_{\Delta,\ell}(z_e,\bar z_e) 
\eea
with the effective $z_e=4|\rho(z)|/(1+|\rho(z)|)^2$. We have then
\bea
&\int_{\Gamma^+} \ud z \int_{\Gamma^+}\ud \bar z |f(z,\bar z)| \sum_{\Delta,\ell} a_{\Delta,\ell} \bigg| F_{\Delta,\ell}(z,\bar z)\bigg| \leq 2 \int_{\Gamma^+} \ud z \int_{\Gamma^+}\ud \bar z |f(z,\bar z)|\mathcal G(z_e,\bar z_e)\left( \frac{z_e \bar z_e}{|z\bar z|}\right)^{\Df}\nonumber\\
&=\int_{\Gamma^+} \ud z \int_{\Gamma^+}\ud \bar z |f(z,\bar z)|\mathcal G(1-z_e,1-\bar z_e)\left( \frac{z_e \bar z_e}{|z\bar z|}\right)^{\Df}\,.
\eea
The dangerous region to be examined is when $z,\bar z$ approach infinity.
Note that
\bea
1-z_e \underset{z\to \infty}{\sim} \frac 1{2|z|}\,,
\eea
so that in this limit we again probe the Euclidean OPE of $\mathcal G$. Following our general argument, convergence will be guaranteed if
\bea
f(\Lambda^\alpha x,\Lambda \bar x)\underset{\Lambda\to\infty}{=}O(\Lambda^{-(1+\alpha)-\epsilon})\qquad \mbox{for all}\quad \alpha\in[0,1]\,.
\eea

\subsection{The $e(z,\bar z)$ kernel}
In this case we have $z\in (\frac 12,1)$ and $\bar z \in \frac 12+i\mathbb R$. On the contour we have
\bea
|F_{\Delta,\ell}(z,\bar z)|\leq \frac{G_{\Delta,\ell}(z,\bar z_e)}{|z \bar z|^{\Df}}+\frac{G_{\Delta,\ell}(1-z,\bar z_e)}{|(1-z)\bar z|^{\Df}}
\eea
where $\bar z_e$ is defined analogously to $z_e$ but with $\bar z$ instead of $z$. Hence in this case we have
\begin{multline}
\int_{\Gamma} \ud \bar z \int_{\frac 12}^1 \ud z|e(z,\bar z)| \sum_{\Delta,\ell} a_{\Delta,\ell} \bigg| F_{\Delta,\ell}(z,\bar z)\bigg|\\
\leq 2\int_{\Gamma^+} \ud \bar z \int_{\frac 12}^1 \ud z|e(z,\bar z)|\left(\frac{\bar z_e}{|\bar z|}\right)^{\Df}\left[\mathcal G(z,\bar z_e)+\mathcal G(1-z,\bar z_e)\right]
\end{multline}
Again we are led to study the asymptotic region where $\bar z\to \infty$ and $z\to 1$. Since $1-\bar z_e\sim 1/(2|\bar z|)$ we see that for the term involving $\mathcal G(z,\bar z_e)=\mathcal G(1-z,1-\bar z_e)$ this becomes a simple Euclidean OPE limit, from which we deduce
\bea
e(1-\mbox{$\frac 1{\Lambda^{\alpha} x}$},\Lambda \bar x)\underset{\Lambda\to \infty}{=} O(\Lambda^{\alpha(1-\Df)-1-\epsilon})\label{eq:conste1}
\eea
As for the term involving $\mathcal G(z_e,1-\bar z)$, going to the boundary of integration now corresponds to taking the double lightcone limit. Making the same assumptions as before on the behaviour of the correlator we find
\bea
e(1-\mbox{$\frac 1{\Lambda^{\alpha} x}$},\Lambda \bar x)\underset{\Lambda\to \infty}{=}\left\{\begin{array}{lcll}
O\left(\Lambda^{\alpha(1-\Df)-1-\epsilon}\right)\,,& d=2\,,& \alpha\in[0,1]\,,& \Df\geq 0\\
O\left(\Lambda^{(\alpha-1)(1-\Df)-\epsilon}\right)\,,& d>2\,,& \alpha\in[0,1)\,,&\Df\geq \frac{d-2}2\\
O\left(\Lambda^{-\epsilon}\right)\,,& d>2\,,& \alpha=1\,,& \Df\leq d-2\\
O\left(\Lambda^{\frac{d-2}2-\Df-\epsilon}\right)\,,& d>2\,,& \alpha=1\,,& \Df\geq d-2
\end{array}
\right.
\label{eq:conste22}
\eea

Overall, the dominant condition is \reef{eq:conste1}, i.e. the one arising from the Euclidean OPE limit for general $d$.

\section{Uplifting 1d functionals}
\label{sec:1duplift}
A special class of higher dimensional functionals is given by setting $z=\bar z$ and acting with any 1d functional. We would like to uplift these functionals to higher dimensions using our ansatz. However, it seems that this cannot be achieved without imposing the existence of a different analyticity structure on the kernels. In particular we must look for kernels which have singularities when $z=\bar z$. We have found that the following ansatz does the trick:
\bea
h_{++}(z,\bar z)=\frac{h(z)-h(\bar z)}{\bar z-z},\qquad h_{+-}(z,\bar z)=\frac{h(z)-h(1-\bar z)}{\bar z-z}
\eea
where $h(z)$ is a 1d functional kernel. Although $h_{+-}(z,\bar z)$ has a singularity when $z=\bar z$ this does not affect the real analyticity properties for $z>1,\bar z<0$ where $h_{+-}$ starts off being originally defined.

To see that this functional does reduce to a 1d functional, we start with the functional action definition and perform a contour deformation
\bea
\omega\left[ \mathcal F\right]&=\int_{++} h_{++}(z,\bar z) \mathcal I_z \mathcal I_{\bar z}\mathcal F(z,\bar z)+\int_{+-} h_{+-}(z,\bar z) \mathcal I_z \mathcal I_{\bar z}\mathcal F(z,\bar z)\\
&=\int_{\Gamma_r}\frac{\ud z}{2\pi i} \int_{\Gamma_r'}\frac{\ud \bar z}{2\pi i} \frac{h(z)-h(\bar z)}{\bar z- z} \mathcal F(z,\bar z)-\int_{\Gamma_r}\frac{\ud z}{2\pi i} \int_{\Gamma_l'}\frac{\ud \bar z}{2\pi i} \frac{h(z)-h(1-\bar z)}{\bar z- z} \mathcal F(z,\bar z)
\eea
The primes on the contours indicate an infinitesimal shift. This is to prevent that the two contours overlap, so as to avoid possible singularities at $z=\bar z$. For definiteness we take this shift in an outward direction, i.e. towards the left and right for the $\Gamma_l$ and $\Gamma_r$ contours respectively. Thanks to these shifts we can split each integral above into two pieces. If we combine the first terms of each we get
\bea
\int_{\Gamma_r}\frac{\ud z}{2\pi i} \int_{\Gamma_r'-\Gamma_l'} \frac{\ud \bar z}{2\pi i} \frac{h(z)}{\bar z- z} \mathcal F(z,\bar z)=\int_{\Gamma_r}\frac{\ud z}{2\pi i}  h(z) \mathcal F(z,z)\,.
\eea
where we have used analyticity of $\mathcal F(z,\bar z)$ for $\bar z\in \mathbb C\backslash (-\infty,0)\cup(1,\infty)$.
As for the other two terms, we can show their contribution is vanishing:
\bea
&-\int_{\Gamma_r}\frac{\ud z}{2\pi i} \int_{\Gamma_r'} \frac{\ud \bar z}{2\pi i} \frac{h(\bar z)}{\bar z-z} \mathcal F(z,\bar z)+\int_{\Gamma_r}\frac{\ud z}{2\pi i} \int_{\Gamma_l'}\frac{\ud \bar z}{2\pi i} \frac{h(1-\bar z)}{\bar z-z} \mathcal F(z,\bar z)\\
=&-\int_{\Gamma_r}\frac{\ud z}{2\pi i} \int_{\Gamma_r'} \frac{\ud \bar z}{2\pi i} \frac{h(\bar z)}{\bar z-z} \mathcal F(z,\bar z)+\int_{\Gamma_l}\frac{\ud z}{2\pi i} \int_{\Gamma_r'}\frac{\ud \bar z}{2\pi i} \frac{h(\bar z)}{\bar z-z} \mathcal F(z,\bar z)\\
=&-\int_{\Gamma'_r} \frac{\ud \bar z}{2\pi i} \int_{\Gamma_r-\Gamma_l} \frac{\ud z}{2\pi i} \frac{h(\bar z)}{\bar z-z} \mathcal F(z,\bar z)=0,
\eea
where in the first step we used $\mathcal F(z,\bar z)=-\mathcal F(1-z,1-\bar z)$ and in the last the known analyticity properties of $\mathcal F(z,\bar z)$. So overall the functional action does reduce to a 1d one.

\section{Why HPPS functionals?}
\label{app:hpps}
The goal of this section is to understand how the HPPS functionals precisely relate to the method introduced in the paper \cite{Heemskerk2009} (see also \cite{Heemskerk2010}). We start from the crossing equation for contact interactions:
\begin{multline}
\sum_{n=0}^\infty\sum_{\ell=0}^{L}\left[ a^{(1)}_{n,\ell} G_{\Delta_{n,\ell},\ell}(z,\bar z|\Df)+a^{(0)}_{n,\ell} \gamma_{n,\ell} \partial_{\Delta} G_{\Delta_{n,\ell},\ell}(z,\bar z|\Df)\right]=\\
=\sum_{n=0}^\infty\sum_{\ell=0}^{L}\left[ a^{(1)}_{n,\ell} G_{\Delta_{n,\ell},\ell}(1-z,1-\bar z|\Df)+a^{(0)}_{n,\ell} \gamma_{n,\ell} \partial_{\Delta} G_{\Delta_{n,\ell},\ell}(1-z,1-\bar z|\Df)\right]
\end{multline}
and specialize to the case $d=2$, for which
\bea
G_{\Delta_{n,\ell},\ell}(z,\bar z|\Df)=\frac 1{2 z^\Df \bar z^\Df}\left[G_{\Df+n}(z)G_{\Df+n+\ell}(\bar z)+G_{\Df+n}(\bar z)G_{\Df+n+\ell}(z)\right].
\eea
Now we note that $z^{-\Df} G_{\Df+k}(z)$ with $k$ integer has no branch cut for negative $z$, and
\bea
\oint \frac{\ud z}{2\pi i}\,
\frac {G_{1-\Df-m}(z)}{z^{2-\Df}}\,  \frac{G_{\Df+n}(z)}{z^{\Df}}=\delta_{m,n} \label{eq:onhpps}
\eea 
where the contour is a small anticlockwise circle around $z=0$. Orthonormality follows since both factors are eigenfunctions of the Casimir operator of the conformal group with generically different eigenvalues, except when $m=n$.

We can think of this as applying a linear functional of the form
\bea
\omega[\mathcal F]:=\oint \frac{\ud z}{2\pi i}\,
\frac {G_{1-\Df-m}(z)}{z^{2-\Df}} \mathcal F(z)
\eea
to the crossing equation. The HPPS procedure amounts to applying the product of two such functionals, in $z$ and $\bar z$ around zero and unity respectively, to obtain equations for the anomalous dimensions appearing in the crossing equation.

To make contact with our approach, we begin by noticing that these are not really good functionals as is, since we cannot apply them to arbitrary crossing equations. Indeed generic conformal blocks will have discontinuities for $z<0$ so the above expression for $\omega$ is in general ill defined.  To make it well defined, we can imagine opening up the contour so that it runs along $z \in \frac 12+i \mathbb R$ together with an arc at infinity. If this arc could be dropped we would obtain:
\bea
\omega[\mathcal F]\overset{?}{=}\int_{\Gamma}\frac{\ud z}{2\pi i}\,
\frac {G_{1-\Df-m}(z)}{z^{2-\Df}} \,  \mathcal F(z)\label{eq:hppstry}
\eea
This is the case if $\Df\in[0,1)$ but not more generally. How to improve the behaviour at large $z$?

The $\beta_{p,q}$ functionals give us the answer. Their action on the crossing vector takes the form

\begin{multline}
\beta_{p-1,q-1}(\Delta,\ell)=\frac 12\left[\int_\Gamma \frac{\ud z}{2\pi i} \frac{1}{z^{1+p}}\,\frac{G_{\frac{\Delta+\ell}2}(z)}{z^{\Df}} \int_\Gamma \frac{\ud \bar z}{2\pi i} \frac{1}{(1-\bar z)^{1+q}}\,\frac{G_{\frac{\Delta-\ell}2}(\bar z)}{\bar z^{\Df}}+(\ell\leftrightarrow -\ell)\right]\\-(p\leftrightarrow q)
\end{multline}
We consider functionals which fall off at infinity at least as fast as $1/z$, so that $p,q\geq 0$ here. Focus on the factor
\bea
\int_\Gamma \frac{\ud z}{2\pi i} \frac{1}{z^{1+p}}\,\frac{G_{h}(z)}{z^{\Df}}\,,
\eea
and set $h=\Df+n$. In this case the block has no branch cut for negative $z$ and we may close the contour on the left, picking up the residue at $z=0$. The expression will be non-zero as long as $p\geq n$. It is natural to demand orthonormality, i.e. that it is only non-zero if $p=n$ exactly. This can be achieved by combining functionals with different $p$. Indeed:
\bea
\int_\Gamma \frac{\ud z}{2\pi i} \frac{1}{z}\,\frac{G_{\Df+n}(z)}{z^{\Df}}=\delta_{n,0}\,,\\
\int_\Gamma \frac{\ud z}{2\pi i} \frac{1}{z^{2}}\left(1-\frac{\Df}2 z\right)\,\frac{G_{\Df+n}(z)}{z^{\Df}}=\delta_{n,1}\,,\\
\int_\Gamma \frac{\ud z}{2\pi i} \frac{1}{z^{3}}\left(1-\frac{1+\Df}2 z+\frac{\Df^2(1+\Df)}{4(1+2\Df)} z^2\right)\,\frac{G_{\Df+n}(z)}{z^{\Df}}=\delta_{n,2}\,,\\
\int_\Gamma \frac{\ud z}{2\pi i} f_m(z|\Df)\,\frac{G_{\Df+n}(z)}{z^{\Df}}=\delta_{m,n}\,,
\eea
with
\bea
f_m(z|\Df)=\frac{1}{z} \frac{(-1)^m}{m!} \frac{(\Df)_m^2}{(2\Df+m-1)_m}\, _3F_2\left(1,-m,-1+2\Df+m;\Df,\Df;\frac 1z\right)
\eea
The orthonormality relations above hold for $n\geq 0$. Note that all kernels $f_m(z)$ behave for large $z$ as $1/z$. We can improve the fall off by relinquishing some of the orthonormality conditions, defining
\bea
f_m^{(p)}(z|\Df)=\frac 1{z^p} f_m(z,\Df+p)\,.
\eea
The modified kernels $f_m^{(p)}$ will then satisfy orthonormality only for $m,n\geq p$. So for instance, if we wanted to orthonormalise the ordinary $\beta_{p,q}$ functionals which fall off at least as fast as $1/z^2$, we would use $f_m^{(1)}$ instead of $f_m^{(0)}$.
 
We now notice that we can rewrite
\bea
f_m(z|\Df)&=\frac {G_{1-\Df-m}(z)- C_m(z|\Df)}{z^{2-\Df}}\\
C_m(z|\Df)&=\frac{ \Gamma (2-\Df)^2 \Gamma
   \left(\frac{3}{2}-\Df-m\right)}{\sqrt{\pi } 2^{2 \Df+2 n-1}}\frac{\, _3F_2\left(1,2-\Df,2-\Df;3-2\Df-m,2+m;z\right)}{\Gamma (2+m) \Gamma (3-2
   \Df-m) \Gamma (1-\Df-m)}\,.
\eea
Overall we see that these functionals take the form of \reef{eq:hppstry} together with a correction. The correction does not change the orthonormality relations \reef{eq:onhpps}, but it is such that  the functional kernel is now better behaved at infinity (since in particular it is $O(z^{-1})$ for all $\Df$). Technically orthonormality follows because we have
\begin{multline}
\left[\mathcal C_2-(1-\Df-m)(\Df-m)\right]\left[z^{2-\Df}\,f_m(z|\Df)\right]=\\
=\left[\mathcal C_2-(1-\Df-m)(\Df-m)\right]C_m(z|\Df)=c_m z^{2-\Df}
\end{multline}
where $\mathcal C_2$ is the Casimir operator of the $d=1$ conformal group\footnote{In detail,%
\bea
\mathcal C_2=(1-z)z^2 \partial_z^2-z^2 \partial z\,.
\eea
}
and $c_m$ is some constant. Even though $C_m(z|\Df)$ is therefore not an eigenfunction of the Casimir operator, it is close enough, since the remainder on the righthand side is purely normal:
\bea
\oint \frac{\ud z}{2\pi i} \frac{G_{\Df+n}(z)}{z^\Df}=0,\qquad \mbox{for all}\quad n\geq 0\,.
\eea

To summarize, the (orthonormalized) $\beta_{p,q}$ functionals are indeed intimately related to the HPPS procedure: they are essentially modifications of the naive HPPS functionals given by \reef{eq:hppstry}, modifications which make them well defined when acting on general CFT correlators.

\section{Comments on product functionals for $d=4$}
\label{app:4dprod}
We now make some comments on why the product functionals can exist for $d=2$ and what happens for $d=4$. We apply our functional ansatz to the crossing vector $F_{\Delta,\ell}$ and try to mimick the 1d procedure as in appendix \ref{sec:bases} so as to get functional actions with integer spaced zeros. The idea is to rotate contours judiciously so as to pick up contributions from those cuts of the conformal blocks which involve phases. Starting from \reef{eq:fgerep} we get:
\begin{equation}
\label{eq:fgeaction}
\begin{split}
\omega[F_{\Delta,\ell}]=&-\int_{--}\,  \mathcal I_{\bar z} \mathcal I_z\left[ f(1-z,1-\bar z) G_{\Delta,\ell}(z,\bar z|\Df)\right]\\
&+\int_{0-}  \, \mathcal I_{\bar z}\left[ e(z,\bar z) G_{\Delta,\ell}(z,\bar z|\Df)\right]
+\int_{-0}  \, \mathcal I_{z}\left[ \bar e(z,\bar z) G_{\Delta,\ell}(z,\bar z|\Df)\right]\\
&+\int_{00} g(z,\bar z) G_{\Delta,\ell}(z,\bar z|\Df)\,.
\end{split}
\end{equation}
Interestingly $\tilde g(z,\bar z)$ drops out from the final expression. The reason why it is possible to get simple functional actions for $d=2$ boils down to the fact
\bea
G_{\Delta,\ell}^{d=2}(z,\bar z)=e^{i\frac{\pi\tau}2} G_{\Delta,\ell}^{d=2}(\mbox{$\frac{z}{z-1}$},\bar z)\,,\qquad \ell\quad \mbox{even},\qquad \mbox{Im}\,z>0\,,
\eea
which allows us to rewrite the terms involving $e,\bar e$ in terms of the blocks themselves up to a transformation in $z$ or $\bar z$ and multiplication by a phase. In general spacetime dimension this identity does not hold. However, from the expression for the blocks themselves,
\bea
G_{\Delta,\ell}^{d=4}(z,\bar z)=\frac{1}{1+\ell}\,\frac{z \bar z}{z-\bar z}\left[k_{\rho}(z) k_{\tau-2}(\bar z)-k_{\tau-2}(z)k_{\rho}(\bar z)\right]\,.
\eea
We see that after multiplying by an appropriate prefactor, it is a sum of two terms, each of which will satisfy a similar relation to the one above, so there is a hope to have the product type functionals working here too.

This is easiest to see from the crossing vectors, which can be written as:
\bea
 F_{\Delta,\ell}(z,\bar z|\Df)=\frac{1}{2(1+\ell)}\frac{H_{\rho}(z|\Df-1) H_{\tau-2}(\bar z|\Df-1)+F_{\rho}(z|\Df-1) F_{\tau-2}(\bar z|\Df-1)-(z\leftrightarrow \bar z)}{z-\bar z}
\eea
From this expression we see that we should take functional kernels which multiply by $z-\bar z$ and then either symmetrize or antisymmetrize in $z$, followed by holomorphic and antiholomorphic actions of $d=1$ functionals in $z$ and $\bar z$. There are hence two sets of product functionals:
\bea
(\omega_{\pm}^{(1)}\otimes \omega_{\pm}^{(2)})(\Delta,\ell):=2\int_{++} \frac{\ud z \ud \bar z}{\pi^2}\,  h_{\pm}^{(1)}(z)h_{\pm}^{(2)}(\bar z)(z-\bar z)\,\left[ \mathcal I_z \mathcal I_{\bar z} F_{\Delta,\ell}(z,\bar z)\pm\mathcal I_z \mathcal I_{\bar z} F_{\Delta,\ell}(z,1-\bar z)\right]\,.\label{eq:4dfunc}
\eea
That is, the functional kernels are chosen as:
\bea
h_{++}(z,\bar z)&=\left[h_{\pm}(z)^{(1)} h_{\pm}^{(2)}(\bar z)-h_\pm^{(1)}(\bar z)h_\pm^{(2)}(z)\right](z-\bar z)\,,\\
h_{+-}(z,\bar z)&=\pm \left[h_{\pm}(z)^{(1)} h_{\pm}^{(2)}(1-\bar z) -h_{\pm}^{(1)}(z) h_{\pm}^{(2)}(1-\bar z)\right](z-\bar z)\,.
\eea
From the form of the 4d crossing vectors we should take 1d functional kernels appropriate for acting on 1d crossing vectors with:
\bea
\Delta_{\phi}^{d=1}=\frac{\Delta_\phi^{d=4}-1}2\,.
\eea 
In the 2d case we had $\Df^{d=1}=\Delta_{\phi}^{d=2}/2$. Accordingly, let us for the purposes of this appendix redefine 1d functional actions as
\bea
\omega\left(\frac{h}2\bigg|\frac{\Df-1}2\right) \rightarrow \omega(h|\Df)\,.
\eea
Then the two sets of functionals have functional actions are given by
\bea
(\omega_{\pm}^{(1)}\otimes \omega_{\pm}^{(2)})(\Delta,\ell|\Df)=\omega^{(1)}_{\pm}(\rho|\Df)\omega^{(2)}_{\pm}(\tau-2|\Df)-\omega^{(2)}_{\pm}(\rho|\Df)\omega^{(1)}_{\pm}(\tau-2|\Df)\label{eq:funcaction4d}
\eea
As in the $d=2$ case, it is natural to take for the 1d functionals the bases discussed in section \ref{sec:review}. However, it is no longer true that any choice of such 1d functionals leads to crossing-compatible functionals in 4d. This is because of the factor $z-\bar z$ in \reef{eq:4dfunc}. The 1d functional kernels $f_{\pm}(z)$ (which are related to $h_{\pm}(z)$, see \ref{sec:bases1}) satisfy
\bea
z f_+(z)&\underset{z\to \infty}{=} O(z^{-2})\\
z f_-(z)&\underset{z\to \infty}{=} O(z^{-1})
\eea
The constraints of section \ref{sec:constraints} imply here that the fall-off must be strictly faster than $z^{-1}$. Hence, while any choice of $+$ type functionals works, this is not so for the $-$ type. The solution in the latter is to take differences of functionals such that the behaviour at infinity is improved:
\bea
\beta_n^-\to \tilde \beta_n^-\equiv \beta_n-b_n \beta_0\,,\qquad \mbox{$b_n$ such that}\quad f_{\tilde \beta_n^-}(z)&\underset{z\to \infty}{=} O(z^{-4})\,,\\
\alpha_n^-\to \tilde \alpha_n^-\equiv \alpha_n-a_n \beta_0\,,\qquad \mbox{$a_n$ such that}\quad f_{\tilde \beta_n^-}(z)&\underset{z\to \infty}{=} O(z^{-4})\,.
\eea
Hence we have six sets of possible product functionals:
\bea
\beta^+_n \otimes \beta^+_m,&\quad n,m \geq 0, \quad n\neq m\\
\alpha^+_n \otimes \alpha^+_m,&\quad n,m \geq 0,\quad n\neq m\\
\beta^+_n \otimes \alpha^+_m,&\quad n,m \geq 0,\\
\tilde \beta^-_n \otimes \tilde \beta^-_m,&\quad n,m \geq 1, \quad n\neq m\\
\tilde \alpha^-_n \otimes \tilde \alpha^-_m,&\quad n,m \geq 0,\quad n\neq m\\
\tilde \beta^-_n \otimes \tilde \alpha^-_m,&\quad n\geq 1,\ m \geq 0\,,
\eea
with functional actions determined by \reef{eq:funcaction4d}. Interestingly, because of the shift in $\tau$ by 2 in that expression, these actions generally have second order zeros on the generalized free field spectrum. Unfortunately the annoying relative minus sign in the same expression means that it is harder to get functional actions with sufficiently nice positivity properties, that would for instance allow us to get interesting bounds. Furthermore the subtraction procedure above also strongly suggests that this set of functionals is not complete. We leave more extensive explorations of these interesting functionals for future work.

\small
\bibliography{biblio}

\providecommand{\href}[2]{#2}\begingroup\raggedright\begin{thebibliography}{10}

\bibitem{Rattazzi:2008pe}
R.~Rattazzi, V.~S. Rychkov, E.~Tonni, and A.~Vichi, ``{Bounding scalar operator
  dimensions in 4D CFT},''
  \href{http://dx.doi.org/10.1088/1126-6708/2008/12/031}{{\em JHEP} {\bfseries
  12} (2008) 031},
\href{http://arxiv.org/abs/0807.0004}{{\ttfamily arXiv:0807.0004 [hep-th]}}.

\bibitem{Hellerman2011}
S.~Hellerman, ``{A Universal Inequality for CFT and Quantum Gravity},''
  \href{http://dx.doi.org/10.1007/JHEP08(2011)130}{{\em JHEP} {\bfseries 08}
  (2011) 130},
\href{http://arxiv.org/abs/0902.2790}{{\ttfamily arXiv:0902.2790 [hep-th]}}.

\bibitem{Poland:2018epd}
D.~Poland, S.~Rychkov, and A.~Vichi, ``{The Conformal Bootstrap: Theory,
  Numerical Techniques, and Applications},''
\href{http://arxiv.org/abs/1805.04405}{{\ttfamily arXiv:1805.04405 [hep-th]}}.

\bibitem{Pappadopulo:2012jk}
D.~Pappadopulo, S.~Rychkov, J.~Espin, and R.~Rattazzi, ``{OPE Convergence in
  Conformal Field Theory},''
  \href{http://dx.doi.org/10.1103/PhysRevD.86.105043}{{\em Phys. Rev.}
  {\bfseries D86} (2012) 105043},
\href{http://arxiv.org/abs/1208.6449}{{\ttfamily arXiv:1208.6449 [hep-th]}}.

\bibitem{Mukhametzhanov:2019pzy}
B.~Mukhametzhanov and A.~Zhiboedov, ``{Modular Invariance, Tauberian Theorems,
  and Microcanonical Entropy},''
\href{http://arxiv.org/abs/1904.06359}{{\ttfamily arXiv:1904.06359 [hep-th]}}.

\bibitem{Mukhametzhanov:2018zja}
B.~Mukhametzhanov and A.~Zhiboedov, ``{Analytic Euclidean Bootstrap},''
\href{http://arxiv.org/abs/1808.03212}{{\ttfamily arXiv:1808.03212 [hep-th]}}.

\bibitem{Qiao:2017xif}
J.~Qiao and S.~Rychkov, ``{A tauberian theorem for the conformal bootstrap},''
  \href{http://dx.doi.org/10.1007/JHEP12(2017)119}{{\em JHEP} {\bfseries 12}
  (2017) 119},
\href{http://arxiv.org/abs/1709.00008}{{\ttfamily arXiv:1709.00008 [hep-th]}}.

\bibitem{Pal2019a}
S.~Pal and Z.~Sun, ``{Tauberian-Cardy formula with spin},''
\href{http://arxiv.org/abs/1910.07727}{{\ttfamily arXiv:1910.07727 [hep-th]}}.

\bibitem{Fitzpatrick2014}
A.~L. Fitzpatrick, J.~Kaplan, and M.~T. Walters, ``{Universality of
  Long-Distance AdS Physics from the CFT Bootstrap},''
  \href{http://dx.doi.org/10.1007/JHEP08(2014)145}{{\em JHEP} {\bfseries 08}
  (2014) 145},
\href{http://arxiv.org/abs/1403.6829}{{\ttfamily arXiv:1403.6829 [hep-th]}}.

\bibitem{Komargodski2013}
Z.~Komargodski and A.~Zhiboedov, ``{Convexity and Liberation at Large Spin},''
  \href{http://dx.doi.org/10.1007/JHEP11(2013)140}{{\em JHEP} {\bfseries 11}
  (2013) 140},
\href{http://arxiv.org/abs/1212.4103}{{\ttfamily arXiv:1212.4103 [hep-th]}}.

\bibitem{Collier2019a}
S.~Collier, Y.~Gobeil, H.~Maxfield, and E.~Perlmutter, ``{Quantum Regge
  Trajectories and the Virasoro Analytic Bootstrap},''
  \href{http://dx.doi.org/10.1007/JHEP05(2019)212}{{\em JHEP} {\bfseries 05}
  (2019) 212},
\href{http://arxiv.org/abs/1811.05710}{{\ttfamily arXiv:1811.05710 [hep-th]}}.

\bibitem{Alday:2015ewa}
L.~F. Alday and A.~Zhiboedov, ``{An Algebraic Approach to the Analytic
  Bootstrap},'' \href{http://dx.doi.org/10.1007/JHEP04(2017)157}{{\em JHEP}
  {\bfseries 04} (2017) 157},
\href{http://arxiv.org/abs/1510.08091}{{\ttfamily arXiv:1510.08091 [hep-th]}}.

\bibitem{Alday:2016njk}
L.~F. Alday, ``{Large Spin Perturbation Theory for Conformal Field Theories},''
  \href{http://dx.doi.org/10.1103/PhysRevLett.119.111601}{{\em Phys. Rev.
  Lett.} {\bfseries 119} no.~11, (2017) 111601},
\href{http://arxiv.org/abs/1611.01500}{{\ttfamily arXiv:1611.01500 [hep-th]}}.

\bibitem{Caron-Huot:2017vep}
S.~Caron-Huot, ``{Analyticity in Spin in Conformal Theories},''
  \href{http://dx.doi.org/10.1007/JHEP09(2017)078}{{\em JHEP} {\bfseries 09}
  (2017) 078},
\href{http://arxiv.org/abs/1703.00278}{{\ttfamily arXiv:1703.00278 [hep-th]}}.

\bibitem{Simmons-Duffin:2016wlq}
D.~Simmons-Duffin, ``{The Lightcone Bootstrap and the Spectrum of the 3d Ising
  CFT},'' \href{http://dx.doi.org/10.1007/JHEP03(2017)086}{{\em JHEP}
  {\bfseries 03} (2017) 086},
\href{http://arxiv.org/abs/1612.08471}{{\ttfamily arXiv:1612.08471 [hep-th]}}.

\bibitem{Simmons-Duffin:2017nub}
D.~Simmons-Duffin, D.~Stanford, and E.~Witten, ``{A spacetime derivation of the
  Lorentzian OPE inversion formula},''
  \href{http://dx.doi.org/10.1007/JHEP07(2018)085}{{\em JHEP} {\bfseries 07}
  (2018) 085},
\href{http://arxiv.org/abs/1711.03816}{{\ttfamily arXiv:1711.03816 [hep-th]}}.

\bibitem{Polyakov:1974gs}
A.~M. Polyakov, ``{Nonhamiltonian approach to conformal quantum field
  theory},'' {\em Zh. Eksp. Teor. Fiz.} {\bfseries 66} (1974) 23--42.
[Sov. Phys. JETP39,9(1974)].

\bibitem{Sen:2015doa}
K.~Sen and A.~Sinha, ``{On critical exponents without Feynman diagrams},''
  \href{http://dx.doi.org/10.1088/1751-8113/49/44/445401}{{\em J. Phys.}
  {\bfseries A49} no.~44, (2016) 445401},
\href{http://arxiv.org/abs/1510.07770}{{\ttfamily arXiv:1510.07770 [hep-th]}}.

\bibitem{Gopakumar:2016wkt}
R.~Gopakumar, A.~Kaviraj, K.~Sen, and A.~Sinha, ``{Conformal Bootstrap in
  Mellin Space},'' \href{http://dx.doi.org/10.1103/PhysRevLett.118.081601}{{\em
  Phys. Rev. Lett.} {\bfseries 118} no.~8, (2017) 081601},
\href{http://arxiv.org/abs/1609.00572}{{\ttfamily arXiv:1609.00572 [hep-th]}}.

\bibitem{Gopakumar:2016cpb}
R.~Gopakumar, A.~Kaviraj, K.~Sen, and A.~Sinha, ``{A Mellin space approach to
  the conformal bootstrap},''
  \href{http://dx.doi.org/10.1007/JHEP05(2017)027}{{\em JHEP} {\bfseries 05}
  (2017) 027},
\href{http://arxiv.org/abs/1611.08407}{{\ttfamily arXiv:1611.08407 [hep-th]}}.

\bibitem{Gopakumar:2018xqi}
R.~Gopakumar and A.~Sinha, ``{On the Polyakov-Mellin bootstrap},''
  \href{http://dx.doi.org/10.1007/JHEP12(2018)040}{{\em JHEP} {\bfseries 12}
  (2018) 040},
\href{http://arxiv.org/abs/1809.10975}{{\ttfamily arXiv:1809.10975 [hep-th]}}.

\bibitem{Mazac2019a}
D.~Mazac and M.~F. Paulos, ``{The analytic functional bootstrap. Part II.
  Natural bases for the crossing equation},''
  \href{http://dx.doi.org/10.1007/JHEP02(2019)163}{{\em JHEP} {\bfseries 02}
  (2019) 163},
\href{http://arxiv.org/abs/1811.10646}{{\ttfamily arXiv:1811.10646 [hep-th]}}.

\bibitem{Mazac:2016qev}
D.~Mazac, ``{Analytic bounds and emergence of AdS$_{2}$ physics from the
  conformal bootstrap},'' \href{http://dx.doi.org/10.1007/JHEP04(2017)146}{{\em
  JHEP} {\bfseries 04} (2017) 146},
\href{http://arxiv.org/abs/1611.10060}{{\ttfamily arXiv:1611.10060 [hep-th]}}.

\bibitem{Mazac:2018}
D.~Mazac and M.~F. Paulos, ``{The analytic functional bootstrap. Part I: 1D
  CFTs and 2D S-matrices},''
  \href{http://dx.doi.org/10.1007/JHEP02(2019)162}{{\em JHEP} {\bfseries 02}
  (2019) 162},
\href{http://arxiv.org/abs/1803.10233}{{\ttfamily arXiv:1803.10233 [hep-th]}}.

\bibitem{Kaviraj2018}
A.~Kaviraj and M.~F. Paulos, ``{The Functional Bootstrap for Boundary CFT},''
\href{http://arxiv.org/abs/1812.04034}{{\ttfamily arXiv:1812.04034 [hep-th]}}.

\bibitem{MazacOPE2018}
D.~Mazac, ``{A Crossing-Symmetric OPE Inversion Formula},''
\href{http://arxiv.org/abs/1812.02254}{{\ttfamily arXiv:1812.02254 [hep-th]}}.

\bibitem{Mazac:2018biw}
D.~Mazac, L.~Rastelli, and X.~Zhou, ``{An Analytic Approach to BCFT$_d$},''
\href{http://arxiv.org/abs/1812.09314}{{\ttfamily arXiv:1812.09314 [hep-th]}}.

\bibitem{Paulos:2019fkw}
M.~F. Paulos and B.~Zan, ``{A functional approach to the numerical conformal
  bootstrap},''
\href{http://arxiv.org/abs/1904.03193}{{\ttfamily arXiv:1904.03193 [hep-th]}}.

\bibitem{Hartman2019b}
T.~Hartman, D.~Mazac, and L.~Rastelli, ``{Sphere Packing and Quantum
  Gravity},''
\href{http://arxiv.org/abs/1905.01319}{{\ttfamily arXiv:1905.01319 [hep-th]}}.

\bibitem{ElShowk:2012hu}
S.~El-Showk and M.~F. Paulos, ``{Bootstrapping Conformal Field Theories with
  the Extremal Functional Method},''
  \href{http://dx.doi.org/10.1103/PhysRevLett.111.241601}{{\em Phys. Rev.
  Lett.} {\bfseries 111} no.~24, (2013) 241601},
\href{http://arxiv.org/abs/1211.2810}{{\ttfamily arXiv:1211.2810 [hep-th]}}.

\bibitem{Gliozzi2013}
F.~Gliozzi, ``{More constraining conformal bootstrap},''
  \href{http://dx.doi.org/10.1103/PhysRevLett.111.161602}{{\em Phys. Rev.
  Lett.} {\bfseries 111} (2013) 161602},
\href{http://arxiv.org/abs/1307.3111}{{\ttfamily arXiv:1307.3111 [hep-th]}}.

\bibitem{El-Showk:2016mxr}
S.~El-Showk and M.~F. Paulos, ``{Extremal bootstrapping: go with the flow},''
  \href{http://dx.doi.org/10.1007/JHEP03(2018)148}{{\em JHEP} {\bfseries 03}
  (2018) 148},
\href{http://arxiv.org/abs/1605.08087}{{\ttfamily arXiv:1605.08087 [hep-th]}}.

\bibitem{Heemskerk2009}
I.~Heemskerk, J.~Penedones, J.~Polchinski, and J.~Sully, ``{Holography from
  Conformal Field Theory},''
  \href{http://dx.doi.org/10.1088/1126-6708/2009/10/079}{{\em JHEP} {\bfseries
  10} (2009) 079},
\href{http://arxiv.org/abs/0907.0151}{{\ttfamily arXiv:0907.0151 [hep-th]}}.

\bibitem{Dolan:2011dv}
F.~A. Dolan and H.~Osborn, ``{Conformal Partial Waves: Further Mathematical
  Results},''
\href{http://arxiv.org/abs/1108.6194}{{\ttfamily arXiv:1108.6194 [hep-th]}}.

\bibitem{Dolan:2003hv}
F.~A. Dolan and H.~Osborn, ``{Conformal partial waves and the operator product
  expansion},'' \href{http://dx.doi.org/10.1016/j.nuclphysb.2003.11.016}{{\em
  Nucl. Phys.} {\bfseries B678} (2004) 491--507},
\href{http://arxiv.org/abs/hep-th/0309180}{{\ttfamily arXiv:hep-th/0309180
  [hep-th]}}.

\bibitem{Dolan:2000ut}
F.~A. Dolan and H.~Osborn, ``{Conformal four point functions and the operator
  product expansion},''
  \href{http://dx.doi.org/10.1016/S0550-3213(01)00013-X}{{\em Nucl. Phys.}
  {\bfseries B599} (2001) 459--496},
\href{http://arxiv.org/abs/hep-th/0011040}{{\ttfamily arXiv:hep-th/0011040
  [hep-th]}}.

\bibitem{Kos:2013tga}
F.~Kos, D.~Poland, and D.~Simmons-Duffin, ``{Bootstrapping the $O(N)$ vector
  models},'' \href{http://dx.doi.org/10.1007/JHEP06(2014)091}{{\em JHEP}
  {\bfseries 06} (2014) 091},
\href{http://arxiv.org/abs/1307.6856}{{\ttfamily arXiv:1307.6856 [hep-th]}}.

\bibitem{Fitzpatrick2012}
A.~L. Fitzpatrick and J.~Kaplan, ``{Unitarity and the Holographic S-Matrix},''
  \href{http://dx.doi.org/10.1007/JHEP10(2012)032}{{\em JHEP} {\bfseries 10}
  (2012) 032},
\href{http://arxiv.org/abs/1112.4845}{{\ttfamily arXiv:1112.4845 [hep-th]}}.

\bibitem{Rychkov:2017tpc}
J.~Qiao and S.~Rychkov, ``{Cut-touching linear functionals in the conformal
  bootstrap},'' \href{http://dx.doi.org/10.1007/JHEP06(2017)076}{{\em JHEP}
  {\bfseries 06} (2017) 076},
\href{http://arxiv.org/abs/1705.01357}{{\ttfamily arXiv:1705.01357 [hep-th]}}.

\bibitem{Kusuki2019}
Y.~Kusuki, ``{Light Cone Bootstrap in General 2D CFTs and Entanglement from
  Light Cone Singularity},''
  \href{http://dx.doi.org/10.1007/JHEP01(2019)025}{{\em JHEP} {\bfseries 01}
  (2019) 025},
\href{http://arxiv.org/abs/1810.01335}{{\ttfamily arXiv:1810.01335 [hep-th]}}.

\bibitem{Penedones2011}
J.~Penedones, ``{Writing CFT correlation functions as AdS scattering
  amplitudes},'' \href{http://dx.doi.org/10.1007/JHEP03(2011)025}{{\em JHEP}
  {\bfseries 03} (2011) 025},
\href{http://arxiv.org/abs/1011.1485}{{\ttfamily arXiv:1011.1485 [hep-th]}}.

\bibitem{Fitzpatrick2013}
A.~L. Fitzpatrick, J.~Kaplan, D.~Poland, and D.~Simmons-Duffin, ``{The Analytic
  Bootstrap and AdS Superhorizon Locality},''
  \href{http://dx.doi.org/10.1007/JHEP12(2013)004}{{\em JHEP} {\bfseries 12}
  (2013) 004},
\href{http://arxiv.org/abs/1212.3616}{{\ttfamily arXiv:1212.3616 [hep-th]}}.

\bibitem{Hogervorst2013}
M.~Hogervorst and S.~Rychkov, ``{Radial Coordinates for Conformal Blocks},''
  \href{http://dx.doi.org/10.1103/PhysRevD.87.106004}{{\em Phys. Rev.}
  {\bfseries D87} (2013) 106004},
\href{http://arxiv.org/abs/1303.1111}{{\ttfamily arXiv:1303.1111 [hep-th]}}.

\bibitem{Heemskerk2010}
I.~Heemskerk and J.~Sully, ``{More Holography from Conformal Field Theory},''
  \href{http://dx.doi.org/10.1007/JHEP09(2010)099}{{\em JHEP} {\bfseries 09}
  (2010) 099},
\href{http://arxiv.org/abs/1006.0976}{{\ttfamily arXiv:1006.0976 [hep-th]}}.

\end{thebibliography}\endgroup
\bibliographystyle{utphys}

\end{document}